\documentclass[floats,floatfix,amssymb,prd,aps,twocolumn,nofootinbib,nolongbibliography,reprint,superscriptaddress]{revtex4-2}

\usepackage{ulem} 

\usepackage{amssymb,amsmath,verbatim,mathtools,needspace,enumitem,etoolbox,graphicx,physics,microtype,afterpage,xspace,tabularx,lmodern,multirow}
\usepackage{gensymb}
\usepackage[dvipappesnames, usenames]{xcolor}
\usepackage[unicode, colorlinks=true, linkcolor=linkcolor, citecolor=linkcolor, filecolor=linkcolor, urlcolor=linkcolor, linktocpage, breaklinks]{hyperref}
\newcommand{\orcid}[1]{\href{https://orcid.org/#1}{\includegraphics[width=10pt]{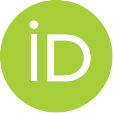}}}
\usepackage[nolist,nohyperlinks]{acronym}
\usepackage{tikz}
\usetikzlibrary{decorations.markings}
\usetikzlibrary{decorations.pathmorphing}

\definecolor{linkcolor}{rgb}{0.0,0.3,0.5}
\definecolor{olive}{rgb}{0.4,0.7,0.2}
\def\RWZ{{\texttt{RWZHyp}}}

\allowdisplaybreaks

\begin{document}

\pagenumbering{arabic}


\title{Inspiral-inherited ringdown tails}

\author{Marina De Amicis \orcid{0000-0003-0808-3026}}
\email{marina.de.amicis@nbi.ku.dk}
\affiliation{Niels Bohr International Academy, Niels Bohr Institute, Blegdamsvej 17, 2100 Copenhagen, Denmark}
\author{Simone Albanesi \orcid{0000-0001-7345-4415}}
\email{simone.albanesi@uni-jena.de}
\affiliation{Niels Bohr International Academy, Niels Bohr Institute, Blegdamsvej 17, 2100 Copenhagen, Denmark}
\affiliation{INFN sezione di Torino, Torino, 10125, Italy}
\affiliation{Dipartimento di Fisica, Universit\`a di Torino, Torino, 10125, Italy}
\affiliation{Theoretisch-Physikalisches Institut, Friedrich-Schiller-Universit{\"a}t Jena, 07743, Jena, Germany}
\author{Gregorio Carullo \orcid{0000-0001-9090-1862}}
\email{g.carullo@bham.ac.uk}
\affiliation{Niels Bohr International Academy, Niels Bohr Institute, Blegdamsvej 17, 2100 Copenhagen, Denmark}
\affiliation{School of Physics and Astronomy and Institute for Gravitational Wave Astronomy, University of Birmingham, Edgbaston, Birmingham, B15 2TT, United Kingdom}

\begin{abstract}

    We study the late-time relaxation of a perturbed Schwarzschild black hole, driven by a source term representing an infalling particle in generic orbits.
    We consider quasi-circular and eccentric binaries, dynamical captures and radial infalls, with orbital dynamics driven by an highly accurate analytical radiation reaction.
    After reviewing the description of the late-time behaviour as an integral over the whole inspiral history, we derive an analytical expression that reproduces the slow relaxation (``tail'') observed in our numerical evolutions, obtained with a hyperboloidal compactified grid, for a given non-circular particle trajectory.
    We find this signal to be a superposition of an infinite number of power-laws, the slowest decaying term being Price's law.
    Next, we use our model to explain the several orders-of-magnitude enhancement of tail terms for binaries in non-circular orbits, shedding light on recent unexpected results obtained in numerical evolutions.
    In particular, we show the dominant terms controlling the enhancement to be activated when the particle is far from the black hole, with small tangential and radial velocities soon before the plunge.
    As we corroborate with semi-analytical calculations, this implies that for large eccentricities the tail amplitude can be correctly extracted even when starting to evolve only from the last apastron before merger.
    We discuss the implications of these findings on the extraction of late-time tail terms in non-linear evolutions, and possible observational consequences.
    We also briefly comment on the scattering scenario, and on the connection with the soft graviton theorem.
    
\end{abstract}

\maketitle

\tableofcontents

\section{Introduction}

%
Binary black hole (BH) mergers constitute an ideal playground to study the many intricate aspects of the general relativistic two-body problem.
Historically, theoretical investigations of the entire evolution of these systems have drawn significant attention due to their intrinsic mathematical complexity and physical relevance.
Nowadays, thanks to the advent of gravitational-wave (GW) astronomy~\cite{LIGOScientific:2016aoc}, such studies have gained extreme importance even from the observational point of view, allowing to predict with high accuracy astrophysical GW signals observed with ever-increasing precision.

In general, it is possible to divide a binary merger in three distinct phases: inspiral, plunge-merger and post-merger.
In the inspiral phase, the two progenitors are in relative motion along a decaying orbit, losing energy and angular momentum due to the emission of GWs. 
The plunge takes place when stable orbits can no longer be sustained due to the reduced angular momentum of the system, and the two BHs fuse together, giving rise to a ``remnant'' BH possessing a dynamical horizon.
After the coalescence, the newly born remnant BH quickly relaxes towards a stationary Kerr configuration.
The relaxation signal is initially dominated by a transient which depends on the binary initial conditions, and then by a superposition of exponentially damped sinusoids, the quasi-normal modes (QNMs) of the system.
Instead, at very late times, the GW strain decays in a power-law fashion, a regime also amenable to a description based on perturbation theory (PT), which will be the focus of our investigations.
Because of the signal morphology, such late time behaviour is commonly denoted as ``tail'', and arises from corrections to the propagation of a signal on a flat light-cone, due to back-scattering taking place against the long-range curvature of the underlying spacetime.

The tail was predicted for the first time by Price~\cite{Price:1971fb}, studying GWs signal associated to dynamical collapse.
In particular, for non-static initial data, a power-law decay of $t^{-2\ell-3}$ is observed at large finite distances, where $\ell$ is the waveform multipole.
Later, Leaver~\cite{Leaver:1986gd} studied the signal observed at future null infinity ($\mathcal{I}^+$) after perturbing a Schwarzschild BH.
At late retarded times, such signal can be described by a single power-law $\tau^{-\ell-2}$ ($\tau^{-\ell-3}$) for stationary (static) initial conditions, with $\tau\equiv t-r_*$ retarded time.
Corrections to this result were computed by Andersson~\cite{Andersson:1996cm}, resulting in a series of decaying power-laws contributing at intermediate times, and suppressed at asymptotically late times. 
These predictions are confirmed by linear numerical experiments performed in vacuum, see e.g. Ref.~\cite{Zenginoglu:2009ey}.
For a more extensive review of past literature on tails, including the rotating case and a large body of numerical studies, see the introduction of Ref.~\cite{Carullo:2023tff}.
In the presence of matter, this back-scattering problem was also studied analytically by Blanchet and Damour~\cite{Blanchet:1992br, Blanchet:1993ec, Blanchet:1994ez} in the context of Multipolar Post-Minkowskian (MPM) theory and by Poisson et al.~\cite{Poisson:1993vp, Poisson:1994yf}.
In particular, the former investigations showed that the tail is an \textit{hereditary} effect carrying information on the entire history of the system.
These works focused on the inspiral stage, while little attention has been paid to hereditary effects on the post-merger phase of a binary merger.

Recently, Ref.~\cite{Albanesi:2023bgi}
performed numerical evolutions of binary mergers in generic orbits within a perturbative setting, incorporating radiation-reaction effects through an analytical expression based on post-Newtonian (PN) results, and resummed according to Effective One Body (EOB) techniques~\cite{Chiaramello:2020ehz}.
This study unexpectedly found an enhancement of several orders of magnitude of the tail amplitude when increasing the progenitors' binary eccentricity, resulting in an earlier transition from a QNMs to a tail-dominated regime.
The search for such enhancement in comparable masses non-linear evolutions was started in Ref.~\cite{Carullo:2023tff}.
This surprising result has not yet found an explanation in the literature, consistent with the fact that, to the best of our knowledge, an explicit modelisation of hereditary contributions to the post-ringdown signal of binary mergers has never been put forward so far.
This is the scope of the present work.

We derive an explicit integral formula capable of predicting the aforementioned late-time enhancement, connecting tail terms to properties of the test-particle non-circular motion in the inspiral, and matching the eccentricity dependence recently found within numerical evolutions. 
Our expressions for the \textit{source-driven} tail are relevant to any kind of nonspinning binary merger, as we showcase by applying them not only to eccentric binaries, but also to dynamical captures and radial infalls.
We find a much more complex behaviour compared to the predictions of source-free PT, with a tail exponent non-monotonic variation at intermediate times, due to a superposition of a large number of exact power-laws.
In the asymptotic $\tau\rightarrow\infty$ limit, homogeneous PT results are instead recovered.
In particular, asymptotic perturbations of systems that become bounded and eventually merge behave as Price's law ($\tau^{-\ell-2}$).
Finally, to single out the reason behind tail terms eccentricity enhancement, we carry out two additional sets of investigations. 
First, we study changes in the tail when integrating over different portions of the inspiral motion, isolating the dominant contribution to the tail excitation, and characterising the key role of the motion around the last apastron.
Second, through an expansion in large $r$ and small tangential velocities, we show how an eccentric binary, which spends a larger fraction of time at large distances just before merger, can emit tail signals that are both enhanced and constructively interfere with each other.

The paper is structured as follows.
First, in Sec.~\ref{sec:RWZ}, we introduce our perturbative framework and discuss the \RWZ{} code, used to evolve the binary system and numerically solve for the emitted GW strain.
Then, in Sec.~\ref{sec:analytic_model}, we present the analytical model of the source-driven tail. 
Sec.~\ref{sec:comparision} is dedicated to test the model predictions against numerical evolutions of eccentric binaries, dynamical captures and radial infalls.
In Sec.~\ref{sec:amplitude_phenomenology}, we identify the mechanism behind the tail enhancement with binary eccentricity, and highlight the key contribution of motion around the last apastron.
Instead, in Sec.~\ref{sec:exponent_phenomenology}, we characterize the tail term as a superposition of a large number of pure power-laws.
In Sec.~\ref{sec:summary} we summarise the results presented in the main text.
We conclude in Sec.~\ref{sec:conclusions}, discussing future directions opened by our findings, both on the theoretical and observational side.
We especially focus on the implications for extracting the post-merger tail signal in fully non-linear simulations of comparable masses mergers.
In the Appendices~\ref{app:appendix}, we give additional details on the analytical computations, show numerical convergence tests, compare with finite distance extractions, include additional results on higher modes tails and asymptotically late time investigations for bounded inspirals and radial infalls, with similar results to the ones shown in the main text. 
We comment on the dependence of the tail amplitude at the time it starts dominating the strain, in terms of the eccentricity at the separatrix and the impact parameter, Eq.~\eqref{eq:impact_parameter_b_DEF}, at the light-ring crossing.
Finally, we briefly discuss exploratory results for scattering systems.

Unless explicitly stated, we work in geometric units $c=G=1$ and assume all quantities rescaled with respect to the central black hole mass $M$. 

\section{Perturbative and numerical framework}
\label{sec:RWZ}
Our analysis focuses on small mass-ratios, thus we linearize Einstein's equations and discard higher order corrections.
We impose both the BH and infalling test-particle to be initially non-spinning. 
Since we are working at linear perturbative order, the remnant (post-merger) BH is also consequently non-spinning.
The background metric is thus the Schwarzshild metric:
\begin{equation}
    \begin{gathered}
           ds^2=-A(r)dt^2+\frac{dr^2}{A(r)}+r^2d\Omega^2 \,,\\
    \end{gathered}
   \label{eq:Schwarzschild_metric}
\end{equation}
with $A(r)=1-2/r$.
We expand the strain in spin-weighted spherical harmonics modes $_{-2}Y_{\ell m}(\Theta,\Phi)$:
\begin{equation}
    h_+-{\rm i}h_\times =  \sum_{\ell}\sum_{m=-\ell}^\ell h_{\ell m}(t) {}_{-2}Y_{\ell m}(\Theta,\Phi) \,.
\end{equation}
From these multipoles, it is possible to build~\cite{Nagar:2005ea} two gauge invariant quantities that transform under parity as $(-1)^{\ell}$ and $(-1)^{\ell+1}$; we refer to the latter as the even/odd master functions $\Psi^{(e/o)}$. 
At large distances from the emitting system, it holds~\cite{Nagar:2005ea}
\begin{equation}
h_{\ell m}=\dfrac{1}{r}\sqrt{\dfrac{\left(\ell+2\right)!}{\left(\ell-2\right)!}}\left(\Psi^{(e)}_{\ell m}+i \Psi^{(o)}_{\ell m}\right)+\mathcal{O}\left(\dfrac{1}{r^2}\right) \,,
\end{equation}
where, depending on the parity of $\ell+m$, only one of the two terms on the right-hand side vanishes.
The functions $\Psi^{(e/o)}$ satisfy two (decoupled) inhomogeneous Schr{\"o}dinger-like equations, the Regge-Wheeler/Zerilli (RWZ) equations
\begin{equation}
        \mathcal{O}^{(e/o)}\Psi_{\ell m}^{(e/o)}(t,r_*) = S_{\ell m}^{(e/o)}(t,r),
    \label{eq:RWZ_equation}
\end{equation}
where
\begin{equation}
    \mathcal{O}^{(e/o)} \equiv \left[\partial_t^2 -\partial_{r_*}^2+V_{\ell m}^{(e/o)}(r_*)\right],
    \label{eq:RWZ_op_def}
\end{equation}
and we have introduced the standard tortoise coordinate $r_*=r+2\log\left(r/2-1\right)$ and the RWZ operator $\mathcal{O}^{e/o}$.
In the following, when not necessary to distinguish the two cases, we will drop the superscripts $(e/o)$.

The potentials in the equation above are the RWZ ones~\cite{Zerilli:1970se,Nagar:2005ea} and the driving source is built from the in-falling particle stress-energy tensor~\cite{Nagar:2005ea}. 
As a consequence, it is localized along the particle trajectory $r(t)$ at all times. This feature can be made explicit by writing:
\begin{equation}
    S^{(e/o)}_{\ell m}= f^{(e/o)}_{\ell m}\delta(r-r(t))+ g^{(e/o)}_{\ell m}\partial_r\delta(r-r(t)) \, .
    \label{eq:source_generic_expr}
\end{equation}
In Appendix~\ref{app:source_expressions} we report the full expressions of the functions $f^{(e/o)}_{\ell m}, g^{(e/o)}_{\ell m}$ for a point-particle, as found in Ref.~\cite{Nagar:2005ea}.

In the present work, we will compute analytical and numerical solutions of the Cauchy problem given by Eq.~\eqref{eq:RWZ_equation}, always using as initial conditions:
\begin{equation}
    \Psi_{\ell m}(t=0,r)=\partial_t\Psi_{\ell m}(t=0,r)=0 \, . 
    \label{eq:InitialCondition}
\end{equation}
These initial conditions are not physical.
In fact, realistic systems  emit gravitational waves from the moment they are created. 
Effectively, Eq.~\eqref{eq:InitialCondition} means neglecting all the history of the system before a certain time and thus imposing a formally incorrect solution.
This implies a partial loss in information, but also an initial transient in which the emitted radiation does not correspond to a real, physical solution of the linearised Einstein equations and, for this reason, is commonly denoted as "junk radiation".
In Appendix~\ref{app:convergence_tests}, we motivate the negligible influence of junk radiation on our results, determining the approximate initial conditions of Eq.~\eqref{eq:InitialCondition} as appropriate for our purposes.

Unless specified, the trajectory of the particle will always be computed numerically, solving the system of Hamiltonian equations~\cite{Nagar:2006xv}
\begin{equation}
\begin{split}
    &\dot{r}=\frac{A}{\hat{H}}p_{r_*} \, ,\\
    &\dot{\varphi}=\frac{A}{r^2\hat{H}}p_{\varphi} ,\\
    &\dot{p}_{r_*}=A\hat{\mathcal{F}}_{r}-\frac{A}{r^2\hat{H}}\left(p^2_{\varphi}\frac{3-r}{r^2}+1 \right) \, ,\\
    &\dot{p}_{\varphi}=\hat{\mathcal{F}}_{\varphi} \, ,
\end{split}
    \label{eq:Hamiltonian_Eqs}
\end{equation}
where $(p_{r_*},p_{\varphi})$ are the $\mu$-rescaled momenta conjugate to the variables $(r_*,\varphi)$,
and $\hat{H}$ is the $\mu$-rescaled Hamiltonian of a test particle in Schwarzschild background
\begin{equation}
    \hat{H}=\sqrt{A\left(1+\frac{p_{\varphi}^2}{r^2}\right)+p_{r_*}^2} \, .
    \label{eq:energy_unit_mu}
\end{equation}
Finally, $\hat{\mathcal{F}}_{r}$ and $\hat{\mathcal{F}}_{\varphi}$ are the components of the dissipative force that drive the dynamics, whose general expression can be found in~\cite{Chiaramello:2020ehz, Albanesi:2021rby}.
These quantities are analytical, built from a PN-based, EOB-resummed analytical expansion for the fluxes of energy and angular momentum observed at infinity, as computed in~\cite{Albanesi:2021rby,Chiaramello:2020ehz}.
Such fluxes have been shown to be consistent with the corresponding numerical expressions in Ref.~\cite{Albanesi:2023bgi}, hence consistent with emission of GWs computed from the evolution.
However, it is important to note that at the operational level, the GW expressions obtained as numerical output of the evolution will \textit{not} directly enter the particle trajectory within our implementation.
Hence, the trajectory is numerically independent from the waveform.
This will be a key point when feeding the particle trajectory as input of our semi-analytical computations used to derive a prediction for the GW strain, ensuring that numerical errors in the numerically-computed GWs (against which we will compare the prediction obtained) cannot  contaminate the semi-analytical predictions.

To solve the problem in Eq.~\eqref{eq:RWZ_equation}-\eqref{eq:InitialCondition}-\eqref{eq:Hamiltonian_Eqs} numerically, we employ the time-domain code \RWZ~\cite{Bernuzzi:2010ty,Bernuzzi:2011aj}.
The software uses a homogeneous grid in tortoise coordinate $r_*$ and, at large distances, a hyperboloidal layer (over which $r_*$ is compactified~\cite{Zenginoglu:2009ey}) is attached to the standard computation domain, where the trajectory evolves.
We define $\rho$ to be the compactified coordinate and $\tau$ the retarded time in the layer.
The coordinates of the layer $(\tau,\rho)$ are connected to those of the standard computational domain $(t,r_*)$ as follows:
\begin{equation}
    \tau-\rho=t-r_* \, .
    \label{eq:RWZcoordinates}
\end{equation}
The hyperboloidal layer allows to extract the GW strain at future null infinity $\mathcal{I}^+$ at a finite location $\rho_+$.
The grid in $r_*$ ends, for negative values, at a finite quantity, in order to keep the horizon outside of the computational domain.
The code uses double precision, hence our computations will have a precision of at most $\sim 10^{-16}$.
Numerical strain values close to this threshold, will be considered dominated by numerical error.
In Appendix~\ref{app:convergence_tests} we also show that the numerical resolution used in all the results discussed below is adequate for our purposes and does not affect any of the results obtained below.

\section{Long-range propagation in curved backgrounds with a source}
\label{sec:analytic_model}

\subsection{General solution}
\label{subsec:general_solution}

The general solution of Eqs.~\eqref{eq:RWZ_equation}-\eqref{eq:InitialCondition}, in terms of the Schwarzschild coordinates $(t,r)$, can be written as the convolution
\begin{equation}
    \Psi_{\ell m}(t,r)=\int_{-\infty}^{t}dt'\int_{-\infty}^{\infty}dr' \, S_{\ell m}(t',r')\, G_{\ell}(t,t';r,r') \, .
    \label{eq:Source_GF_convolution_DEF}
\end{equation}
Where $G_{\ell}(t,t';r,r')$ is the Green's function, defined as solution to the impulsive problem
\begin{equation}
    \mathcal{O}|_{t,r_*}\, G_{\ell}(t,t';r,r')=\delta(t-t')\delta(r-r') \, .
    \label{eq:GreensFun_time}
\end{equation}
Note that we assume homogeneous boundary conditions on the Green's function, for all times, at $r'_*\rightarrow\pm\infty$.
In Appendix~\ref{app:computations_GF}, we review the derivation of the propagator controlling the tail, which can be obtained in the limit of large $r$ and small frequencies\,\footnote{In this section exclusively, we use explicit units of $M$ to highlight the relevant scales.} $\omega M\ll 1$. 
The former approximation is connected to the tail being due to the corrections to the flat light-cone propagator, arising from the long range spacetime curvature~\cite{Leaver:1986gd,Blanchet:1992br,Andersson:1996cm}. 
The latter approximation encodes the fact that small frequency waves are the ones interacting the most with the curved geometry on large scales~\cite{Price:1971fb,Leaver:1986gd,Andersson:1996cm}, and implies that the propagator we derive is the retarded Green's function only in the limit of large retarded times $\tau$ compared with the source retarded time, $\tau\gg t'+\rho_+$. 
The result for the propagator in these limits, assuming that the observer is located at $\mathcal{I}^+$, is 
\begin{equation}
\begin{gathered}
        G_{\ell}(\tau,t';\rho_+,r')=
  \theta(\tau-t'-\rho_+)\cdot \\ \frac{\left(-1\right)^{\ell}2^{\ell+1}\ell!(\ell+1)!}{\left(2\ell+1\right)!}\frac{\left(r'\right)^{\ell+1}}{\left(\tau-t'-\rho_+\right)^{\ell+2}} \,,
\end{gathered}
\label{eq:Retarded_GF_result_time_domain}
\end{equation}
where the Heaviside function serves to impose causality, since we are considering the retarded Green's function.
Plugging-in the result above in the general expression for the tail strain Eq.~\eqref{eq:Source_GF_convolution_DEF}, together with the point-particle source expression Eq.~\eqref{eq:source_generic_expr}, and considering an observer located at $(\tau,\rho_+)$, yields
\begin{widetext}
\begin{equation}
    \Psi_{\ell m}(\tau,\rho_+)=c_{\ell}\int_{-\infty}^{\tau-\rho_+}dt'\frac{r^{\ell}(t')\left\lbrace r\left[f_{\ell m}(t',r)-\partial_r g_{\ell m}(t',r)\right]-\left(\ell+1\right)g_{\ell m}(t',r)\right\rbrace_{r=r(t')} }{\left(\tau-t'-\rho_+\right)^{\ell+2}} \ ,\ \ \ c_{\ell}=\frac{\left(-1\right)^{\ell}2^{\ell+1}\ell!(\ell+1)!}{\left(2\ell+1\right)!} \,,
    \label{eq:tail_signal_analytic_expr}
\end{equation}
\end{widetext}
where we have denoted as $r(t')$ the value of $r$ along the 

\noindent point-particle trajectory. 
Note that $f_{\ell m }$ and $g_{\ell m}$ in the above are computed along the trajectory as well.
In Appendix~\ref{app:source_expressions}, we show the full expressions of the functions $f_{\ell m},g_{\ell m}$ for a point-particle, as computed in~\cite{Nagar:2005ea,Nagar:2006xv}.
The failure of our model for $\tau-\rho_+\approx t'$ is made manifest by the upper limit of integration in Eq.~\eqref{eq:tail_signal_analytic_expr}, since the integrand is singular at this point. 
We can interpret this by stating that our model can describe signals travelling well inside the light-cone, but fails to describe signals marginally close to it.
In the present work, we will focus our attention on systems that become bounded after a certain timescale, and we will limit our analysis to the post-merger signal.
Since the source contribution to Eq.~\eqref{eq:tail_signal_analytic_expr} dies exponentially after the light-ring crossing, we expect our results to not be influenced by the singularity in $t'\approx \tau-\rho_+$.

We briefly discuss systems that are unbounded at all times, e.g. scattering scenarios, in Appendix~\ref{app:scattering}, highlighting future research directions for this case in Sec.~\ref{sec:conclusions}.

\subsection{Intermediate vs asymptotic behaviour}
\label{subsec:inter_asymp_behaviour}
%
The analytic model for the tail, Eq.~\eqref{eq:tail_signal_analytic_expr}, is an integral over the entire past history of the source.
For this reason, we expect the tail to show a more complicated phenomenology compared to what is predicted by source-free PT~\cite{Price:1971fb,Andersson:1996cm,Zenginoglu:2009ey}.
In general, since the source is an oscillating function, also the real and imaginary part of the late-times waveform Eq.~\eqref{eq:tail_signal_analytic_expr} are non-monotonic functions.
Moreover, we cannot sort out from the integral a single power-law behaviour in the observer retarded time $\tau$, since the integral is rather a superposition of power-laws in $\tau$ with location of the asymptote (corresponding to the zero in the denominator) depending on the integration variable $t'$. 
%
%
Since the source decays exponentially after the light-ring crossing~\cite{Albanesi:2023bgi}, there exist a certain timescale after which the source information will not affect the signal anymore, leaving place to a single pure power-law in $\tau$ dictating the asymptotic decay of the perturbation.

%
%
Two question arises from the above intuition. 
The first concerns the timescale that an observer at $\mathcal{I}^+$ has to wait in order for the tail to be a single power-law,
the second is relative to the value of the power-law exponent in this asymptotic limit.
%
%
To answer the above questions, we start considering initially unbounded systems originating at a time $T_{\rm in}$, that become bounded after a certain timescale $T_{\rm bound}$ due to radiation-reaction. For the moment, we assume that the observer is located at very late times $\tau\gg T_{\rm bound}$, after the merger has occurred.
Then we can separate the integration domain in Eq.~\eqref{eq:tail_signal_analytic_expr} accordingly, in an interval during which the system is unbounded, $\left(T_{\rm in},T_{\rm bound}\right)$, and one over which the system is bounded $\left(T_{\rm bound},\tau-\rho_+\right)$.
We focus first on the contribution to the late-time signal of the dynamics in the interval $\left(T_{\rm in},T_{\rm bound}\right)$, during which
we assume the test-particle to be in the far away region $r'\gg M$, moving slowly.
As a consequence, the space-time curvature can be neglected
and the test-particle trajectory can be approximated as $x_i(t)\simeq v_i t$ with $v_i$ constant velocity, i.e. we expand the source Eq.~\eqref{eq:source_generic_expr} neglecting all terms $\mathcal{O}(G)$ or higher, and work at lowest PN order.
The source for the $(\ell,m)=(2,2)$ mode can be written in terms of the $(00)$ component of the particle stress energy tensor $T_{00}=\mu\delta^3(x_i-v_i t)$, as
\begin{equation}
    S^{(e)}_{22}\propto r T^{(2,2)}_{00}=\mu\frac{\delta(r-|v|t)}{r} \, .
    \label{eq:source22_largeR_smallV}
\end{equation}
Plugging this expression in Eq.~\eqref{eq:tail_signal_analytic_expr}, yields
\begin{equation}
    \psi_{\rm unbound}(\tau,\rho_+)\propto\int_{T_{\rm in}}^{T_{\rm bound}}dt'\frac{\mu|v|^2t'^2}{(\tau-t'-\rho_+)^4} \, .
    \label{eq:psi22_tail_unbounded_approx}
\end{equation}
The integral above can be carried out analytically and it reads
\begin{equation}
   \begin{gathered}
       \psi_{\rm unbound}\propto
        \frac{\left(\rho_+-\tau\right)^2+3\left(-\rho_++\tau\right)T_{\rm in}+3T_{\rm in}{^2}}{3\left(\rho_{+}-\tau-T_{\rm in}\right)^3}\\
         -\frac{\left(\rho_+-\tau\right)^2+3\left(-\rho_++\tau\right)T_{\rm bound}+3T_{\rm bound}{^2}}{3\left(\rho_{+}-\tau-T_{\rm bound}\right)^3} \, .
   \end{gathered}
   \label{eq:tail_unbounded}
\end{equation}
When considering the limit $\tau\gg T_{\rm in},T_{\rm bound}$, each of the terms in Eq.~\eqref{eq:tail_unbounded} give a leading power-law contribution $\propto \tau^{-1}$, equal in modulo but opposite in sign. A similar cancellation can be found for the $\propto\tau^{-2}, \tau^{-3}$ contributions, leaving a dominant power-law $\propto\tau^{-4}$.

We now analyze the contribution to the late-times signal of the bounded dynamics.
As mentioned above, after the light-ring crossing the source decays exponentially~\cite{Albanesi:2023bgi}, hence does the integrand in Eq.~\eqref{eq:tail_signal_analytic_expr}.
If we let $T_{\rm f}$ be the time at which the source can be considered zero (up to a given precision), when performing an observation at times $\tau>\rho_++T_{\rm f}$, we can replace the upper limit of integration with $T_{\rm f}$.
Then, we Taylor expand the integrand, assuming $\tau\gg T_{\rm f}+\rho_+$
\begin{equation}
\begin{gathered}
        \psi_{\rm bound}(\tau,\rho_+)=\frac{c_{\ell}}{\tau^{\ell+2}}\cdot\\ \int_{T_{\rm bound}}^{T_{\rm f}}dt'S_{\ell}(t')\left[1+\sum_{n=1}^{\infty}\frac{\left(\ell+1+n\right)!}{n!\left(\ell+1\right)!}
\left(\frac{t'+\rho_+}{\tau}\right)^n\right] \, ,
\end{gathered}
\label{eq:tail_analytic_expansion_power_laws}
\end{equation}
where we denoted as $\psi_{\rm bound}(\tau,\rho_+)$ the contribution to the late-time signal of the bounded dynamics.
The result in Eq.~\eqref{eq:tail_analytic_expansion_power_laws} is a superposition of power-law decays\,\footnote{Note that this result is fundamentally different from the one obtained in Ref.~\cite{Andersson:1996cm}. 
In the latter, power-laws corrections to the propagator giving rise to Price's law, were computed. 
Instead, the propagator we consider is the same as the one through which Price's law can be derived.}, with the smallest decay being $\propto \tau^{-\ell-2}$. 
We expect the importance of faster decaying terms to depend on the pre-merger dynamics, apart from the observer retarded time $\tau$. 
As we move $\tau$ to progressively late times, the faster decaying contributions will eventually die-off, leaving Price's law as dominant component.
The transient regime characteristic timescale depends on the \textit{excitation coefficients} of each power-law contribution: the more enhanced are these coefficients, the longer the intermediate regime will be.
From Eq.~\eqref{eq:tail_analytic_expansion_power_laws} these coefficients depend on an integral of the source $S(t')$ multiplied by a factor $\left(t'+\rho_+\right)^n$. 
Hence, the excitation coefficient of each power-law correction to Price's law depends both on the specific orbital dynamics under consideration, and on the amount of inspiral history included in the evolution, for timescales at which the source is still appreciably excited.
Instead, at very early past times, the source suppression will cutoff the contribution of higher-order terms.
We refer the reader to Sec.~\ref{sec:exponent_phenomenology} for a quantitative discussion on the latter point. 

To summarize, the above results show that, even if the system is initially in an unbounded configuration, the asymptotic relaxation is dominated by a $\tau^{-2-\ell}$ power-law, while the $\tau^{-1}$, $\tau^{-2}$, $\tau^{-3}$ contributions cancel out. 
This result is in agreement with the homogeneous PT literature~\cite{Price:1971fb,Leaver:1986gd,Andersson:1996cm}.
The intermediate behaviour of the tail in the post-merger phase can instead be approximated by a superposition of exact power-laws in $\tau$, with expansion coefficients depending on the source history.
In the following, we will apply this expansion from $T_{\rm in}$, for both initially and dynamically bounded systems.
We briefly discuss in Appendix~\ref{app:scattering} the analytical prediction for the tail signal emitted in a scattering scenario.


%
\begin{table}[h]
\begin{tabular}{cccccccc}
\hline \hline
$e_0$ & $\hat{H}_0$ & $p_{\varphi,0}$ &  $r_0$ & $b_{LR}$ & $e_{\rm sep}$ & $t_{LR}$ & $n_{\rm api}$\\ 
\hline 
0.9 & 0.9890 & 3.9170 & 83.000 & 3.9315 & 0.869 & 14555 & 9 \\ 
0.8 & 0.9791 & 3.8313 & 40.000 & 3.8881 & 0.778 & 5386 & 7 \\ 
0.7 & 0.9713 & 3.7671 & 26.667 & 3.8429 & 0.670 & 6921 & 13 \\ 
0.6 & 0.9649 & 3.7139 & 20.000 & 3.7998 & 0.563 & 8955 & 21 \\ 
0.5 & 0.9587 & 3.6502 & 15.400 & 3.7699 & 0.483 & 5654 & 15 \\ 
0.4 & 0.9538 & 3.6001 & 12.500 & 3.7400 & 0.393 & 4856 & 14 \\ 
0.3 & 0.9525 & 3.6103 & 11.429 & 3.7075 & 0.276 & 14895 & 47 \\ 
0.2 & 0.9484 & 3.5514 & 9.375 & 3.6909 & 0.201 & 7996 & 25 \\ 
0.1 & 0.9453 & 3.5044 & 7.778 & 3.6766 & 0.114 & 3773 & 11 \\ 
0.0 & 0.9449 & 3.5000 & 7.000 & 3.6693 & 0.000 & 4308 & 40* \\  
\hline \hline
\end{tabular}
\caption{From left to the right: initial eccentricity, initial energy and angular momentum, initial radius, impact parameter at the light-ring crossing, eccentricity at the separatrix crossing, time of the light-ring crossing and number of apastri. 
The results are relative to the eccentric and quasi-circular simulations.
For eccentric orbits, we initialize the trajectory at an apastron, hence $r_0$ is the coordinate of the first apastron.
For the quasi-circular case, we do not show the number of apastra, but show instead the number of orbits before the plunge.
Note that the eccentricity decreases during the inspiral, but because of its definition in terms of radial turning points, it can increase close to the separatrix, as exemplified in Fig.~1 of~\cite{Albanesi:2023bgi}.
}
\label{tab:sims_ecc}
\end{table}
\begin{table}[h]
\begin{tabular}{ccccccccc}
\hline \hline
$n_{\rm enc}$ & $\hat{H}_0$ & $p_{\varphi,0}$ &  $r_0$ & $b_{LR}$ & $e_{\rm sep}$ & $t_{LR}$\\ 
\hline 
 1 & 1.000001 & 3.9980 & 300.0 & 3.9687 & ... & 2650 \\ 
 2 & 1.000001 & 4.0065 & 300.0 & 3.9457 & 0.955 & 8059 \\ 
 3 & 1.000001 & 4.0150 & 300.0 & 3.9327 & 0.927 & 15947 \\ 
 4 & 1.000001 & 4.0235 & 300.0 & 3.9232 & 0.907 & 23827 \\ 
 5 & 1.000001 & 4.0320 & 300.0 & 3.9155 & 0.892 & 31878 \\ 
 6 & 1.000001 & 4.0405 & 300.0 & 3.9075 & 0.876 & 40228 \\ 
 7 & 1.000001 & 4.0447 & 300.0 & 3.8968 & 0.851 & 45037 \\ 
 8 & 1.000001 & 4.0490 & 300.0 & 3.8737 & 0.819 & 49786 \\  
\hline \hline
\end{tabular}
\caption{From left to the right: number of encounters, initial energy and angular momentum, initial radius impact parameter at the light-ring crossing, eccentricity at the separatrix and time of the light-ring crossing. The results are relative to the dynamical captures simulations.
The case $n_{\rm enc}=1$ corresponds to a direct capture. In this case, there are no eccentric orbits before the merger, hence we do not report $e_{\rm sep}$.}
\label{tab:sims_enc_scatt}
\end{table}
\begin{figure*}[t]
\includegraphics[width=1.0\textwidth]{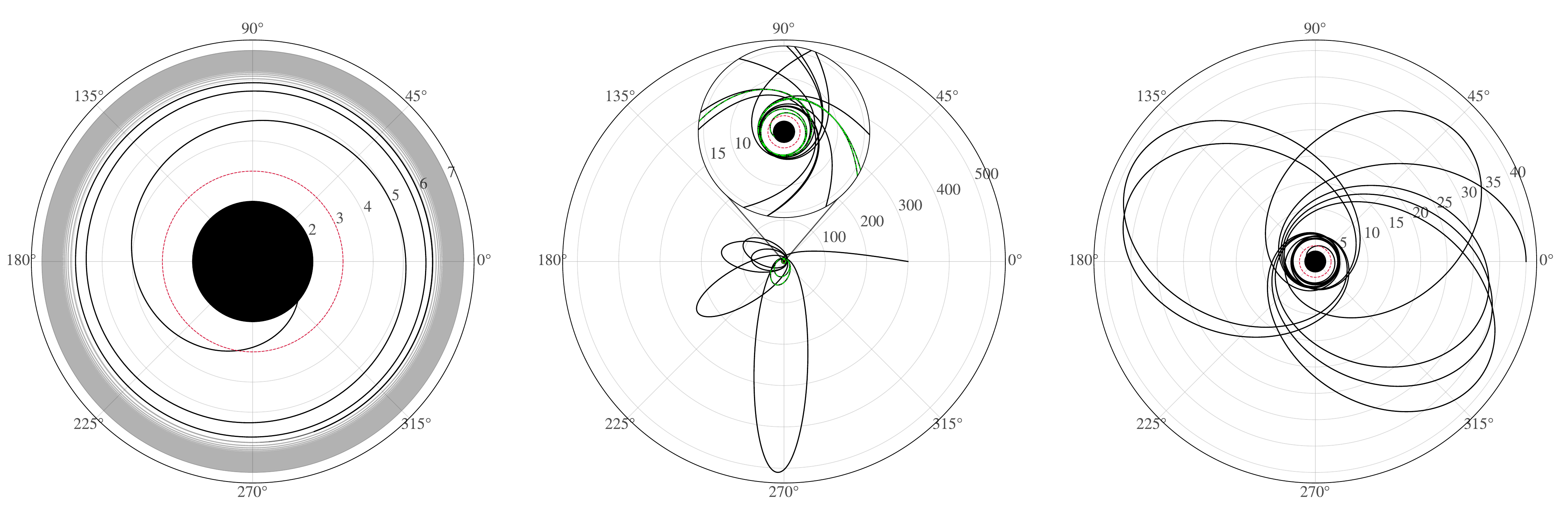}
\caption{Left: quasi-circular inspiral and plunge, with test-particle initialized at a distance $r_0=7.0$.
Center: Dynamical capture configuration, with initial angular momentum $p_{\varphi,0}=4.0405$. A zoomed-in view of the dynamics at smaller radii is shown in the inset.
Right: trajectory of a test-particle moving in an orbit with initial eccentricity $e_0=0.8$.
The red line marks the light ring.
The green line highlights the portion of the trajectory used in Fig.~\ref{fig:A-p22_vs_t-tpeak_theta9.45_child_power-laws_expansion}.
\label{fig:polar_plots}}
\end{figure*}
\section{Comparison with numerical results}
\label{sec:comparision}

We now analyze  the $(2,2)$ multipolar component of the waveform, $h_{22}$, produced by a particle orbiting around a Schwarzschild BH.
In particular, we focus our attention on bounded orbits with varying eccentricities, on dynamical captures (i.e. initially unbounded orbits becoming bounded after some time due to radiation reaction), and finally on radial infalls from different distances.
In Fig.~\ref{fig:polar_plots}, we report examples of these different dynamics.
In Table~\ref{tab:sims_ecc} and~\ref{tab:sims_enc_scatt} we show the relevant parameters for each configuration considered.
Note that we always impose the initial polar and azimuthal angles to be $\theta=\pi/2$ and $\varphi=0$ respectively.
 For bounded orbits, we report the initial eccentricity and the number of apastri before merger. For dynamical captures, we show the number of encounters. 
For simulations ending in an eccentric merger,  we report the eccentricity at the separatrix crossing time\,\footnote{
During the inspiral, the test-particle can be assumed to move along eccentric stable orbits identified through the eccentricity $e$ and the semi-latus rectum $\iota$, as long as $\iota-2e\geq6$ is satisfied. Values of $e$ and $\iota$ such that $\iota_{\rm sep}=6+2e_{\rm sep}$ identify the last stable orbit, denoted as separatrix.
} .
For all simulations,
we show the initial energy and angular momentum, the initial distance from the BH in terms of the coordinate $r_0$, the time of the light-ring crossing and the impact parameter computed at the light-ring crossing $b_{LR}$, where we define $b$ as the ratio~\cite{Albanesi:2023bgi,Carullo:2023kvj}

\begin{equation}
    b=p_{\varphi}/\hat{H} \, .
    \label{eq:impact_parameter_b_DEF}
\end{equation}

The eccentricity is defined (for bounded systems) through its relation with the apastron and the periastron coordinates $r_{\pm}$ of the orbit~\cite{Chiaramello:2020ehz,Albanesi:2021rby,Albanesi:2023bgi}

\begin{equation}
    e=\frac{r_+-r_-}{r_++r_-} \, .
\label{eq:eccentricity_DEF}
\end{equation}
Note that the initial eccentricity alone is not enough to predict the dynamical evolution of a binary. Another parameter is necessary, e.g. the initial semilatus rectum which can be computed from initial energy and angular momentum. 
For bounded orbits, we select a test-particle mass $\mu=10^{-3}$, while, 
for simplicity, when simulating dynamical captures we set the test-particle mass to be $\mu=10^{-2}$. 
This is because given certain initial conditions $(E_0,p_{\varphi,0})$, the test-particle can either be directly captured by the central BH, have multiple close encounters before the merger, or scatter away. 
As discussed in Refs.~\cite{Nagar:2020xsk,Albanesi:2024xus}, the region within the parameter space $(E_0,p_{\varphi,0})$ for which captures involving multiple encounters are possible, decreases with the increase of the mass ratio.
Hence, the larger is the mass of the test-particle, the easier it is to obtain different multiple encounters simulations.
Below, we will always show mass-rescaled quantities, in such a way that this choice will not affect our results.
Finally, we analyze five different radial infalls, with the test-particle of mass $\mu=10^{-2}$, initial energy $E_0=1.00$ and angular momentum $p_{\varphi,0}=0.0$, from different initial distances $r_0=\lbrace 100,200,300,400,500 \rbrace$. The time of the light-ring crossing in these configurations is $t_{LR}=\lbrace 496, 1370, 2495, 3825, 5331 \rbrace$ respectively.
The trajectory of the test-particle in the aforementioned settings is computed numerically by means of the \RWZ~code, as detailed in Sec.~\ref{sec:RWZ}.
We use the same code to obtain the numerical (linear) waveform produced by the motion of said systems, as observed at $\mathcal{I}^+$, with the aim to test the model introduced in the previous section.
In particular, we plug the numerical trajectory solved by \RWZ~in the integral form Eq.~\eqref{eq:tail_signal_analytic_expr} and use a trapezoidal method\,\footnote{
Note that we tested also the built-in function implementing Simpson's rule, yielding the same results.
} built in the \texttt{scipy}\,\footnote{
Specifically, we use \texttt{integrate.trapz}$[f(t), dx=dt']$ (or \texttt{integrate.simps}$[f(t), dx=dt']$), where $f(t)$ is the integrand in Eq.~\eqref{eq:tail_signal_analytic_expr} computed along the numerical trajectory, while $dt'$ is the spacing between the time steps of the latter.}~\cite{scipy} library to compute the integration.
Since we are interested in the tail part of the signal, we focus on two quantities of interest that can be extracted from $h_{22}$, the amplitude $A_{22}(\tau) = |h_{22}|$ and the tail exponent
\begin{equation}
    p\equiv\frac{d\ln A_{22}(\tau)}{d\ln\tau} \, .
    \label{eq:p_def}
\end{equation}
In the present section, we always shift the axis of the retarded time $\tau$ to have a zero at the time of the light-ring crossing\,\footnote{The reference retarded time enters the definition of $p$: choosing a different value implies assuming a different functional form for the tail, since it moves its asymptote. This does not affect the asymptotic value of $p$, but only its intermediate behaviour.}.
We have checked that this is close to the peak of $A_{22}$ for all the configurations considered, so that it is possible, from our results, to estimate correctly the order of magnitude of the tail amplitude when it starts to dominated over the ringdown, with respect to the peak amplitude of the whole signal.
We remind to Table~I of Ref.~\cite{Albanesi:2023bgi} for an estimate of the delay between the peak of the orbital frequency and the quadrupolar amplitude for eccentric and quasi-circular orbits.

\subsection{Initially bounded case: eccentric and quasi-circular binaries}
\begin{figure*}[t]
\includegraphics[width=1.0\textwidth]{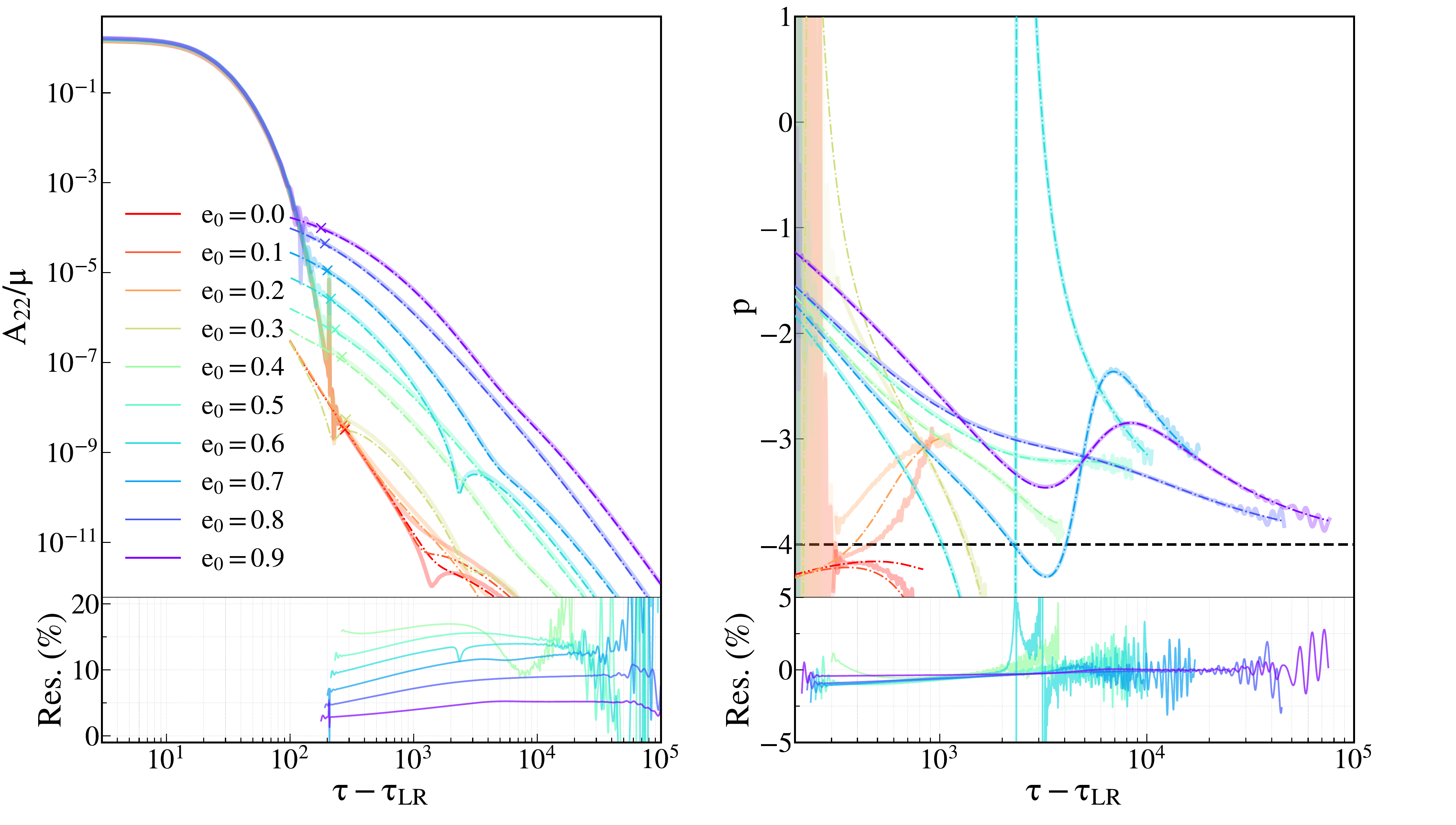}
\caption{Left: Mass-rescaled amplitude of the $(2,2)$ waveform multipole vs the observer retarded time, rescaled with respect to the time $\tau_{LR}$ at which the test-particle crosses the light-ring. 
Right: value of the tail exponent, Eq.~\eqref{eq:p_def}.
The thick solid lines are the numerical experiments, computed integrating Eq.~\eqref{eq:RWZ_equation} with the \RWZ{}  code. The thin dot-dashed lines are the analytical prediction for the late-time behaviour Eq.~\eqref{eq:tail_signal_analytic_expr}.
The dashed black horizontal line on the right, is Price's law.
These results are relative to the eccentric and quasi-circular simulations of Table~\ref{tab:sims_ecc}, each labeled by the initial eccentricity $e_0$.
We cut the simulations for values of the amplitude $A_{22}/\mu=10^{-12}$, four orders of magnitude before the numerical precision threshold dictated by double precision and when numerical noise become noticeable (high frequency oscillations in the plot on the right).
Below each plot, for simulations with $e_0>0.3$, the residuals between numerical results and analytical predictions are shown, in $\%$, to quantify the agreement/mismatch.
\label{fig:A22-p_vs_t-tLR_eccs}}
\end{figure*}
\begin{figure*}[t]
\includegraphics[width=1.0\textwidth]{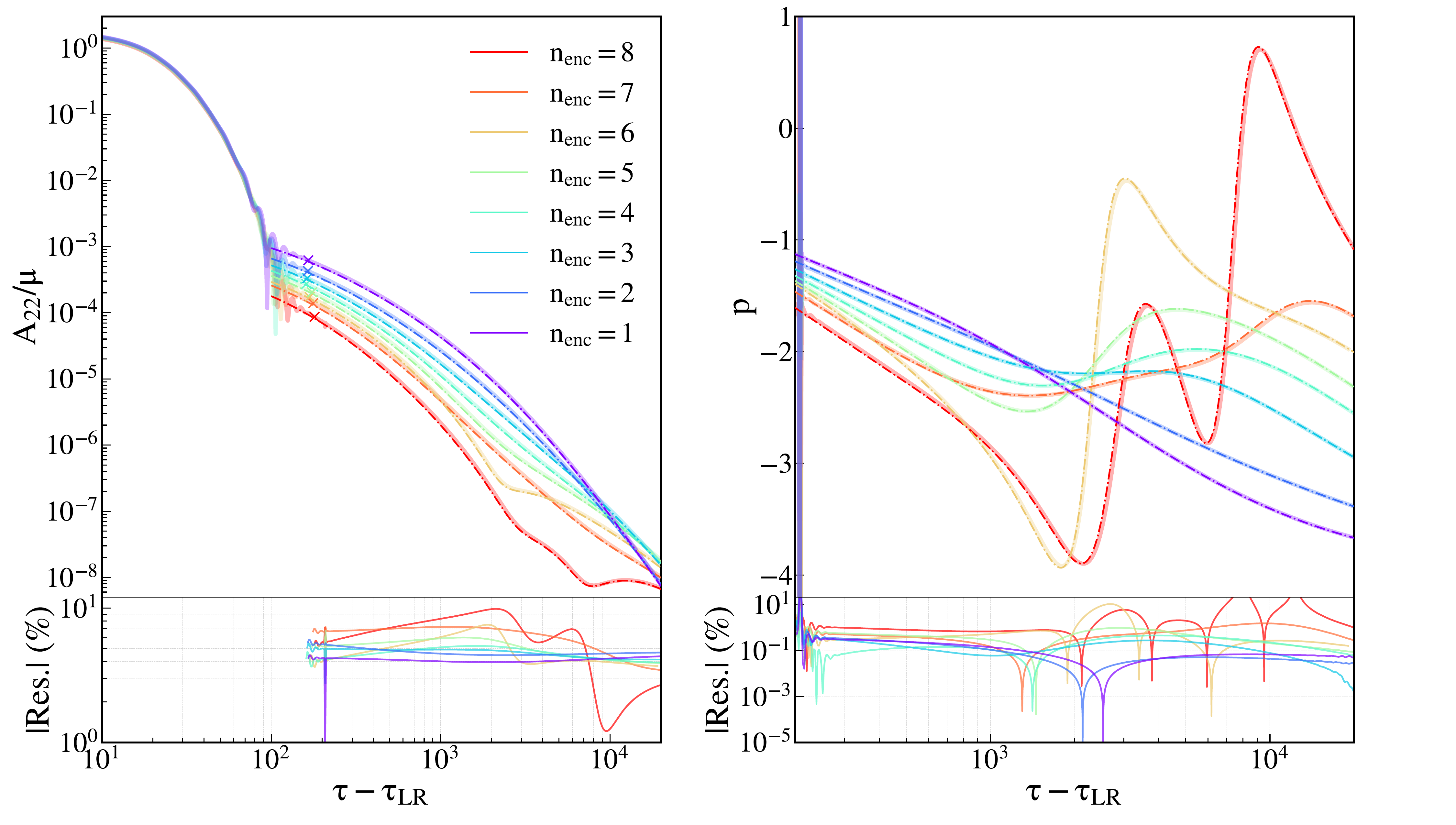}
\caption{Left: Mass-rescaled amplitude of the $(2,2)$ waveform multipole vs the observer retarded time rescaled with respect to the time of light-ring crossing $\tau_{LR}$. 
Right: value of the tail exponent, Eq.~\eqref{eq:p_def}.
The thin dot-dashed lines are the analytical predictions for the late-time behaviour,  Eq.~\eqref{eq:tail_signal_analytic_expr}, while the thick solid lines are numerical experiments obtained integrating Zerilli equation with the \RWZ~code.
The results are relative to the dynamical captures in Table~\ref{tab:sims_enc_scatt}, each simulation is labeled by the number of encounters $n_{\rm enc}$ between the test-particle and the BH.
Below each plot, the absolute value of the residuals between numerical results and analytical predictions are shown, to quantify the agreement.
\label{fig:A22-p_vs_t-tLR_dyn_capt}}
\end{figure*}
\begin{figure*}[t]
\includegraphics[width=1.0\textwidth]{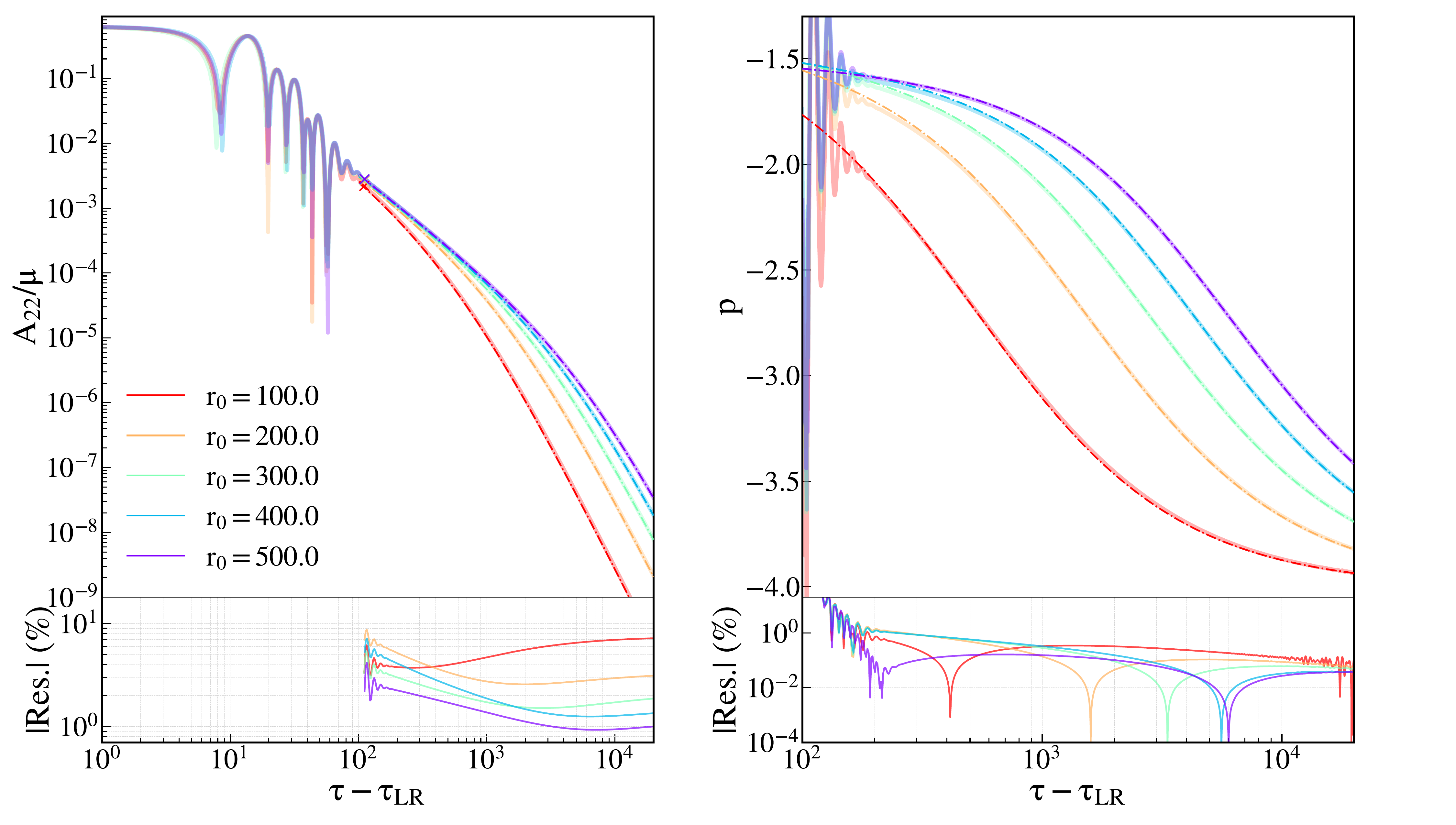}
\caption{Left: Mass-rescaled amplitude of the $(2,2)$ waveform multipole vs the observer retarded time rescaled with respect to the time of light-ring crossing $\tau_{LR}$. 
Right: value of the tail exponent, Eq.~\eqref{eq:p_def}.
The thin dot-dashed lines are the analytical predictions for the late-time behaviour, Eq.~\eqref{eq:tail_signal_analytic_expr},  while the thick solid lines are numerical experiments obtained integrating with the \RWZ~code.
The results are relative to radial infalls starting from the distances $r_0$ in the labels, with initial energy $E_0=1.00$. 
Note that the particle is infalling in the $xy$-plane, along the $x$ axis.
Below each plot, the absolute value of the residuals between numerical results and analytical predictions are shown, to quantify the agreement.
\label{fig:A22-p_vs_t-tLR_rad_infall}}
\end{figure*}

We now focus on systems initialized as bounded, starting by considering initial data for the particle trajectory on a quasi-circular binary, and then increase the eccentricity. 
In Table~\ref{tab:sims_ecc} we report the initial conditions and eccentricities of all the systems we have considered.
To compute the numerical evolutions with the \RWZ~code, we have multiplied the source in Eq.~\eqref{eq:RWZ_equation} by a factor $\mu^{-1}$. This does not change the results in the waveform if not for an overall multiplicative factor, allowing to circumvent the threshold given by double precision.
In Fig.~\ref{fig:A22-p_vs_t-tLR_eccs} we show the results of our numerical experiments, together with the analytical prediction for the late-time signal in Eq.~\eqref{eq:tail_signal_analytic_expr}. 
In particular, we show the amplitude of the $(2,2)$ mode, rescaled with respect to the test-particle mass $\mu=10^{-3}$, and the tail exponent $p$ as defined in Eq.~\eqref{eq:p_def}.
Below each plot, for simulations with $e_0>0.3$, we report the residuals, defined as $100*(X_{\rm numerical}-X_{\rm analytical})/X_{\rm numerical}$, quantifying the agreement level between numerical and analytical results, $X_{\rm numerical}$ and $X_{\rm analytical}$ respectively.

As already noted in~\cite{Albanesi:2021rby,Carullo:2023tff}, the time at which the tail starts to dominate on the ringdown strongly depends on the eccentricity of the progenitors' binary, and is due to a different amplitude of the tail at these intermediate times.
In particular, the higher the eccentricity, the more the tail is excited, and the sooner it starts to dominate.
The first test of our model is to reproduce this scaling in the amplitudes.
From the left panel of Fig.~\ref{fig:A22-p_vs_t-tLR_eccs}, it can be seen how for eccentricities larger than $e_0 \sim 0.3$ the model reproduces the amplitude of the tail, from the moment it starts to dominate over the ringdown, to asymptotically late times.
In particular the agreement is good for $e_0\geq 0.8$, for which the residuals are $\leq 10 \%$. As the eccentricity decreases, the residuals increase approximately by an overall constant factor, but remain $\leq 17\%$.
For small eccentricities, our model is able only to infer the order of magnitude of the amplitude at the transition.
In the right panel of Fig.~\ref{fig:A22-p_vs_t-tLR_eccs}, we report the tail exponent $p$ extracted from numerical experiments and from our model. 
The model reproduces with very high accuracy the numerical experiments for $e_0>0.3$, but not for lower eccentricities.
For $e_0>0.3$, the residuals are in the interval $[-2.5,2.5]\%$, when not taking into account the high-frequency oscillations at late-times in the numerical waveforms (due to numerical noise). Also, note that for the simulation  $e_0=0.6$, the residuals diverge at $\tau\simeq \tau_{LR}+3\cdot 10^3$, due to the numerical value of $p$ crossing zero.
These residuals do not show any clear trend in the eccentricity.
The high accuracy in the prediction of the exponent $p$ is consistent with the mismatch between analytical predictions and numerical results for $A_{22}$ being well approximated by a constant factor through the evolution, since by definition $p$ is not sensitive to an overall rescaling of $A_{22}$.
The mismatch for low eccentricities is expected and consistent with the fact that Eq.~\eqref{eq:tail_signal_analytic_expr} was derived always assuming a source localized at large $r$ with respect to the BH.
Hence, the longer the test-particle spends far away from the BH during the inspiral, the better agreement we can expect.
In a medium/high eccentric binary, this condition is satisfied up to times close to the merger, while, for low eccentricities, the test-particle trajectory receives support at small $r$ for a longer time during the last stage of the inspiral, as can be seen in Fig.~\ref{fig:dynamics_ecc} of Sec.~\ref{sec:amplitude_phenomenology}.
Instead, we do not have a clear understanding as to why the residuals seem to be approximately constant along the tail evolution. We leave the investigation of this behaviour and related model improvements to future work.

Note that we cut the results in  Fig.~\ref{fig:A22-p_vs_t-tLR_eccs} for values of the amplitude $A_{22}/\mu=10^{-12}$, four orders of magnitude above the numerical double precision threshold.
Right before the simulations are cut, high frequency oscillations are already present in the numerical results. 
For low eccentricity configurations, the tail starts to dominate when the signal is close to this strain value. 
Hence, we could partially impute the mismatch between our model and the experiments to limitations in numerical precision.

As mentioned in the previous section, our model predicts the tail to be a hereditary effect that depends on the entire inspiral history.
In particular, it is an integral over the source that, for generic orbits, is an oscillating function. 
We thus expect a more complex behaviour than a monotonic relaxation to a single power-law as in e.g. Ref.~\cite{Zenginoglu:2009ey}.
For instance, destructive interference among various components of the back-scattered signal can result in the amplitude $A_{22}$ nearly going to zero before increasing again (as dictated by the very late-time behaviour), implying a quasi-divergence in the tail exponent, which depends on the amplitude derivative. 

This is confirmed by the numerical evolutions, as shown in the right panel of Fig.~\ref{fig:A22-p_vs_t-tLR_eccs}, where for $e_0=0.6$ the cusp in the amplitude is reflected in an almost singular behaviour\,\footnote{We do not fully include its evolution in the plot for visualisation reasons. Since it spans a wide range, capturing it completely would make the visualisation of other results significantly harder.} of the tail exponent $p$.


\subsection{Dynamically bounded case}
\label{sbsec:dynamical_bounded_case}

We now analyze systems which are initially unbounded and, after a certain time, become bounded due to radiation-reaction, eventually merging.
In Table~\ref{tab:sims_enc_scatt} we show the initial conditions used for all of the simulations and, for each one of them,  the number of close encounters between the test-particle and the BH. Some of these configurations have also been studied in Ref.~\cite{Albanesi:2024xus}.
In Fig.~\ref{fig:A22-p_vs_t-tLR_dyn_capt}, the results of the numerical evolutions computed integrating  Eq.~\eqref{eq:RWZ_equation},~\eqref{eq:InitialCondition} with the \RWZ~code are compared with the analytical model Eq.~\eqref{eq:tail_signal_analytic_expr}.
Below each plot, we show the behaviour of the residuals absolute value.
From this comparison we see a good agreement for all of the simulations considered, from intermediate to late times. 
In particular, the absolute value of the residuals is $\leq 10\%$ for the amplitude $A_{22}$ and $\leq 1\%$ for the tail exponent, for all simulations in Table~\ref{tab:sims_enc_scatt}.
This implies that our model is able to reproduce both the amplitude of the tail, as well as the non-trivial evolution of the exponent $p$, from the time it starts to dominate over the QNMs, up to very-late times.
We note that the amplitude of the tail at the transition time increases the smaller the number of encounters is.
We elaborate this point in further detail in Sec.~\ref{sec:amplitude_phenomenology}, where we discuss which inspiral trajectory feature is able to enhance or suppress the tail.
Here, we just point out that this scaling in the amplitude is consistent with what found in Fig.~\ref{fig:A22-p_vs_t-tLR_eccs}.
In fact, GWs are mainly emitted at turning points along the trajectory, hence a larger number of encounters during the inspiral phase implies that the test-particle orbit loses more energy and angular momentum before the merger, resulting in a progressive circularization of the orbit, Fig.~\ref{fig:dynamics_dyn_capt}.
Quantitatively, for the dynamical capture configurations with $n_{\rm enc} > 2$ under consideration, it holds $e> 0.95$ after the first encounter.
The larger the initial angular momentum, the higher the eccentricity after the first encounter, since less radiation is emitted. 
However, systems with large angular momenta will undergo multiple close encounters before plunging, so that the final eccentricity at the separatrix-crossing will be lower. 
Indeed, the configuration with $n_{\rm enc} = 8$ results in the lowest eccentricity at separatrix-crossing, having 
$e_{\rm sep} =0.797$.

From the discussion in Sec~\ref{subsec:inter_asymp_behaviour}, we expect that the tail exponent $p$ will relax towards a $-\ell-2$ value at asymptotically late times. In fact, we show that the slower decaying terms, led by $\sim\tau^{-1}$, vanish at asymptotically late-times, for systems ending in a merger, at first order in perturbation theory.
The results depicted in Fig.~\ref{fig:A22-p_vs_t-tLR_dyn_capt} seem to confirm these predictions: for simulations with $n_{\rm enc}=1,2$ number of encounters, the exponent $p$ is relaxing towards $p=-4$.
Simulations with larger $n_{\rm enc}$ take longer time to merge, and as result of a more prolonged history there is a longer intermediate behaviour in the post-merger tail (see Sec.~\ref{sec:exponent_phenomenology} for more details).
In particular, the relaxation of $p$ towards its asymptotic limit is not monotonic for $n_{\rm enc}>2$. 
As already discussed in the previous section, this happens because the source is oscillating, hence destructive interference between tail signals generated at different times can gives rise to such non-monotonic behaviour.
In Sec.~\ref{sec:exponent_phenomenology} we will study in more detail the case with $n_{\rm enc}=8$, by means of a numerical evolution long enough to recover Price's law, and indeed will characterize the non-monotonic intermediate behaviour of $p$ as a superposition of a large number of power-laws in $\tau$, with different decay rates.


\subsection{Radial infall}
\label{subsec:radial_infall}
In Fig.~\ref{fig:A22-p_vs_t-tLR_rad_infall} we compare numerical experiments against the analytical prediction Eq.~\eqref{eq:tail_signal_analytic_expr} for radial infalls from different initial distances $r_0=\lbrace 100,200,300,400,500 \rbrace$.
In the plots, we show the absolute value of the residuals to quantify the agreement.
The analytical prediction matches very accurately all the numerical evolutions, with the absolute value of the residuals being $\leq 10\%$ ($\leq 1\%$) for the amplitude (tail exponent).
It should also be noted how the amplitude at the transition from a QNMs to a tail-dominated behaviour is larger than all of the configurations previously analyzed; we will explain this phenomenon in Sec.~\ref{sec:amplitude_phenomenology}.
For what concerns the intermediate behaviour, defined as the relaxation to the asymptotic limit, i.e. Price's law, there are two important considerations to be made.
The further from the BH is the initial location of the test-particle, the longer is the intermediate behaviour of the tail, before approaching $\sim\tau^{-4}$.
Moreover, this relaxation is monotonic. 
This is consequence of the source being non oscillating, since $\varphi$ is fixed along all the trajectory. 
As mentioned above, this removes the destructive interference among tail signals emitted close to each other, yielding a monotonic relaxation.

In Ref.~\cite{Bernuzzi:2012ku}, it is shown the post-merger tail generated by a geodesic radial infall from $r_0=7$, when observed at $\mathcal{I}^+$. 
As in our case, the relaxation of $p$ therein depicted is monotonic. However, in Ref.~\cite{Bernuzzi:2012ku} the tail exponent $p$ reaches the asymptotic value from below, i.e. from smaller values. 
In our case, the value $p\rightarrow -4$ is reached from above.
We have verified that such apparent discrepancy stems from the different definitions adopted in Eq.~\eqref{eq:p_def}, in particular in the choice of a reference time.
As mentioned above, in the present work, unless explicitly stated, we report all results with $\tau$ rescaled with respect to $\tau_{LR}$, the time at which an observer at $\mathcal{I}^+$ sees the test-particle crossing the light-ring.
In Ref.~\cite{Bernuzzi:2012ku}, the time is instead rescaled with respect to the radial infall starting time.
Such choice can change the intermediate behaviour of $p$ and yield the observed inversion of the tail exponent relaxation towards a constant value.


\section{Tail amplitude: last apastron contributions}
\label{sec:amplitude_phenomenology}

In Fig.~\ref{fig:A22-p_vs_t-tLR_eccs}, it is shown that the time of transition from a QNM to a tail-dominated behaviour depends on the eccentricity of the progenitors' binary.
Similarly, Figs.~\ref{fig:A22-p_vs_t-tLR_dyn_capt},~\ref{fig:A22-p_vs_t-tLR_rad_infall} show a similar behaviour for other classes of non-circular orbits.
In the present section we investigate which specific features of the non-circular orbits are causing the tail enhancement.

First, we isolate the portion of the point-particle inspiral trajectory which contributes the most to the tail amplitude.
To do so, we compare the tail amplitude obtained from the numerical evolution and our model, and study how this comparison evolves as we change the initial time of the integration in our semi-analytical computation, to include progressively less inspiral history.
Beyond understanding which portion of the trajectory is determining the tail, an additional byproduct of this analysis is learning ``how much history'' needs to be included in order to obtain a reasonable estimate of the tail amplitude, within some accuracy threshold.
This information is useful e.g. when aiming to extract tail terms from simulations of comparable-mass mergers, in which only a limited  number of cycles is available (see Sec.~\ref{sec:conclusions} for a more detailed discussion).
Then, based on the intuition drawn from the above investigation, we derive an expansion that allows to deduce which specific orbital features are determining the tail behaviour.

Throughout the present section, we refer to $A_{\rm tail} \coloneq A_{22}(\tau_{\rm trans})$ as ``tail amplitude'', where $\tau_{\rm trans}=\bar{\tau}+5\tau_{220}$,
and $\bar{\tau}$ is the time of the flex in the frequency $\omega_{22}$ of the $(2,2)$ multipole, when transitioning from the fundamental mode frequency $\omega_{220}$ to a zero value, i.e. the one corresponding to the tail regime.
The factor $5\tau_{220}$ serves to exclude the QNMs portion and we found it to be a reasonable approximation for the time at which the tail starts to dominate, see also Ref.~\cite{Carullo:2023tff}.

\subsection{Eccentric binaries}

\begin{figure*}[t]
\includegraphics[width=0.9\textwidth]{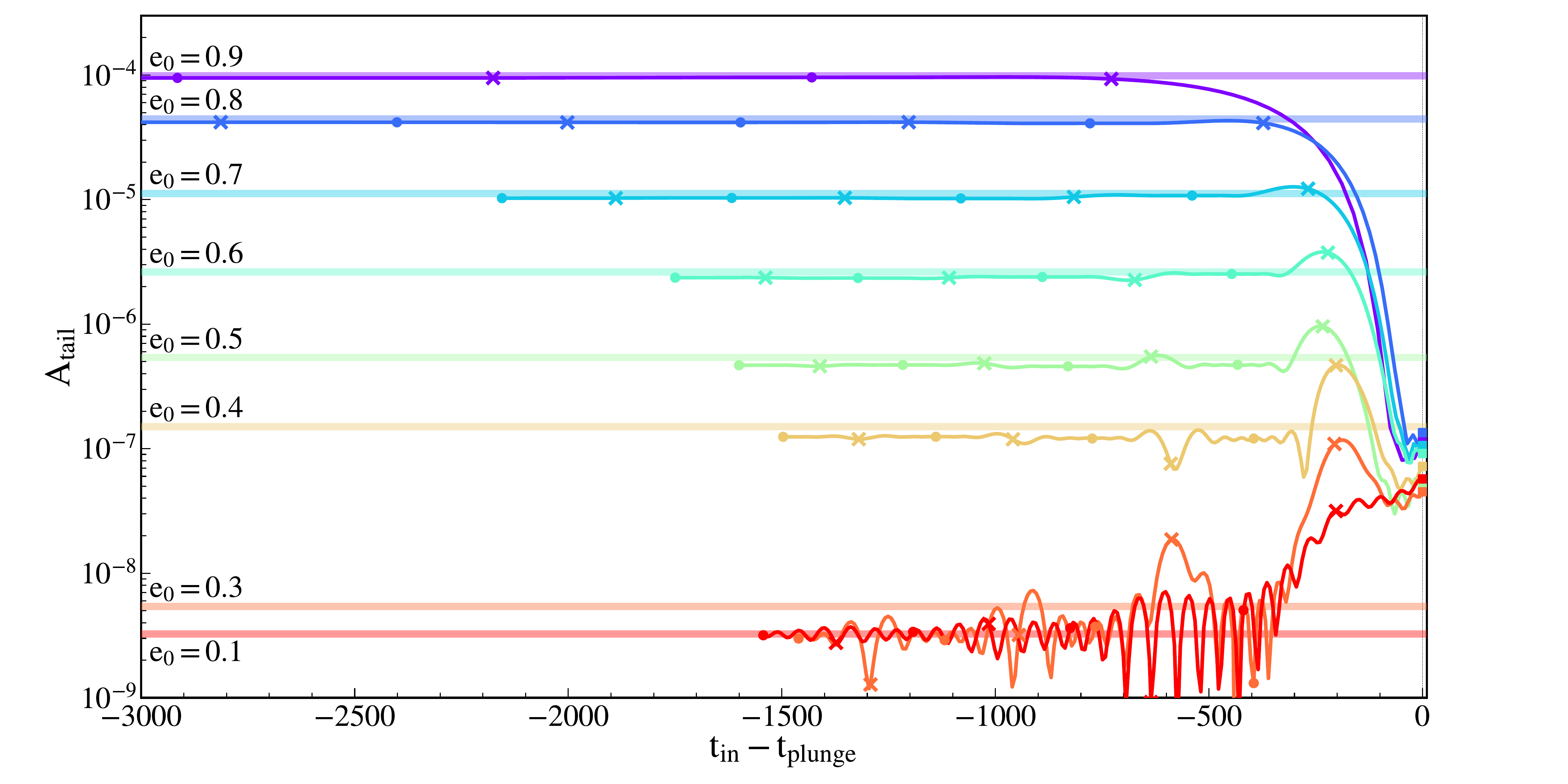}
\caption{
Horizontal opaque lines: amplitude of the $(2,2)$ multipole, $A_{22}$ at the transition between QNM and tail-dominated regime, obtained integrating the full problem in Eq.~\eqref{eq:RWZ_equation},~\eqref{eq:InitialCondition} with the \RWZ~code. 
Each color correspond to one of the simulations in Table~\ref{tab:sims_ecc}, labeled by the value of the initial eccentricity $e_0$. 
For readability of the plot, we do not display the results relative to $e_0=0.2$, characterized by the same oscillatory behaviour of $e_0=0.1$.
The thick lines represent the values of $A_{22}$ at $\tau_{\rm trans}$ computed with the model Eq.~\eqref{eq:tail_signal_analytic_expr}, by changing the initial time $t_{\rm in}$ of integration on the $x$-axis.
The $x$-axis is also rescaled with respect to the plunge time, i.e. the last time at which $\ddot{r}\equiv 0$ before the merger.
Crosses (dots) indicate the time of apastra (periastra).
\label{fig:AT_vs_t-tmerg_eccs}}
\end{figure*}
\begin{figure*}[t]
\includegraphics[width=0.99\textwidth]{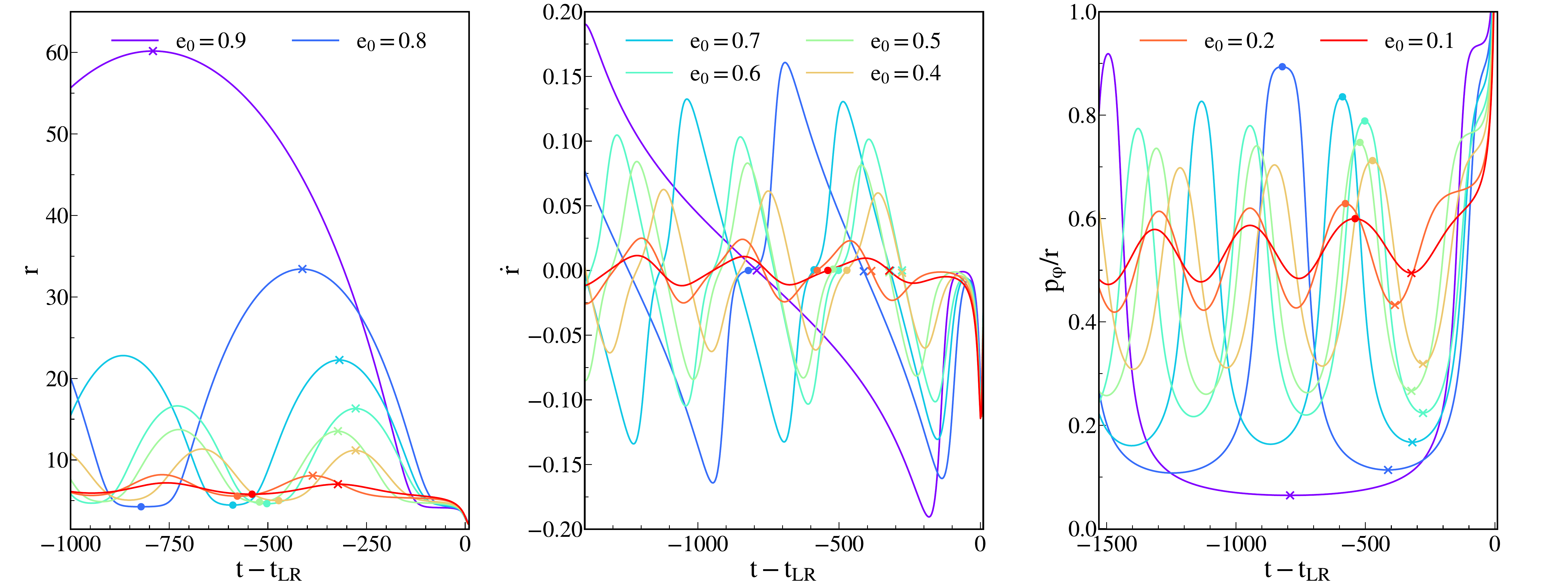}
\caption{Radius (left), radial velocity (center) and angular momentum per unit $r$ (right) vs the time rescaled with respect to the time of the light ring crossing. 
Different colours identify the different eccentric simulations of Table~\ref{tab:sims_ecc}, labeled by the initial eccentricity $e_0$.
The crosses (dots) indicate the last apastron (periastron).}
\label{fig:dynamics_ecc}
\end{figure*}

The results of the analysis described above are depicted in Fig.~\ref{fig:AT_vs_t-tmerg_eccs}, for the simulations in Table~\ref{tab:sims_ecc}.
Note that, for all the eccentricities available, we start the integration in the analytical model Eq.~\eqref{eq:tail_signal_analytic_expr} from the fourth periastron before the merger.
Numerical evolutions include more orbits, however, we did not considered relevant to include additional past history in the analysis, due to the converging behaviour of $A_{\rm tail}$ vs $t_{\rm in}$.

For intermediate to high eccentricities, the motion around the last apastron is the part of the trajectory that mostly determines $A_{\rm tail}$.
In particular, for high eccentricities, one could consider only the motion from the last apastron, and still correctly determine $A_{\rm tail}$.
Instead, when considering intermediate eccentricities, an oscillatory behaviour arises.
We interpret these oscillations as due to ingoing and outgoing motion near the last apastron, generating tail terms that are comparable in modulo but opposite in sign, leading to cancellations.
From this picture, we get the heuristic intuition that the tail is enhanced by a motion happening near an apastron, at large distances from the BH, $r\gg 2$.
As can be seen in Fig.~\ref{fig:dynamics_ecc}, this regime corresponds to small radial and angular velocities $\dot{r},p_{\varphi}/r\ll 1$. 
To verify such intuition, we expand Eq.~\eqref{eq:tail_signal_analytic_expr} according to these conditions, starting from the large distance ($r\gg 2$) approximation, yielding
\begin{equation}
\begin{gathered}
        \Psi_{\ell m}(\tau,\rho_+)=\int_{T_{in}}^{T_{f}}dt'\frac{r^{\ell}(t')e^{-im\varphi(t')}P_{\ell m}\left(\cos\theta_0\right)}{\left(\tau-t'-\rho_+\right)^{\ell+2}}\cdot\\
        \cdot\hat{H}\left[a_1\sqrt{1-\dot{r}^2}+a_2\dot{r}\frac{p_{\varphi}}{r\hat{H}}+a_3\frac{p^2_{\varphi}}{r^2\hat{H}^2}\right] \, ,
\end{gathered}
\label{eq:tail_expanded_large_R}
\end{equation}
where the coefficients  $a_{1,2,3}$ are given by
\begin{equation}
    \begin{split}
        &a_1=a_0\left(\ell+1\right)\left(\ell+2\right) \, ,\\
        &a_2=a_04i m \, ,\\
        &a_3=a_0\left(\lambda-2m^2-2\right) \, ,\\
        &a_0=c_{\ell}\frac{8\pi\mu}{\lambda\left(\lambda-2\right)} \, .
    \end{split}
\end{equation}
We now expand in small $\dot{r},p_{\varphi}/r\ll 1$.
Considering the expression for the energy per unit mass Eq.~\eqref{eq:energy_unit_mu} in these limits, we obtain
\begin{equation}
    \begin{gathered}
        \Psi_{\ell m}(\tau,\rho_+)=\int_{T_{in}}^{T_{f}}dt'\frac{r^{\ell}(t')e^{-im\varphi(t')}P_{\ell m}\left(\cos\theta_0\right)}{\left(\tau-t'-\rho_+\right)^{\ell+2}}\cdot\\
        \cdot\left[a_1-\frac{a_1}{2}\dot{r}^2+a_2\dot{r}\frac{p_{\varphi}}{r}+\left(a_3+\frac{a_1}{2}\right)\frac{p^2_{\varphi}}{r^2}\right] \, .
\end{gathered}
\label{eq:tail_expanded_apo}
\end{equation}
This integral form confirms our previous intuition. 
The overlap between the propagator and the source is enhanced for large distances $r$, since low frequencies signals (the ones contributing to the tails) not only are scattered the most by the background, but are also emitted by a motion on large scales.
It is important to note the oscillatory term in the integrand of Eq.~\eqref{eq:tail_expanded_apo} for $m\neq 0$ modes.
This term implies that the faster $\varphi$ varies, the more destructive interference will be present between tail signals emitted close to each other.
Hence, the further from the BH is the location of the last apastron $r_{\rm apo}$ and the more time the test-particle spends near it, the more enhanced the post-merger tail will be.
Both these features are related to the eccentricity; systems with higher eccentricity have larger $r_{\rm apo}$ and smaller $\dot{\varphi}_{\rm apo}$, due to Kepler's second law.
Note that this implies that the expansion in small $p_{\varphi}/r$ in Eq.~\eqref{eq:tail_expanded_apo} is an expansion in the eccentricity\,\footnote{This intuition is in agreement with the Newtonian limit, in which $\left(p^2_{\varphi}/r  -1 \right) \sim e$, see for instance the discussion in~\cite{Placidi:2021rkh}.}, as implied by Fig.~\ref{fig:dynamics_ecc}.
If for very high eccentricities, near the apastra, we can neglect the last two terms in the square parenthesis, instead these become relevant for intermediate eccentricities.
In particular, the third term depends on the sign of $\dot{r}$,  and is the one responsable for the cancellations among tails emitted close to the apastron during outgoing and ingoing motion, observed in Fig~\ref{fig:AT_vs_t-tmerg_eccs}.
The second term in Eq.~\eqref{eq:tail_expanded_apo} depends as well on $\dot{r}$, but not on its sign. It does not imply cancellation among ingoing and outgoing motion, but is part of the expansion around the apastron. 
In particular, as we move away from it, this factor, opposite in sign to the leading order, suppresses the tail.

The approximation in Eq.~\eqref{eq:tail_expanded_apo} does not hold for low-eccentricities since, in these cases, the test-particle is located near the BH during the whole last stage of the inspiral, see Fig.~\ref{fig:dynamics_ecc}.

\begin{figure*}[t]
\includegraphics[width=0.9\textwidth]{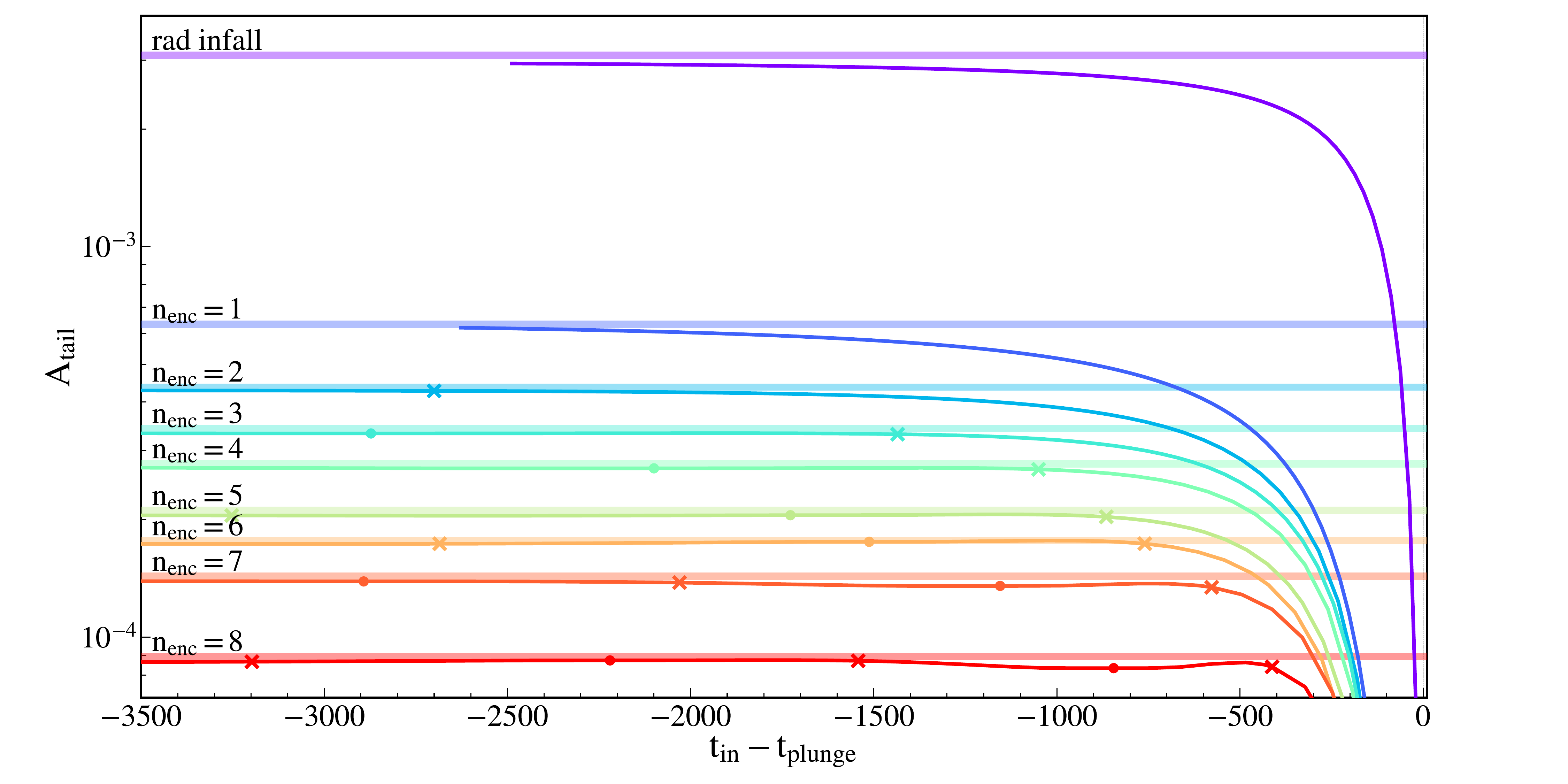}
\caption{Horizontal opaque lines: amplitude of the $(2,2)$ multipole $A_{22}$ at the transition between QNM and tail-dominated regime, obtained integrating the full RWZ problem in Eq.~\eqref{eq:RWZ_equation},~\eqref{eq:InitialCondition} with the \RWZ~code. Each color correspond to one of the captures in Table~\ref{tab:sims_enc_scatt}, labeled by the value of the number of encounters $n_{\rm enc}$, plus a radial infall from the same distance $r_0=300$.
The solid lines represent the values of $A_{22}$ at $\tau_{\rm trans}$ computed with the model Eq.~\eqref{eq:tail_signal_analytic_expr}, by changing the initial time $t_{in}$ of integration that is on the $x$-axis, rescaled with respect to the plunge time, i.e. the last time at which $\ddot{r}\equiv 0$ before the merger.
Crosses (dots) indicate that the starting time of integration is a turning point far from (close to) the BH.
\label{fig:AT_vs_t-tmerg_dyn_capts}}
\end{figure*}
\begin{figure*}[t]
\includegraphics[width=0.99\textwidth]{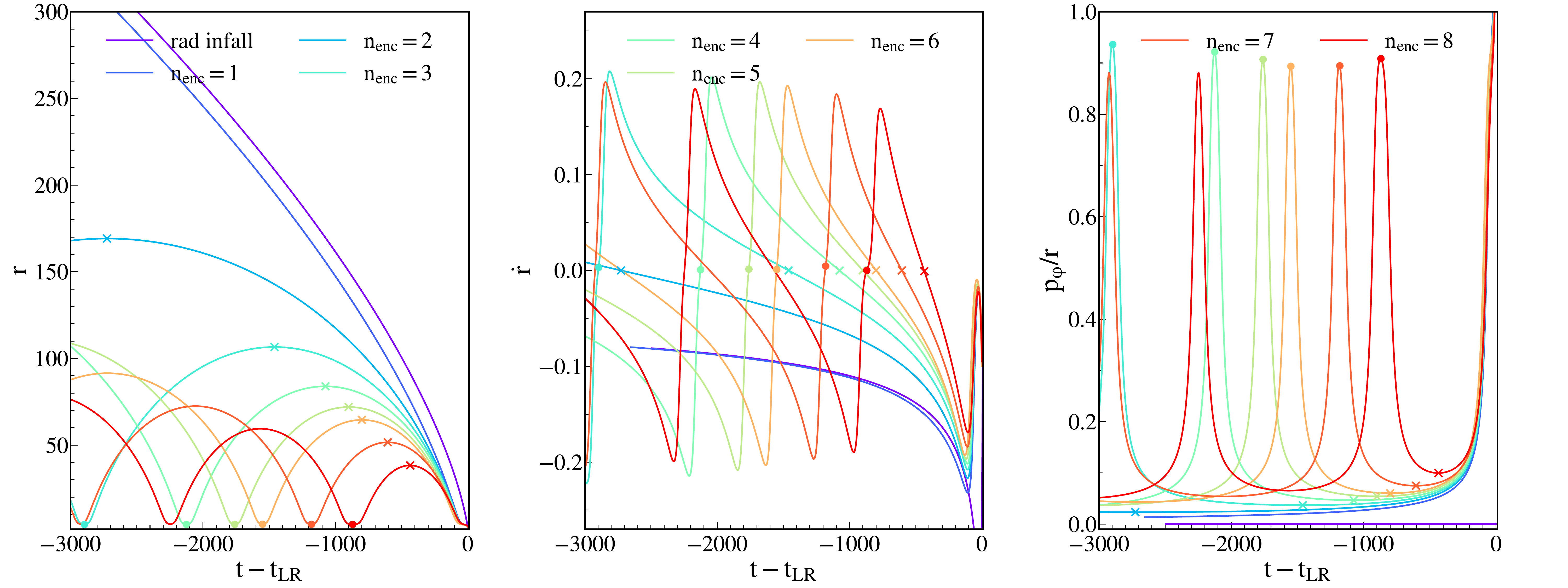}
\caption{Radius (left), radial velocity (center) and angular momentum per unit $r$ (right) vs the time rescaled with respect to the time of the light ring crossing. 
Different colours identify the different simulations of dynamical captures listed in Table~\ref{tab:sims_enc_scatt}, labeled by the number of encounters $n_{\rm enc}$, plus a radial infall from the same distance $r_0=300$.
The crosses (dots) indicate the last turning points, further (closer) to the BH.
\label{fig:dynamics_dyn_capt}}
\end{figure*}

From the analysis in this section, it emerged that it is not practical to describe $A_{\rm tail}$ in term of the initial eccentricity $e_0$.
As already discussed at the beginning of Sec.~\ref{sec:comparision}, the eccentricity evolves during the inspiral in a non-trivial way. Hence the initial eccentricity, taken alone, it is not sufficient to determine $e$ at late stages of an orbit.
The two parameters that have been previously used in the literature to parametrize the merger/ringdown waveform are the eccentricity at the separatrix $e_{\rm sep}$,~\cite{Albanesi:2021rby}, and the impact parameter at the light-ring crossing/merger~\cite{Albanesi:2023bgi,Carullo:2023kvj,Carullo:2024smg}.
In Appendix~\ref{app:Atail_parametrization} we present an investigation of the behaviour of $A_{\rm tail}$ as function of $e_{\rm sep}$ and $b_{LR}$ for bounded orbits and dynamical captures, to explore the dependence on these parameters.
However, it is not straightforward to connect these quantities to the motion near the last apastron; we leave a more detailed study on the parametrization of $A_{\rm tail}$ to future work.

\subsection{Dynamical captures and radial infalls}
We analyze the trajectory of the test-particle in the dynamical captures listed in Table~\ref{tab:sims_enc_scatt}.
These systems are initialized as unbounded, and, during the first encounter, become bounded due to emission of gravitational radiation resulting in highly eccentric orbits that eventually merge.
As shown by the results in Fig~\ref{fig:dynamics_dyn_capt}, if the number of encounters $n_{\rm enc}>1$, near the last apastron the particle is far away from the BH with small tangential velocity $p_{\varphi}/r$.
As the number of encounters $n_{\rm enc}$ increases, the distance from the BH at the last apastron decreases, while the tangential velocity increases resulting in less time spent around this location. 
This is due to the fact that the GWs are emitted mainly at the turning points, thus the more encounters are present, the more the orbit loses energy and evolve towards a more "circularized" setting.
Thus, we expect a reasoning similar to the one in the previous section to hold, considering that an higher $n_{\rm enc}$ implies smaller eccentricity of the last stable orbit, as discussed in Sec.~\ref{sbsec:dynamical_bounded_case}.

We repeat the experiment of the previous section, i.e. we compare $A_{\rm tail}$ from the numerical evolutions with the one computed from the model Eq.~\eqref{eq:tail_signal_analytic_expr}, varying the starting time of integration, $t_{\rm in}$.
The part of history relevant to determine $A_{\rm tail}$ is the motion from the last apastron, in agreement with what was found in the previous section for bounded orbits.
In fact, from the comparision of the trajectories in Fig.~\ref{fig:dynamics_ecc} and Fig.~\ref{fig:dynamics_dyn_capt}, we see that, even for the larger value of $n_{\rm enc}$ considered in Table~\ref{tab:sims_enc_scatt}, the last orbit has features compatible with an eccentricity close to the two most eccentric simulations in Table~\ref{tab:sims_ecc} (see also discussion in Sec.~\ref{sec:comparision}).
In such a setting, as mentioned above, the last two terms in Eq.~\eqref{eq:tail_expanded_apo} can be neglected and the influence of in/out-going motion near the last apastron on $A_{\rm tail}$ is negligible.

\begin{figure*}[t]
\includegraphics[width=1.0\textwidth]{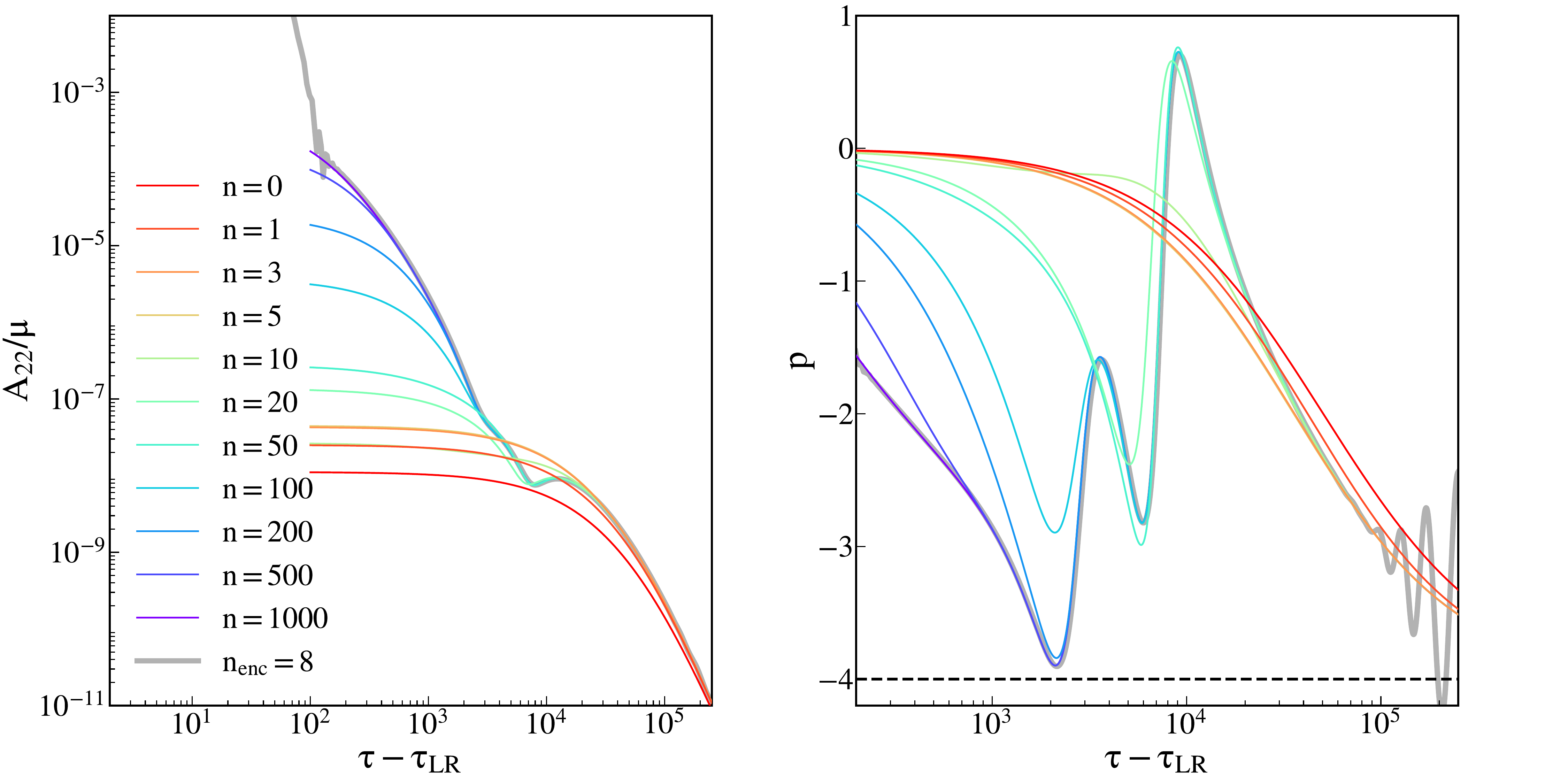}
\caption{
Left: Mass-rescaled amplitude of the $(2,2)$ waveform multipole against the observer retarded time, rescaled with respect to the time $\tau_{LR}$ at which the test-particle crosses the light-ring. 
Right: value of the tail exponent, Eq.~\eqref{eq:p_def}.
The system under study is the dynamical capture with $n_{\rm enc}=8$ in Table~\ref{tab:sims_enc_scatt}.
The time between initializing the test-particle and the light-ring crossing is $\sim 5 \cdot 10^4$.
The gray thick line correspond to the numerical experiment obtained integrating Eqs.~\eqref{eq:RWZ_equation},\eqref{eq:InitialCondition} with the \RWZ~code. High-frequency oscillations in the plot on the right for very late times ($\gtrsim 10^5$) are due to numerical noise.
The coloured lines are computed through the expansion in power-laws in the retarded time $\tau$, Eq.~\eqref{eq:tail_analytic_expansion_power_laws}. %
The label $n$ specify how many power-laws have been added to Price's law (horizontal line in the right panel).
\label{fig:A-p22_vs_t-tpeak_theta9.45_power-laws_expansion}
}
\end{figure*}

\textcolor{white}{.}

An interesting limiting case is $n_{\rm enc}=1$, for which the orbit does not have a turning point.
Consistently with the intuition developed above, we find that  contributions from all times are relevant in this case, as depicted in Fig.~\ref{fig:AT_vs_t-tmerg_dyn_capts}.
Similar considerations also holds for a radial infall starting from the same initial distance of $r_0=300$.
A curious feature to note is that in the $n_{\rm enc}=1$ case, the amplitude is suppressed with respect to a radial infall from the same distance.
This is puzzling at first since, as depicted in Fig.~\ref{fig:dynamics_dyn_capt},  the $n_{\rm enc}=1$ simulation dynamics is close to the radial infall one, except in the plunge phase that, however, does not seem to influence $A_{\rm tail}$, as shown in Fig.~\ref{fig:AT_vs_t-tmerg_dyn_capts}.
The reason of this can be traced back to $p_{\varphi}/r$, that remains small for the whole 
orbit, allowing us to consider the expansion of Eq.~\eqref{eq:tail_expanded_apo}.
The third term in the expansion, proportional to $\dot{r}$ and $p_{\varphi}/r$, acts as a small negative contribution with respect to the leading one in the $n_{\rm enc}=1$ case, due to the ingoing nature of the motion.
At the same time, the oscillating factor $\sim e^{i m\varphi}$ will induce interference among subsequent tail terms.
These terms are not present for a radial infall (the latter being a constant), explaining  the amplitude suppression in the  $n_{\rm enc}=1$ case, compared to a radial infall from the same distance.

%

\begin{figure*}[t]
\includegraphics[width=1.0\textwidth]{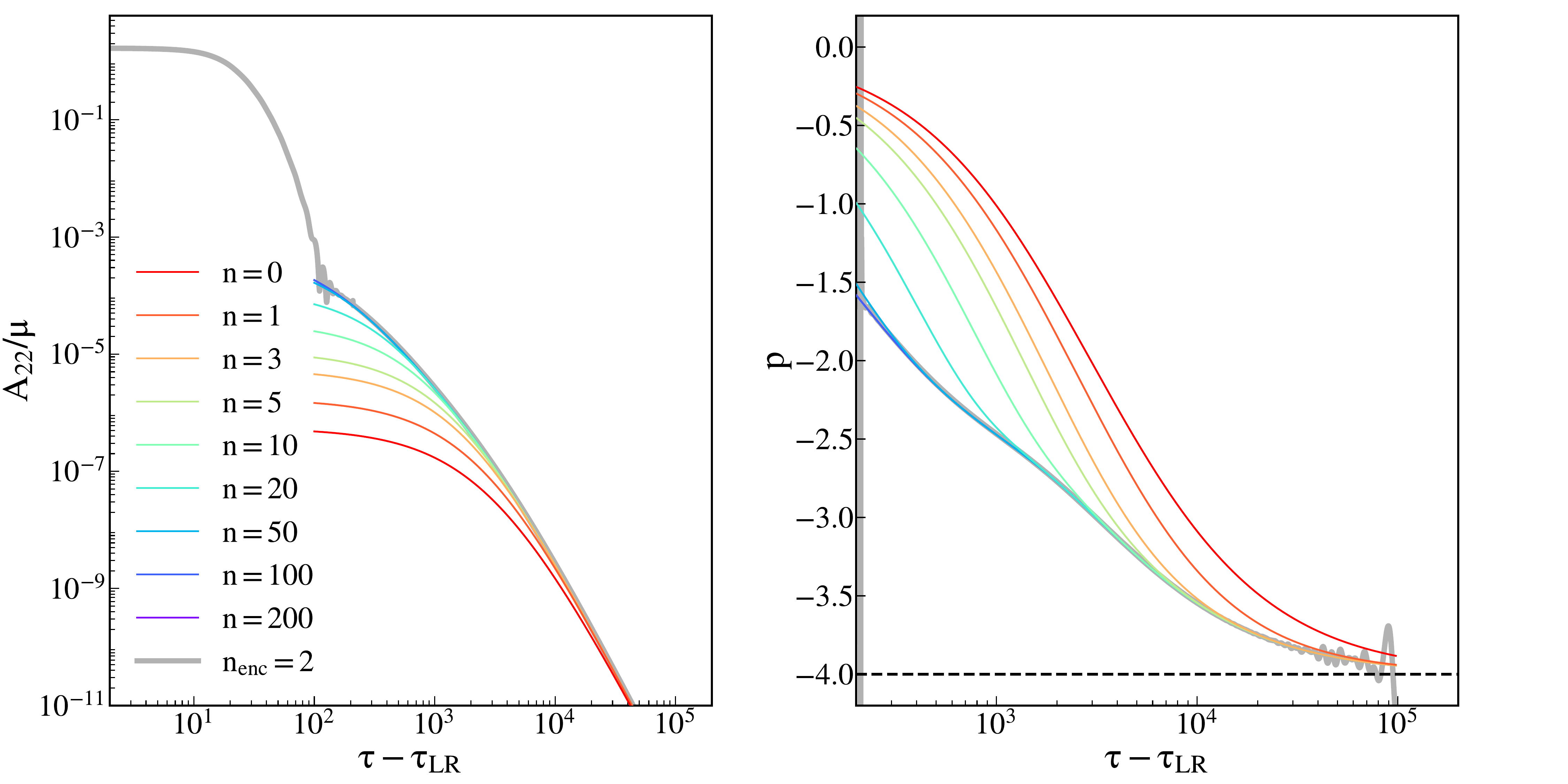}
\caption{
Left: Mass-rescaled amplitude of the $(2,2)$ waveform multipole against the observer retarded time, rescaled with respect to the time $\tau_{LR}$ at which the test-particle crosses the light-ring. 
Right: value of the tail exponent, Eq.~\eqref{eq:p_def}.
The system under study follows the same trajectory of the dynamical capture with $n_{\rm enc}=8$ in Table~\ref{tab:sims_enc_scatt}.
However, the integration included only the last $n_{\rm enc}=2$ encounters of the same evolution.
The time between initializing the test-particle and the light-ring crossing is $\sim 2 \cdot 10^3$.
The gray thick line correspond to the numerical experiment obtained integrating Eq.~\eqref{eq:RWZ_equation},\eqref{eq:InitialCondition} with the \RWZ~code. High-frequency oscillations in the plot on the right for very late times ($\gtrsim 3\cdot10^4$) are due to numerical noise.
The coloured lines are computed through the expansion in power-laws in the retarded time $\tau$, Eq.~\eqref{eq:tail_analytic_expansion_power_laws}. 
The label $n$ specify how many power-laws have been added to Price's law.
\label{fig:A-p22_vs_t-tpeak_theta9.45_child_power-laws_expansion}}
\end{figure*}

\section{Power-laws superposition}
\label{sec:exponent_phenomenology}

In the previous section, we discussed the amplitude of the tail around the time at which it starts to dominate over the QNMs-driven regime. 
Now, we focus on the phenomenology of the tail after this time. 
As derived in Secs.~\ref{sec:analytic_model} and~\ref{sec:comparision}, the tail is initially dominated by an intermediate transient, leaving place after some time to Price's law. 
Here, we characterize the decaying behaviour of the transient regime and present a very long-lived selected simulation, to explicitly show that Price's law is recovered both numerically and analytically, as expected.
In particular, we analyze the dynamical capture of Table~\ref{tab:sims_enc_scatt} with $n_{\rm enc}=8$. 
In Appendix~\ref{app:superposition_power_laws_qc_dyn-capt}, we report the same analysis for a radial infall from $r_0=300$ and for the inspiral with initial eccentricity $e_0=0.9$ of Table~\ref{tab:sims_ecc}.

Our model predicts that the intermediate behaviour can be described by a superposition of power-laws in observer retarded time $\tau$, Eq.~\eqref{eq:tail_analytic_expansion_power_laws}. 
The lowest-order of this expansion corresponds to Price's law, and thus will dominate at asymptotically late times.
This prediction is in agreement with the aforementioned numerical evolutions, as shown by the results in Fig.~\ref{fig:A-p22_vs_t-tpeak_theta9.45_power-laws_expansion} and Fig.~\ref{fig:A-p22_vs_t-tpeak_rad_infall_R0300_ecc0.9_power-laws_expansion} for the other configurations analyzed in Appendix~\ref{app:superposition_power_laws_qc_dyn-capt}.
Note that, in deriving Eq.~\eqref{eq:tail_analytic_expansion_power_laws}, we considered as upper limit of integration in the analytical model, a time $T_f$ large enough so that the source has vanished. In all of the configurations analyzed in the present section, and in Appendix~\ref{app:superposition_power_laws_qc_dyn-capt}, we study the tail from the time at which it starts dominating the strain. By this time, in the limits of double precision, the test-particle has effectively already crossed the horizon and the source has long vanished. Hence, we do not set an upper limit for the integral in Eq.~\eqref{eq:tail_analytic_expansion_power_laws}, but we integrate over all $t'<\tau-\rho_+$. However, as explained above, the integrand has vanished before the earliest time $\tau$ at which we compute the analytical tail as expansion in power-laws superposition. For instance, for the radial infall analyzed in Fig.~\ref{fig:A-p22_vs_t-tpeak_rad_infall_R0300_ecc0.9_power-laws_expansion} of Appendix~\ref{app:superposition_power_laws_qc_dyn-capt}, the source has vanished and $r=2$ at a time $\sim 54$ after the light-ring crossing.

The intermediate regime relevance can be quantified by how many power-laws are necessary to reach convergence, which in the above figure corresponds to $n \sim 1000$.
The excitation coefficient of each Price's law correction term $\sim\tau^{-n-\ell-2}$ is given by an integral over the source $S_{\ell m}(t')$, multiplied by a factor $(t'+\rho_+)^n$.
The term $(t'+\rho_+)^n$ seems to imply that the longer is the inspiral's past history, the more enhanced the excitation coefficient of $\tau^{-n-\ell-2}$ is.
However, moving $T_{\rm in}$ further and further in the past will yield a convergent behaviour of the waveform, due to the presence of the weight $S(t')$ which is suppressed in this limit.
In fact, as discussed in Sec.~\ref{subsec:inter_asymp_behaviour}, $S(t')$ can either vanish in the asymptotic past, or give rise to a suppressed $1/\tau$ tail that does not propagate at asymptotically late times (see also the discussion in Appendix~\ref{app:scattering}).

To test these predictions, we turn to the dynamical capture analyzed in Fig.~\ref{fig:A-p22_vs_t-tpeak_theta9.45_power-laws_expansion}.
At the beginning of this simulation, the system is unbounded. Going earlier in time with respect to the history considered would result in a suppressed source, hence a suppressed tail contribution. 
Instead, the source is not suppressed once the system becomes bounded, during subsequents encounters.
Hence, ``excluding" some past encounters from the integral in Eq.~\eqref{eq:tail_analytic_expansion_power_laws}, would significantly change the excitation coefficients of the power-laws therein, giving rise to a different intermediate regime of $p$.
We show this point explicitly by running the same analysis as in Fig.~\ref{fig:A-p22_vs_t-tpeak_theta9.45_power-laws_expansion}, but changing the initial time of integration in Eq.~\eqref{eq:tail_analytic_expansion_power_laws}, $T_{\rm in}$. 
We compare these results with a numerical evolution obtained initializing the test-particle (hence starting the integration) along a different point of the same inspiral trajectory.
The initial conditions on the emitted radiation are still imposed as in Eq.~\eqref{eq:InitialCondition}, while the initial conditions of the test-particle are such that the trajectory remains unchanged.
The starting time for the numerical integration is fixed to match the initial time of the analytical one.
The section of the trajectory considered is highlighted (green) in Fig.~\ref{fig:polar_plots}, in particular, we now consider a motion including only the last two encounters. 
In the original simulation the test-particle orbited around the BH for a time $\sim 5 \cdot 10^4$ before the light-ring crossing while we have now reduced this time to $\sim 2 \cdot 10^3 $.
The results of the analysis, shown in Fig.~\ref{fig:A-p22_vs_t-tpeak_theta9.45_child_power-laws_expansion}, confirm the model's prediction: faster decaying power-laws are less excited when considering a reduced amount of history. 
As a consequence, the system reaches Price's law on a shorter timescale. 
In particular, when considering a longer fraction of the inspiral, as in Fig.~\ref{fig:A-p22_vs_t-tpeak_theta9.45_power-laws_expansion}, Price's law is approached well further than a time $\sim 2\cdot 10^5 $ after the light-ring crossing (estimate based on the amplitude). 
Instead, when considering a trajectory including only the last two encounters, Price's law is recovered at $\sim 10^4 $.
In agreement with the model, the number of power-law terms required to recover the numerical result has now significantly decreased to $n\sim200$.
These analyses further stress the impact of initial conditions on the extraction of the tail exponent, in stark contrast to the amplitude at transition which is far less dependent on the trajectory integration, as shown in the previous section.

To conclude, all the simulations considered are consistent with Price's law at asymptotically late times, and slower decaying terms led by the $\tau^{-1}$ tail, discussed in Sec.~\ref{subsec:inter_asymp_behaviour}, are not present. 
Such result holds for systems that are originally unbounded as in Fig.~\ref{fig:A-p22_vs_t-tpeak_theta9.45_power-laws_expansion}, as well as for systems directly initialized as bounded, Figs.~\ref{fig:A-p22_vs_t-tpeak_theta9.45_child_power-laws_expansion},~\ref{fig:A-p22_vs_t-tpeak_rad_infall_R0300_ecc0.9_power-laws_expansion}.
This is consistent with the picture of Sec.~\ref{sec:analytic_model} and linear perturbation theory~\cite{Price:1971fb,Leaver:1986gd,Andersson:1996cm}.

\section{Summary}
\label{sec:summary}
We have investigated the late-time relaxation of a Schwarzschild BH perturbed by an infalling test-particle.
We worked in non-homogeneous perturbation theory, with a source representing the matter content of the test-particle, and an orbit driven by an highly accurate EOB-resummed analytical radiation reaction.
Analysing the late-times propagation of low-frequency signals, we derived an analytical formula for the late-time perturbations, Eq.~\eqref{eq:tail_signal_analytic_expr}.
This model is an integral over the entire past history of the system, revealing that the late-time relaxation of a BH carries imprints of the system's information in the far past. 
We tested this model against numerical evolutions obtained solving the full Regge-Wheeler/Zerilli equations, for different inspiral configurations. 
Our model is in good agreement with these non-circular results, as shown in Fig.~\ref{fig:A22-p_vs_t-tLR_eccs}, Fig.~\ref{fig:A22-p_vs_t-tLR_dyn_capt} and Fig.~\ref{fig:A22-p_vs_t-tLR_rad_infall}, respectively for bounded elliptical binaries, dynamical captures and radial infalls.

Our results shed light on the nature of tails in the presence of a source, somehow hidden in homogeneous PT~\cite{Price:1971fb,Leaver:1986gd, Andersson:1996cm}, in which the integral over the ``history'' of the system is reduced to a local expression computed on a Cauchy hypersurface.
In the non-homogeneous case, the tail is in fact due to the interaction of a time-varying quadrupole source, with the long-range structure of the background. 
Low-frequency signals emitted by the source will interact the most with the background, resulting in their scattering; as a consequence, an observer at large distances from the BH will not see the signal as travelling on the light-cone, but with all the velocities inside it.
This is the heuristical intuition behind the formal result in Eq.~\eqref{eq:tail_signal_analytic_expr}, that explains how the late-time relaxation of a BH is, in fact, a ``memory'' effect analogous to the hereditary tail of multipolar-post-Minkowskian theory~\cite{Blanchet:1992br}.

We have found that $A_{\rm tail}$, the amplitude of the tail at the transition between the QNM and tail dominance $\tau_{\rm trans}$, depends mainly on the motion near the last apastron for eccentric binaries or dynamical captures Figs.~\ref{fig:AT_vs_t-tmerg_eccs},~\ref{fig:AT_vs_t-tmerg_dyn_capts}.
In particular, we have shown that the tail is enhanced by the motion at large distances with small tangential velocity; the first condition guarantees that the overlap between the source and the tail-propagator is large, while the second is necessary to minimize destructive interference among tail signals emitted closes to each other.
For this reason, $A_{\rm tail}$ is maximized for a radial infall starting at the same distance than a capture.
For a radial infall from small distances $r_0\lesssim 200$, $A_{\rm tail}$ is larger the further the infall starts, while it saturates to a maximum value for $r_0\sim 200$, Fig.~\ref{fig:A22-p_vs_t-tLR_rad_infall}.
These results are able to explain the scaling of the tail amplitude with the progenitors' binary eccentricity observed in Ref.~\cite{Albanesi:2023bgi} in the test-mass case and discussed in Ref.~\cite{Carullo:2023tff} for comparable masses.

We have also proposed an expansion of the tail expression Eq.~\eqref{eq:tail_signal_analytic_expr}, valid at late retarded times $\tau\gg\rho_++t'$, as superpositions of power-laws in $\tau$.
This approximation allows to sort out the complicated behaviour of the tail in a contribution that scales as the leading homogeneous PT tail (the slowest power-law that dominates the asymptotic limit), with faster decaying terms whose excitation coefficients depend on the nature of the source, that eventually die out of the signal.
In particular, for the systems considered in the main text, that eventually become bounded and merge, we recover the asymptotic decay $\sim\tau^{-\ell-2}$ of Refs.~\cite{Price:1971fb,Leaver:1986gd,Andersson:1996cm}.
Slower decaying contributions, led by $\sim\tau^{-1}$, emitted during the initial unbounded stage (in the case of a dynamical capture) 
cancels out at asymptotically late times.
We have tested the expansion Eq.~\eqref{eq:tail_analytic_expansion_power_laws} against numerical experiments, to understand the relevance of the fast decaying contributions.
The results of Fig.~\ref{fig:A-p22_vs_t-tpeak_theta9.45_power-laws_expansion} and Appendix~\ref{app:superposition_power_laws_qc_dyn-capt} show that a large number of fast-decaying power-laws is necessary to correctly reproduce the numerical experiments, starting from the time $\tau_{\rm trans}$ at which the tail starts to dominate over the QNMs.

For completeness, we have also analyzed the behaviour of higher multipoles of the waveform: an odd mode, the $(3,2)$, and the $(4,4)$, in Appendix~\ref{app:higher_modes}. 
We observe for both modes a similar enhancement of the tail amplitude with the initial eccentricity, Fig.~\ref{fig:A32-p_vs_t-tLR_eccs},~\ref{fig:A44-p_vs_t-tLR_eccs}, and a similar level of agreement with the numerical evolutions.
These results are particularly relevant for the $(4,4)$ mode, since this is the lowest mode in which quadratic QNMs significantly appear for a binary merger.
A complete description of the ringdown in a non-linear setting would in fact benefit from the inclusion of the tail, when considering a generic planar orbit.

All our results are expressed in terms of the radiative coordinate $\tau$, as observed at $\mathcal{I}^+$, acting as a very good approximation to what would be observed in a real detection on Earth.  
However, when performing numerical simulations in a fully non-linear setting the signal is often extracted at finite distance in terms of the time coordinate $t$.
Hence, to connect with these latter studies, we have studied a configuration extracted at finite distance in Appendix~\ref{app:finite_distance}; as shown in Fig.~\ref{fig:A22-p_vs_t-tpeak_finite_distance}, such settings preserve the scaling of the tail with the progenitors' binary eccentricity, however the amplitude of the tail is suppressed of at least one order of magnitude, also for a radial infall.

In Appendix~\ref{app:Atail_parametrization} we show the behaviour of the tail amplitude by the time it starts to dominate the strain, $A_{\rm tail}$, as function of the eccentricity at the separatrix $e_{\rm sep}$ and the impact parameter in Eq.~\eqref{eq:impact_parameter_b_DEF} at the light-ring crossing $b_{LR}$. From our results, it emerges that $e_{\rm sep}$ is more suited to describe $A_{\rm tail}$ than $b_{LR}$. In particular, $A_{\rm tail}$ is not a function of $b_{LR}$, since it exhibits a double-valued behaviour with respect to this parameter.

Finally, in Appendix~\ref{app:scattering} we have studied a scattering case. In this setting, the source is present at all times, hence it will continue to emit signals at asymptotically late times. 
Our model fails to describe such system, however, when appropriately regularised, it still predicts a $\tau^{-1}$ tail signal travelling marginally close to the flat light-cone, in agreement with the classical soft graviton theorem~\cite{Saha:2019tub,Sahoo:2021ctw}.

\section{Future developments}
\label{sec:conclusions}

The present work paves the way to many further investigations.
First, the large tail enhancement predicted by our model might render the tail a phenomenon of observational interest, and a dedicated study is needed to asses the observability of this low-frequency component.
To our knowledge, tail observability studies have been proposed only in the inspiral regime, studying the signal dephasing~\cite{Blanchet:1993ec,Blanchet:1994ez} or as a correction to the signal luminosity~\cite{Poisson:1993vp,Poisson:1994yf}. 
A targeted investigation would also be needed to understand what kind of information could be extracted from the tail, if observed.
For instance, due to the dependence of $A_{\rm tail}$ on the last orbit motion, the tail could provide stronger constraints on the inspiral orbital parameters, additionally breaking parameters degeneracies.

Our work gives insights on how to extract a tail signal from a fully non-linear numerical evolution, useful for investigations following the first exploratory results of Ref.~\cite{Carullo:2023tff}.
Such evolutions usually extract the signal at finite distance and we showed that in such case, for a highly eccentric binary (radial infall), the tail starts to dominate when the signal is at most $\sim10^{-5}$ ($10^{-3}$) times smaller than $A_{22}$ peak, at least in the test-mass limit. 
If the signal is correctly extrapolated at $\mathcal{I}^+$, for instance using a Cauchy-characteristic extraction method~\cite{Bishop:1996gt}, the tail is enhanced. 
In particular, its amplitude at the time it starts to dominate over the ringdown, $A_{\rm tail}$, is approximately one order of magnitude larger than the aforementioned results for a signal extracted at finite distance.
These results suggest that a very eccentric inspiral or radial infall, correctly extracted at $\mathcal{I}^+$, would yield the optimal configuration to observe a post-merger tail.
Relatedly, even if in principle the tail depends on the entire inspiral history, $A_{\rm tail}$ is chiefly determined by the motion in the last orbit.
Assuming that this feature correctly extrapolates to the comparable masses case (as supported by many similar examples in the two-body problem, see e.g. Refs.~\cite{Wardell:2021fyy, Islam:2023aec,Nagar:2022icd}), simulations with a short inspiral should be sufficient to preserve the $A_{\rm tail}$ scaling with the eccentricity.
Albeit we believe that the underlying physical effect discussed by some of us in Ref.~\cite{Carullo:2023tff} is correct, certain considerations highlighted above were not accounted for in that exploratory study.
The latter will soon be revisited using the knowledge we have now gained in the test-mass limit in this study, as well as in second-order perturbation and dynamical settings studied in Ref.~\cite{Cardoso:2024jme}.
Specifically, a more careful waveform extrapolation and metric reconstruction procedure will certainly improve the robustness of the result.

The agreement we found in the transient regime of the tail signal could not instead be predicted using the methods of Ref.~\cite{Cardoso:2024jme}, which however still yield the correct asymptotic limit.
Hence, according to the classification put forward in the introduction of Ref.~\cite{Cardoso:2024jme}, our results  also settle the ``transient scenario'' case, responsible for the tail enhancement in binary mergers, not considered in their investigations.
Interestingly, Ref.~\cite{Cardoso:2024jme}, found that the outgoing motion yields a distinctive new tail signature in dynamical scenarios in which one of the object escapes at infinity, a case which we did not consider here.
In fact, as mentioned in Sec.~\ref{subsec:inter_asymp_behaviour} and Appendix~\ref{app:scattering}, our model fails to accurately describe the relaxation of systems not ending in a merger. 
The reason is that if the source does not decay exponentially anywhere in time, contributions of near/on the flat light-cone propagation will be relevant at all times.
Instead, Eq.~\eqref{eq:tail_signal_analytic_expr} is only able to make predictions on hereditary effects travelling well inside the flat light-cone.
We leave for future work investigation of these components, and their application to scenarios such as scatterings, reporting preliminary investigations in Appendix~\ref{app:scattering}.
These analysis would complement what found in Ref.~\cite{Cardoso:2024jme}, which instead considered a flat light-cone propagation. 

Albeit for simplicity here we focused on non-spinning black holes, a natural extension to our analysis would be to compute corrections to our analytical model in a small spins expansion or in the full rotating case.
Results from homogeneous perturbation theory predict a power-law asymptotic relaxation of Kerr BHs, with an exponent that depends on the class of initial conditions considered~\cite{Hod:1999rx}.
It would then be important to compare these prediction with realistic settings in which the perturbation is produced by an infalling test-particle and to investigate how the scaling of the tail amplitude at $\tau_{\rm trans}$ is affected by the central BH spin. 

In the present work, we investigated the dependence of the tail amplitude by the time it starts dominating over the QNMs, $A_{\rm tail}$, vs the eccentricity at the separatrix $e_{\rm sep}$ and the impact parameter of Eq.~\eqref{eq:impact_parameter_b_DEF} at the light-ring crossing. This analysis hints $e_{\rm sep}$ to be well suited to describe $A_{\rm tail}$.
An interesting continuation of this work would be an in-depth study of the tail amplitude $A_{\rm tail}$ parametrization, using the results of Appendix~\ref{app:Atail_parametrization} as starting point.

Finally, we stress that a correct understanding of late-time relaxation of perturbed spacetimes in the most general settings, even if not directly connected to observational signatures, is a foundational problem in general relativity due to its connection with spacetime stability~\cite{Cardoso:2024jme} and to spacetime asymptotic symmetries~\cite{Choi:2024ygx}. 

\textit{Software:} The version of the \texttt{RWZHyp} code bears the tag \texttt{tails}, on the  \texttt{rwzhyp{\_}eccentric} branch.

\textit{Addendum --} Closely afterwards our study appeared in preprint, independent numerical investigations of the tail generation and eccentricity enhancement were also presented in Ref.~\cite{Islam:2024vro}. Their results are in complete agreement with the model and numerical evolutions presented here. The study, which additionally characterises the negligible impact of different prescriptions for radiation reaction on the numerical orbital evolutions, shows how the picture presented here also holds in the case of a Kerr black hole.

\acknowledgments
We thank Sebastiano Bernuzzi, together with Alessandro Nagar and Anil Zenginoglu, for the continuous development and maintenance of the \texttt{RWZHyp} software, without which this work would have not been possible.
We thank the Institut des Hautes Études Scientifiques for its warm hospitality, where part of this work was developed.
We are indebted to Thibault Damour for sharp and extremely instructive discussions, pointing us to the relevant multipolar post-Minkowskian literature and to soft graviton theorem results, and for suggesting an expansion in power-laws. 
The present research was also partly  supported by the ``\textit{2021 Balzan Prize for Gravitation: Physical and Astrophysical Aspects}'', awarded to Thibault Damour.
We are also indebted to Alessandro Nagar and Piero Rettegno for fruitful interactions on the subject of this article.
We are grateful to Emanuele Berti, Tousif Islam and Andrea Puhm for stimulating conversations, and to Maarten Van De Meent for suggesting a power-law expansion as well.
G.C. thanks Emanule Berti and the Bloomberg Center for Physics and Astronomy of The Johns Hopkins University for generous hospitality during the last stages of this work.
G.C. acknowledges funding from the European Union’s Horizon 2020 research and innovation program under the Marie Sklodowska-Curie grant agreement No. 847523 ‘INTERACTIONS’.
S.A. acknowledges support from the Deutsche Forschungsgemeinschaft (DFG) project ``GROOVHY'' 
(BE 6301/5-1 Projektnummer: 523180871).
We acknowledge support from the Villum Investigator program by the VILLUM Foundation (grant no. VIL37766) and the DNRF Chair program (grant no. DNRF162) by the Danish National Research Foundation.
This project has received funding from the European Union's Horizon 2020 research and innovation programme under the Marie Sklodowska-Curie grant agreement No 101131233.\\

\appendix
\label{app:appendix}

\section{Tail propagator}
\label{app:computations_GF}
In order to solve Eq.~\eqref{eq:GreensFun_time} with homogeneous boundary conditions, we move to the frequency domain, through a Fourier transform (FT).
We define as $\mathcal{F}$ the Fourier transform operator, adopting the following convention on the sign
\begin{equation}
    \tilde{G}(\omega;r,r')=\int_{t'}^{\infty}dt\ e^{i\omega(t-t')}G(t-t';r,r'),
\end{equation}
where $\omega$ is, for the moment, a real parameter. We will later consider its analytic continuation.
Applying $\mathcal{F}$ to both sides of Eq.~\eqref{eq:GreensFun_time}, yields
\begin{equation}
    \left[-\omega^2-\partial_{r_*}^2+V(r)\right]\tilde{G}(\omega;r,r')=\delta(r-r').
    \label{eq:GreensFun_freq}
\end{equation}
The frequency-domain retarded Green's function $\tilde{G}(\omega;r,r')$ can be computed as the combination
\begin{equation}
\begin{gathered}
    \hat{G}(\omega;r,r')=\frac{1}{W(\omega)}\left[ \theta(r-r')\hat{u}^{in}(\omega,r')\hat{u}^{out}(\omega,r)+\right.\\
    \left.\theta(r'-r)\hat{u}^{in}(\omega,r)\hat{u}^{out}(\omega,r')\right],
\end{gathered}
\end{equation}
with $W(\omega)\equiv \hat{u}^{in}\partial_{r}\hat{u}^{out}-\hat{u}^{out}\partial_{r}\hat{u}^{in}$ Wronskian of $\hat{u}$, independent solutions of the homogeneous problem associated to Eq.~\eqref{eq:GreensFun_freq}.
Here, $u^{in}$ ($u^{out}$) is a purely ingoing (outgoing) mode at $r_*\rightarrow-\infty$ ($r_*\rightarrow\infty$), limits in which the potential vanish and the homogeneous equation reduces to a plane wave equation in $(t,r_*)$.
In the following, we will always assume a situation in which the observer is located far away with respect to the source of the signal $r\gg r'$. 
For this reason, in the expression above, we only consider the first term.

We are interested in the signal generated by the interaction among the source and the long-range structure of the potential. 
Hence, we focus on the long-range propagator; following~\cite{Andersson:1996cm}, we define the function
\begin{equation}
    \psi(\omega,r) = A^{1/2}\hat{u}(\omega,r),
\end{equation}
and expand in large $r$ the differential equation it satisfies~\cite{Andersson:1996cm,Asada:1997zu}
\begin{equation}
   \left[\partial^2_\rho+1+\frac{2\eta}{\rho}-\frac{\lambda}{\rho^2}\right]\psi(\omega,r)=0,
    \label{eq:Coulomb_WaveFunction}
\end{equation}
where
\begin{equation}
    \lambda=\ell(\ell+1) \ \ \  \ \ \ \rho = \omega r \ \ \  \ \ \ \eta =  2 \omega M.
\end{equation}
The above is the Coulomb wave equation, characterized by the following independent solutions, written in terms of the confluent hypergeometric function $M(a,b,z)$
\begin{equation}
\begin{gathered}
    F_{\ell}\left(-\eta,\rho\right)=2^{\ell}e^{\pi\eta/2}\rho^{\ell+1}\frac{|\Gamma(\ell+1+i\eta)|}{(2\ell+1)!}\\ \cdot e^{-i\rho} M\left(\ell+1+i\eta,2\ell +2 , 2i\rho\right),\\
    H_{\ell}(-\eta,\rho)=-2\eta e^{-\pi\eta}F_{\ell}\left(-\eta,\rho\right) \ln(2\rho)
    + \left(\genfrac{}{}{-.2pt}{} {\mathrm{single \ valued}}{\mathrm{terms}} \right).
\end{gathered}
\end{equation}

Note that we will consider an analytical continuation of $\omega$, which means that $\rho$ will be complex, thus the first part of the solution $H_{\ell}$ is not single-valued due to the complex logarithm, and this give rise to a branch-cut in the complex frequency plane. 

Neglecting the prefactors $A(r)^{-1/2}$ in the definition Eq.~\eqref{eq:uin_uout_DEF}, we build $\hat{u}^{in/out}$ using $F_{\ell}$ and $H_{\ell}$ as:
\begin{equation}
    \begin{gathered}
        \hat{u}^{in}\left(\eta,\rho\right)=F_{\ell}\left(-\eta,\rho\right),\\
        \hat{u}^{out}\left(\eta,\rho\right)=H_{\ell}\left(-\eta,\rho\right)+ i F_{\ell}\left(-\eta,\rho\right).
    \end{gathered}
    \label{eq:uin_uout_DEF}
\end{equation}

With the construction above, $\hat{u}^{out}$ has the desired asymptotic behaviour, in fact, this solution is an outgoing wave in the limit $r\rightarrow\infty$~\cite{abramowitz1968handbook}\,\footnote{Note that we have approximated $\omega M( r/M+\log(r/M))\approx \rho $.}:
\begin{equation}
    \begin{gathered}
            \hat{u}^{out}\left(\eta,\rho\right)\approx \exp\left[-\frac{i \pi l}{2} +i\rho + i \eta\log\left(\eta\right)+\right. \\
    \left.i \arg \Gamma\left(\ell+1+i\eta\right)\right].
    \end{gathered}
    \label{eq:uout_larger}
\end{equation}
Since this expansion is probing the large scale structure, the presence of an even horizon is not anymore manifest in the Coulomb equation, Eq.~\eqref{eq:Coulomb_WaveFunction}. Thus, we do not require $\hat{u}^{in}$ to be a purely ingoing plane wave at the horizon, but simply require regularity in the limit $r\ll 1$~\cite{abramowitz1968handbook}:
\begin{equation}
    \begin{gathered}
            \hat{u}^{in}\left(\eta,\rho'\right)\approx \frac{2^{l}e^{\pi\eta/2}}{\Gamma\left(2\ell+2\right)}\left(\rho'\right)^{\ell+1}
            |\Gamma\left(\ell+1+i\eta\right)| \, .
    \end{gathered}
    \label{eq:uin_smallr'}
\end{equation}
We approximate the full solution $\hat{u}^{out}$ with its asymptotic expansion Eq.~\eqref{eq:uout_larger}, motivated by the fact that this solution is evaluated at the observer location, assumed to be $\mathcal{I}^+$ in the present work.
On the other hand, $\hat{u}^{in}$ is computed along the source, suggesting that the large $r$ approximation above will hold along the entire inspiral dynamics, for which $r>2M$, failing only in the plunge-merger, when the source is very close to the horizon $r\sim M$.
For $u^{in}$, we use the intuition that the tail is generated by the back-scattering of the low-frequency signal. Since the position of the source is finite, $r'<\infty$, and we are in the small $\omega \ll 1$ limit, we assume $\rho'\ll 1$ and approximate $u^{in}$ with Eq.~\eqref{eq:uin_smallr'}.
Following the same reasoning, we expand both solutions in small $\omega M=\eta\ll 1$, keeping only up to the first order correction~\cite{Asada:1997zu,Andersson:1996cm}. 
This yields:
\begin{equation}
    \begin{gathered}
        \hat{u}^{out}\left(\eta,\rho\right)\simeq e^{i\rho-i\ell \pi/2}\left[1+i\eta\left(\log \eta+\gamma-\sum_{k=1}^{\ell}\frac{1}{k}\right)\right] \, ,
    \end{gathered}
\end{equation}
where $\gamma$ is Euler's constant, and
\begin{equation}
    \begin{gathered}
        \hat{u}^{in}\left(\eta,\rho'\right)\simeq \frac{2^{\ell}\ell ! }{\left(2\ell+1\right)!}\left(1+\pi \eta/2\right)\left( \rho'\right)^{\ell+1} \, ,
    \end{gathered}
\end{equation}
Finally, note that the Wronskian for the solutions defined in Eq.~\eqref{eq:uin_uout_DEF} is $W=-\omega$.
Now we have all the necessary pieces to build the Green's function in frequency domain:
\begin{equation}
\begin{gathered}
        \hat{G}_{\ell}(\omega;r,r')=\frac{1}{W(\omega)}\hat{u}^{in}\left(\omega,r'\right)\hat{u}^{out}\left(\omega,r\right)\\
        =\frac{\left(-i\right)^{\ell}2^{\ell}\ell!}{\left(2\ell+1\right)!}e^{i\omega r}\omega^{\ell}\left( r'\right)^{\ell+1}\cdot\\
        \left[1+\pi M\omega+2i M \omega\left(\log 2M\omega+\gamma-\sum_{k=1}^{\ell}\frac{1}{k}\right)\right] \, .
\end{gathered}
\label{eq:GreensFun_freq_approx}
\end{equation}
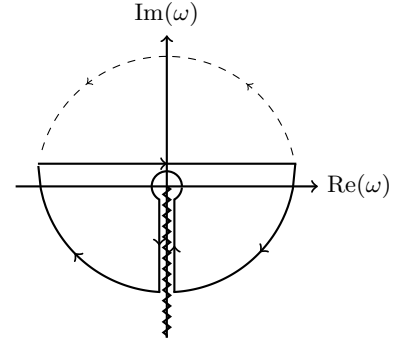
\begin{figure}[t]
\centering
\begin{tikzpicture}[decoration={
    markings,
    mark=at position 0.5 with {\arrow{>}}}
]

\def\epsilon{0.2} 
\def\R{1.7} 

\draw[->, thick] (-2,0) -- (2,0)   node[right] {$\mathrm{Re}(\omega)$};
\draw[->, thick] (0,-2) -- (0,2)  node[above] {$\mathrm{Im}(\omega)$};

\draw[thick,decorate,decoration={zigzag,segment length=4pt,amplitude=1.3pt}] (0,0) -- (0,-2);

\draw[thick, postaction={decorate}] (-\R-0.005,1.5*\epsilon) -- (\R+0.013,1.5*\epsilon);

\draw[thick] (\epsilon,0) arc (0:180:\epsilon);
\draw[thick] plot[domain=\epsilon/2:\epsilon, samples=50] ({\x},{-sqrt(\epsilon*\epsilon - \x*\x)});
\draw[thick] plot[domain=-\epsilon:-\epsilon/2, samples=50] ({\x},{-sqrt(\epsilon*\epsilon - \x*\x)});

\draw[thick, postaction={decorate}] plot[domain=(\R):(\epsilon/2-0.013), samples=50] ({\x},{1.5*\epsilon-sqrt(\R*\R - \x*\x)});

\draw[thick, postaction={decorate}] plot[domain=-\epsilon/2+0.013:-\R, samples=50] ({\x},{1.5*\epsilon-sqrt(\R*\R - \x*\x)});

\draw[thick, postaction={decorate}](\epsilon/2,-\R-0.004+1.5*\epsilon) -- (\epsilon/2,-\epsilon+0.037) ;

\draw[thick, postaction={decorate}]   (-\epsilon/2,-\epsilon+0.037)--
(-\epsilon/2,-\R-0.004+1.5*\epsilon);

\draw[dashed, thin, postaction={decorate}] plot[domain=0.0:-\R+0.03, samples=100] ({\x},{sqrt((\R+0.02)*(\R+0.02) - \x*\x)});
\draw[dashed, thin, postaction={decorate}] plot[domain=\R-0.03:0.0, samples=100] ({\x},{sqrt((\R+0.02)*(\R+0.02) - \x*\x)});

\end{tikzpicture}
\caption{Complex frequency plane. The zig-zag line represents the branch cut, the thick (dashed) line is the integrating contour appropriate to obtain the retarded (advanced) Green's function.}
\label{fig:contour_plot}
\end{figure}

If the particle would have been in a simple geodesic trajectory, e.g. a circular orbit, such that the stress energy tensor (hence the source) could be written as superposition of $\delta$-functions in values of $\omega$ finite and different from zero, then we could have solved for the signal in the real frequency domain and use the $\delta$ distributions to go back to time domain, as in~\cite{Poisson:1993vp,Poisson:1994yf}.
However, in our case we are dealing with orbits that decay in time, and there are no fixed real frequencies associated, at all times, with the evolution of the system.
In this more generic setting, we solve for the time-domain Green's function and then use the convolution Eq.~\eqref{eq:Source_GF_convolution_DEF} to solve for the signal, in time domain.
Then, to solve the inverse Fourier transform integral, it is necessary to perform an analytical continuation in $\omega$
\begin{equation}
    G_{\ell}(t-t';r,r')=\frac{1}{2\pi}\int_{\Gamma}d\omega \hat{G}_{\ell}(\omega;r,r') e^{-i\omega(t-t')}
\end{equation}
where $\Gamma$ is the path on the complex frequencies plane along which the integration is performed.
After considering an analytical continuation of $\omega$, the integrand is a polydromous function due to the presence of a logarithm with complex argument.
To deal with the logarithm it is necessary to introduce a branch-cut in the complex plane, that we fix along the negative imaginary axis, as can be seen in Fig.~\ref{fig:contour_plot}.
There are at this point two possible contours for the integration, the paths are depicted in Fig.~\ref{fig:contour_plot}, with thick and dashed lines.
We fix $\Gamma$ by requiring convergence of the integrand. 
In particular, we focus on the exponential $\exp\left[-i\omega\left(t-t'-r\right)\right]$. For an observer at $\mathcal{I}^+$, i.e. $\rho\equiv\rho_+$, recalling definition Eq.~\eqref{eq:RWZcoordinates}, we can write
\begin{equation}
    e^{-i\omega\left(t-t'-r\right)}=e^{-i\omega\left(\tau-t'-\rho_{+}\right)}.
\end{equation}
We are interested in the late-time signal\,\footnote{The contribution of the small frequency part of the signal dominates at late times.} $\tau\gg t'+\rho_+$, for which the exponential above is well behaved in the lower half plane. 
With this prescription on the contour, the only non-zero contribution to the full integral comes from the branch-cut, i.e. from the two lines running along the imaginary real axis.
The contour along these lines has opposite orientations.
As a consequence, only the multi-valued part of Eq.~\eqref{eq:GreensFun_freq_approx} gives a non-zero contribution, while the monodromous parts of the integral cancels out.
The difference between the complex logarithm, evalued on the different sides of the branch, gives a $2\pi i$ contribution.
Then, computing the integral $\omega\in(-i\infty,0]$, we obtain the radiative tail propagator\,\footnote{The $\theta$ function here serves as a reminder of the fact that we are considering the situation $\tau>t'+\rho_+$. 
However, Eq.~\eqref{eq:Retarded_GF} is the retarded propagator only in the limit $\tau\gg t'+\rho_+$, since we derived this result in frequency space assuming the limit of $\omega M\ll 1$.}~\cite{Leaver:1986gd}:
\begin{equation}
\begin{gathered}
        G_{\ell}(\tau,t';\rho_+,r')=-\theta(\tau-t'-\rho_+)  \frac{\left(-i\right)^{\ell}2^{\ell+1}\ell!}{\left(2\ell+1\right)!}\cdot \\
        \left(r'\right)^{\ell+1}\int_{0}^{-i\infty}d\omega
  e^{-i\omega (\tau-t'-\rho_+)}\omega ^{\ell+1}=\\
  \theta(\tau-t'-\rho_+) \frac{\left(-1\right)^{\ell}2^{\ell+1}\ell!(\ell+1)!M}{\left(2\ell+1\right)!}\frac{\left(r'\right)^{\ell+1}}{\left(\tau-t'-\rho_+\right)^{\ell+2}} \, .
\end{gathered}
\label{eq:Retarded_GF}
\end{equation}

\section{Source Expression} 
\label{app:source_expressions}
In the following, we show the explicit form of the source functions $f^{e/o}_{\ell m}$, $g^{e/o}_{\ell m}$ used in our paper, taken from~\cite{Nagar:2005ea}.
For the even sector

\begin{equation}
\begin{gathered}
    f^e_{\ell m}=-\frac{16\pi\mu A^2(r)Y^*_{\ell m}}{r\hat{H}\lambda\left[r(\lambda-2)+6M\right]}\left\lbrace -2im \hat{p}_{r_*}\hat{p}_{\varphi}+\right.\\
    \left. 3 M\left(1+\frac{4\hat{H}^2r}{r(\lambda-2)+6M} \right)-\frac{r\lambda}{2}+\right.\\
    \left.\frac{\hat{p}_{\varphi}^2}{r^2(\lambda-2)}\left[r(\lambda-2)(m^2-\lambda-1)+ 2M(3m^2-\lambda-5)\right]+\right.\\
    \left.\left(\hat{p}_{\varphi}^2+r^2\right)\frac{4M}{r^2}\right\rbrace \, ,
\end{gathered}
\end{equation}
\begin{equation}
    g^e_{\ell m}=-\frac{16\pi\mu A^3(r)Y^*_{\ell m}}{r\hat{H}\lambda\left[r(\lambda-2)+6M\right]}\left(\hat{p}_{\varphi}^2+r^2\right) \, .
\end{equation}

For the odd sector

\begin{equation}
\begin{gathered}
    f^o_{\ell m}=\frac{16\pi\mu \partial_{\theta}Y^*_{\ell m}}{r\lambda(\lambda-2)}\left\lbrace A(r)\left(\frac{\hat{p}_{r_*}\hat{p}_{\varphi}}{\hat{H}}\right)_{,t}-\frac{2\hat{p}_{\varphi}}{r}A^2(r)+
    \right.\\
    \left.\frac{2M}{r^2}A(r)\hat{p}_{\varphi}\left(1-\frac{p_{r_*}^2}{\hat{H}}\right)-\frac{i m }{r^2}A^2(r)\frac{\hat{p}_{r_*}\hat{p}_{\varphi}^2}{\hat{H}} \right\rbrace \, .
\end{gathered}
\end{equation}

\begin{equation}
    g^o_{\ell m}=\frac{16\pi\mu A^2(r)\partial_{\theta}Y^*_{\ell m}}{r\lambda(\lambda-2)}\hat{p}_{\varphi}\left(1-\frac{p_{r_*}^2}{\hat{H}}\right) \, .
\end{equation}

\section{Power-laws superposition: additional configurations}
\label{app:superposition_power_laws_qc_dyn-capt}
\begin{figure*}[t]
\includegraphics[width=1.0\textwidth]{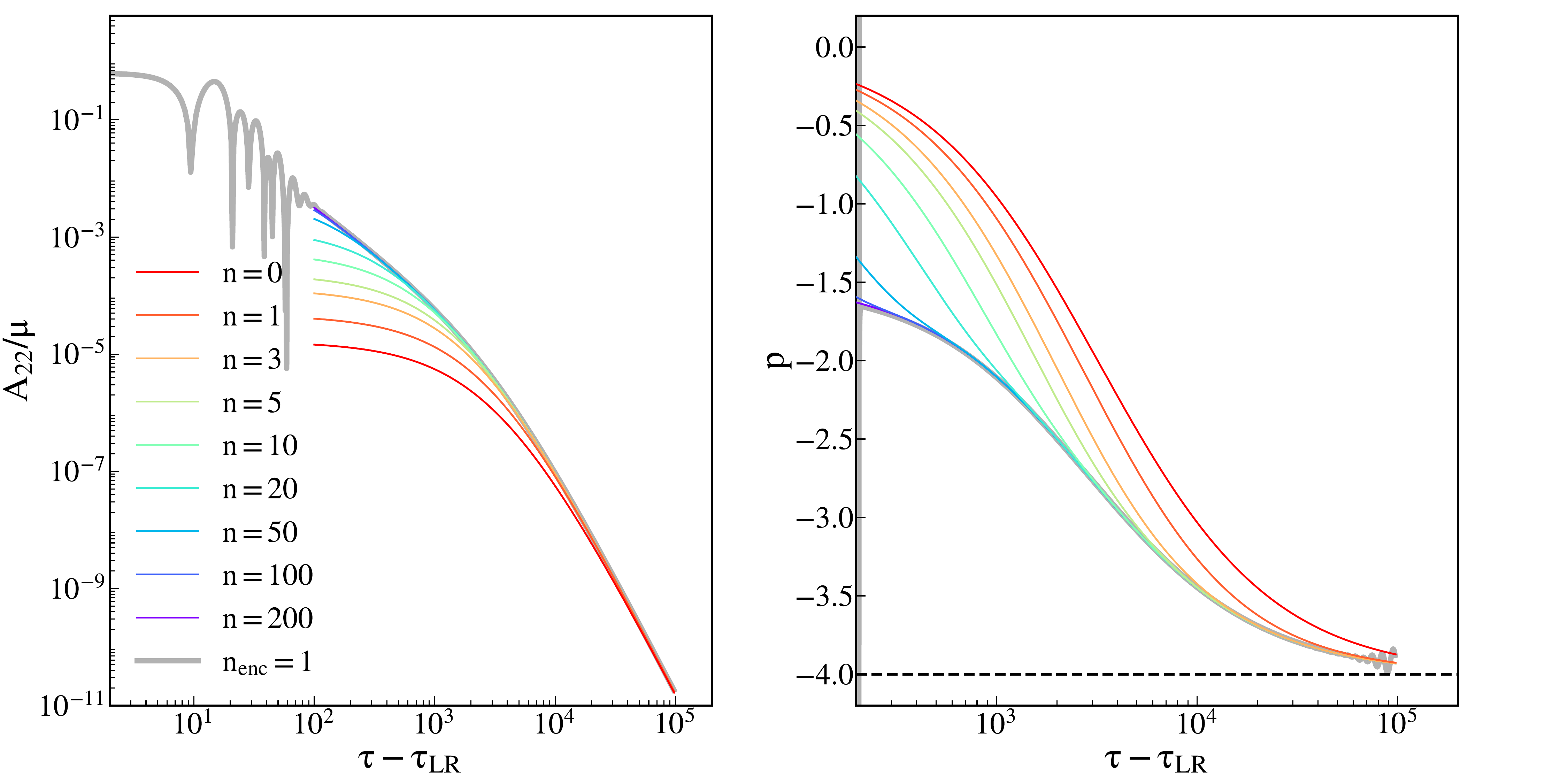}
\includegraphics[width=1.0\textwidth]{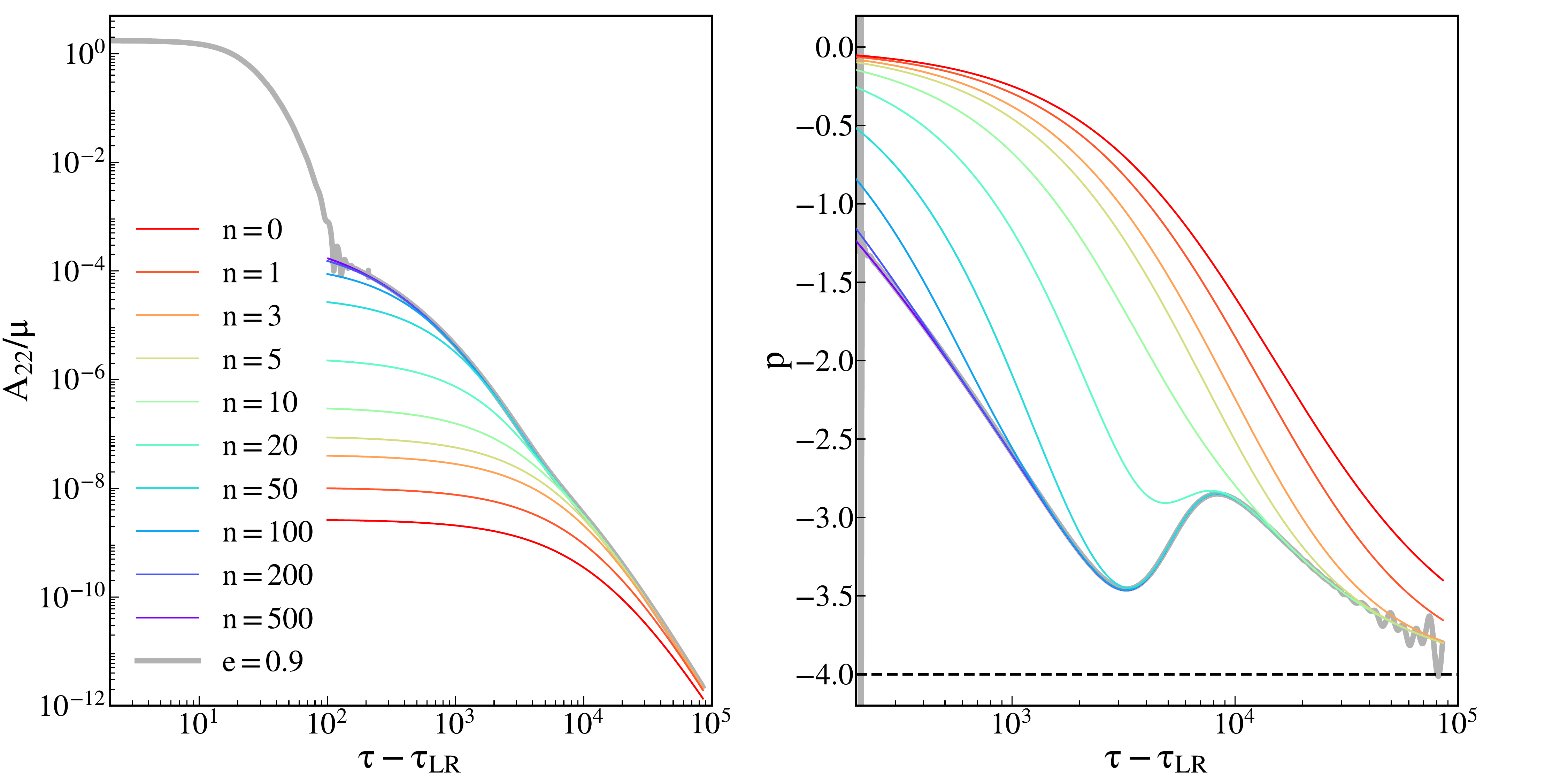}
\caption{Left: Mass-rescaled amplitude of the $(2,2)$ waveform multipole against the observer retarded time, rescaled with respect to the time $\tau_{LR}$ at which the test-particle crosses the light-ring. 
Right: value of the tail exponent, Eq.~\eqref{eq:p_def}.
Top row: the system under study is the radial infall from $r_0=300$ in Table~\ref{tab:sims_enc_scatt}.
The time between initializing the test-particle and the light-ring crossing is $\sim 2 \cdot 10^3 $.
Bottom row: the system under study is the orbit with initial eccentricity $e_0=0.9$ in Table~\ref{tab:sims_ecc}. 
The time between initializing the test-particle and the light-ring crossing is $\sim 1.5 \cdot 10^4 $.
The gray thick line correspond to the numerical experiment obtained integrating Eq.~\eqref{eq:RWZ_equation},\eqref{eq:InitialCondition} with the \RWZ~code. 
High-frequency oscillations in the plots on the right for very late times ($\sim 10^5$) are due to numerical noise.
The coloured lines are computed through the expansion in power-laws in the retarded time $\tau$, Eq.~\eqref{eq:tail_analytic_expansion_power_laws}. 
The label $n$ specify how many power-laws have been added to Price's law (horizontal line in the right panel).
\label{fig:A-p22_vs_t-tpeak_rad_infall_R0300_ecc0.9_power-laws_expansion}}
\end{figure*}
\begin{figure*}[t]
\includegraphics[width=1.0\textwidth]{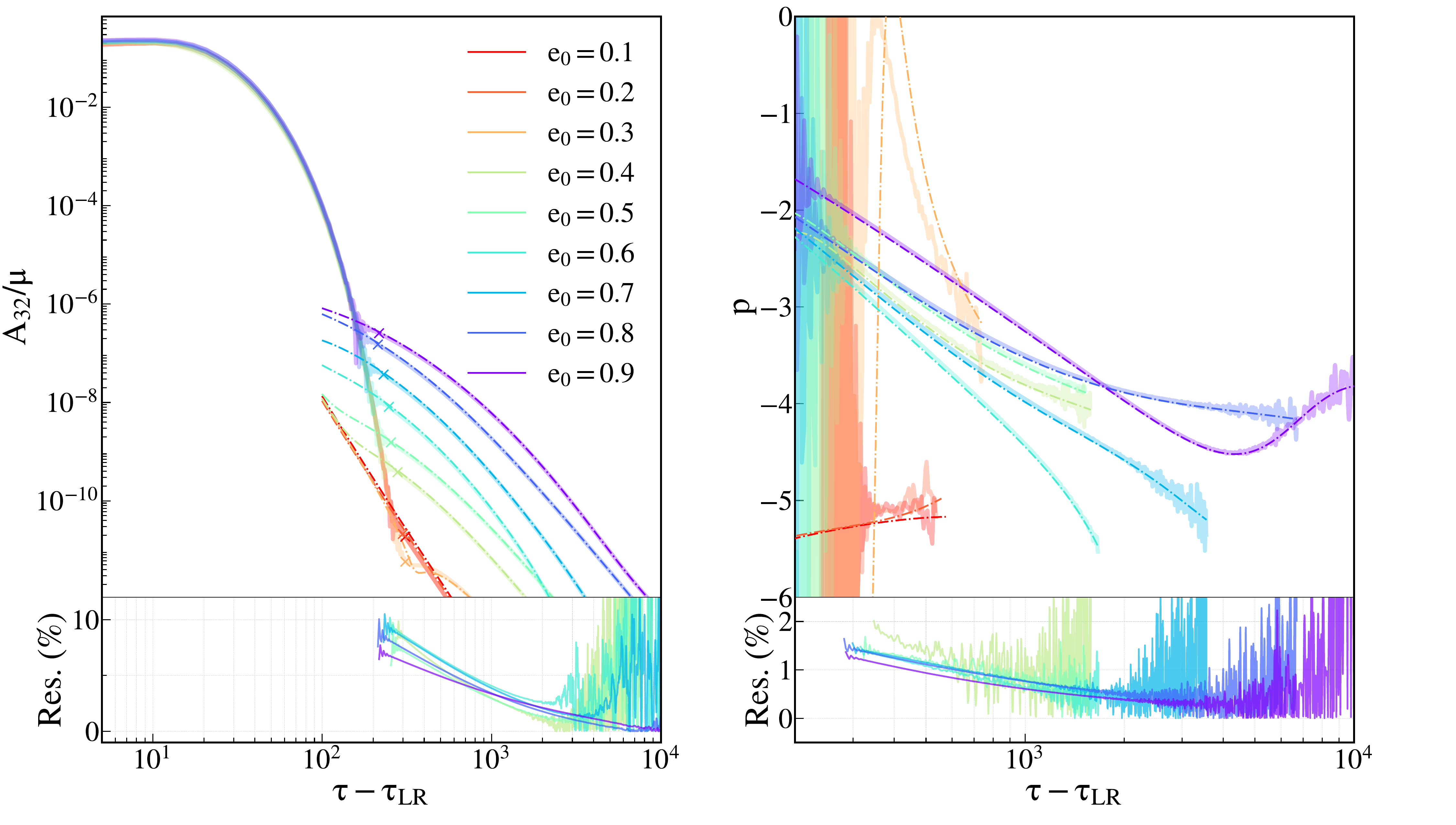}
\caption{Left: Mass-rescaled amplitude of the $(3,2)$ waveform multipole against the observer retarded time, rescaled with respect to the time $\tau_{LR}$ at which the test-particle crosses the light-ring. 
Right: value of the tail exponent, Eq.~\eqref{eq:p_def}.
The thick solid lines are the numerical evolutions, computed integrating Eq.~\eqref{eq:RWZ_equation} with the \RWZ~code. The thin dot-dashed lines are the analytical prediction for the late-time behaviour, Eq.~\eqref{eq:tail_signal_analytic_expr}.
These results are relative to the eccentric simulations of Table~\ref{tab:sims_ecc}, each labeled by the initial eccentricity $e_0$.
We cut the simulations for values of the amplitude $A_{32}=10^{-16}$, corresponding to the double precision numerical threshold.
Below each plot, for simulations with $e_0>0.3$, the absolute value of the residuals between numerical results and analytical predictions are shown, in $\%$, to quantify the agreement/mismatch.
\label{fig:A32-p_vs_t-tLR_eccs}}
\end{figure*}

Here, we test the expansion of the analytical model Eq.~\eqref{eq:tail_signal_analytic_expr} in $n$ exact power-laws Eq.~\eqref{eq:tail_analytic_expansion_power_laws} valid at large retarded times $\tau$, against numerical evolutions of the radial infall from $r_0=300$ and the  orbit with initial eccentricity $e_0=0.9$ of Table~\ref{tab:sims_ecc} of Table~\ref{tab:sims_enc_scatt}.
The results are shown in Fig.~\ref{fig:A-p22_vs_t-tpeak_rad_infall_R0300_ecc0.9_power-laws_expansion}, top and bottom row respectively.

The results are in perfect agreement with what already discussed in Sec.~\ref{sec:exponent_phenomenology}.
In particular, the history considered for the radial infall has approximately the same length ($\sim 2 \cdot 10^3 $) as the simulation in Fig.~\ref{fig:A-p22_vs_t-tpeak_theta9.45_child_power-laws_expansion}. 
As a result, the number of faster-decaying terms necessary to reach convergence ($n\sim 200$) in the post-merger tail is also of the same order as in Fig.~\ref{fig:A-p22_vs_t-tpeak_theta9.45_child_power-laws_expansion}, as is the timescale after which Price's law (i.e. agreement with the $n=0$ term) is recovered in the amplitude ($\sim 10^4 $).
The eccentric simulation in Fig.~\ref{fig:A-p22_vs_t-tpeak_rad_infall_R0300_ecc0.9_power-laws_expansion} (bottom row) has a longer inspiral before the merger, $\sim 1.5 \cdot 10^4 $. 
As a consequence, a larger number of faster-decaying terms are necessary compared to the radial plunge simulation to reach Price's law, which happens on a longer timescale ($\sim 10^5 $) consistently with the case of Fig.~\ref{fig:A-p22_vs_t-tpeak_theta9.45_power-laws_expansion}, as expected.

\begin{figure*}[t]
\includegraphics[width=1.0\textwidth]{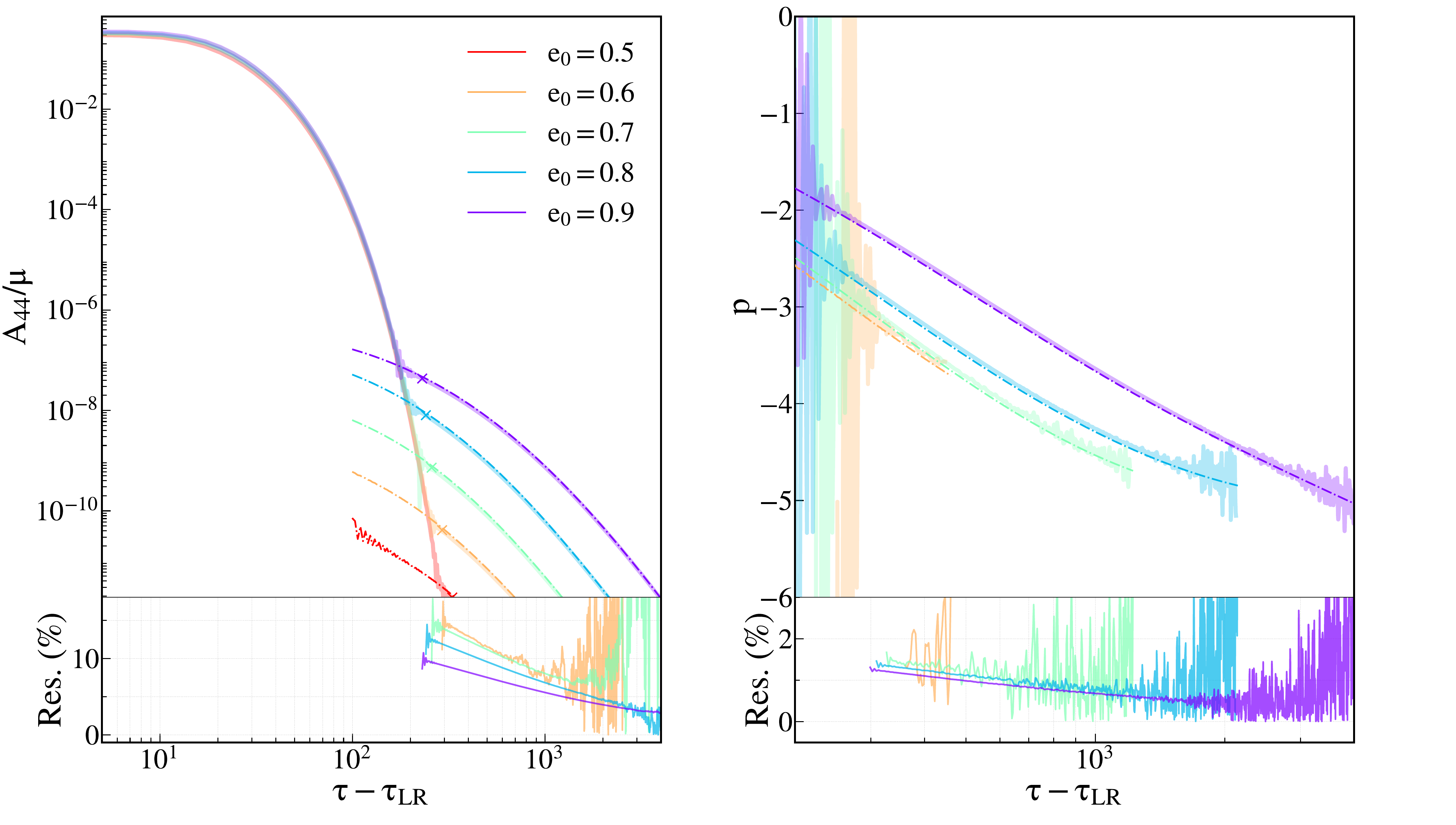}
\caption{Left: Mass-rescaled amplitude of the $(4,4)$ waveform multipole against the observer retarded time, rescaled with respect to the time $\tau_{LR}$ at which the test-particle crosses the light-ring. 
Right: value of the tail exponent, Eq.~\eqref{eq:p_def}.
The thick solid lines are the numerical experiments, computed integrating Eq.~\eqref{eq:RWZ_equation} with the \RWZ~code. The thin dot-dashed lines are the analytical prediction for the late-time behaviour Eq.~\eqref{eq:tail_signal_analytic_expr}.
These results are relative to the eccentric simulations of Table~\ref{tab:sims_ecc}, each labeled by the initial eccentricity $e_0$.
We cut the simulations for values of the (non-rescaled) amplitude $A_{44}=10^{-16}$, corresponding to the double precision numerical threshold.
Below each plot, the absolute value of the residuals between numerical results and analytical predictions are shown, in $\%$, to quantify the agreement/mismatch.
\label{fig:A44-p_vs_t-tLR_eccs}}
\end{figure*}

\section{Higher modes}
\label{app:higher_modes}
We investigate the late-time decay of other waveform multipoles $(\ell,m)=(3,2)$ and $(4,4)$.
The mode $(3,2)$ allows to test our model in the odd sector.
Moreover, such mode becomes important for spinning systems, an interesting subject of future investigations.
The mode $(4,4)$ is instead of interest since it is the lowest mode in which quadratic QNMs appear. 
These modes are long lived, thus a non-linear analysis of the late-time relaxation for generic inspirals, could benefit from correctly including the tail.

The results of the comparison between model Eq.~\eqref{eq:tail_signal_analytic_expr} and the numerical evolutions, focusing on eccentric binaries of Table~\ref{tab:sims_ecc}, is reported in Fig.~\ref{fig:A32-p_vs_t-tLR_eccs} and Fig.~\ref{fig:A44-p_vs_t-tLR_eccs} for the $(3,2)$ and $(4,4)$ mode respectively.
Note that we cut both numerical evolutions and analytic results for value of the (non-rescaled) amplitude smaller than the double precision threshold $10^{-16}$. 
This implies that, for the $(4,4)$ mode, we can only study the late-time tail in configurations with initial eccentricity $e_0\geq0.5$ of Table~\ref{tab:sims_ecc}.

We find a scaling in the amplitude of the tail with eccentricity, similar to what is found for the $(2,2)$ mode, Fig.~\ref{fig:A22-p_vs_t-tLR_eccs}. 
The model is in good agreement with the numerical experiments for high eccentricities, while it performs worst for $e_0\leq 0.3$, for the $(3,2)$ mode.
We attribute these discrepancies to the fact that the tail starts to dominate the signal very close to the double precision threshold.
Moreover, as already stated in the text, we expect the analytical model to fail for small eccentricities, since in these system the test-particle spends a greater amount of time at small distances from the BH, close to the merger, where our model is not formally valid.
To quantify the agreement/mismatch between the numerical evolutions $X_{\rm numerical}$ and the analytical results $X_{\rm analytical}$ we have shown in Figs.~\ref{fig:A32-p_vs_t-tLR_eccs},~\ref{fig:A44-p_vs_t-tLR_eccs}, for $e_0\geq 0.4$, the residuals, defined as $100*(X_{\rm numerical}-X_{\rm analytical})/X_{\rm numerical}$.
For both modes considered, the residuals of the tail exponent are in the interval $(0,2)\%$, while for the amplitudes the residuals are in the interval $(0,10)\%$, $(0,15)\%$ for the $(3,2)$ and the $(4,4)$ mode respectively.

\section{Tail observed at finite distances}
\label{app:finite_distance}
\begin{figure*}[t]
\includegraphics[width=1.0\textwidth]{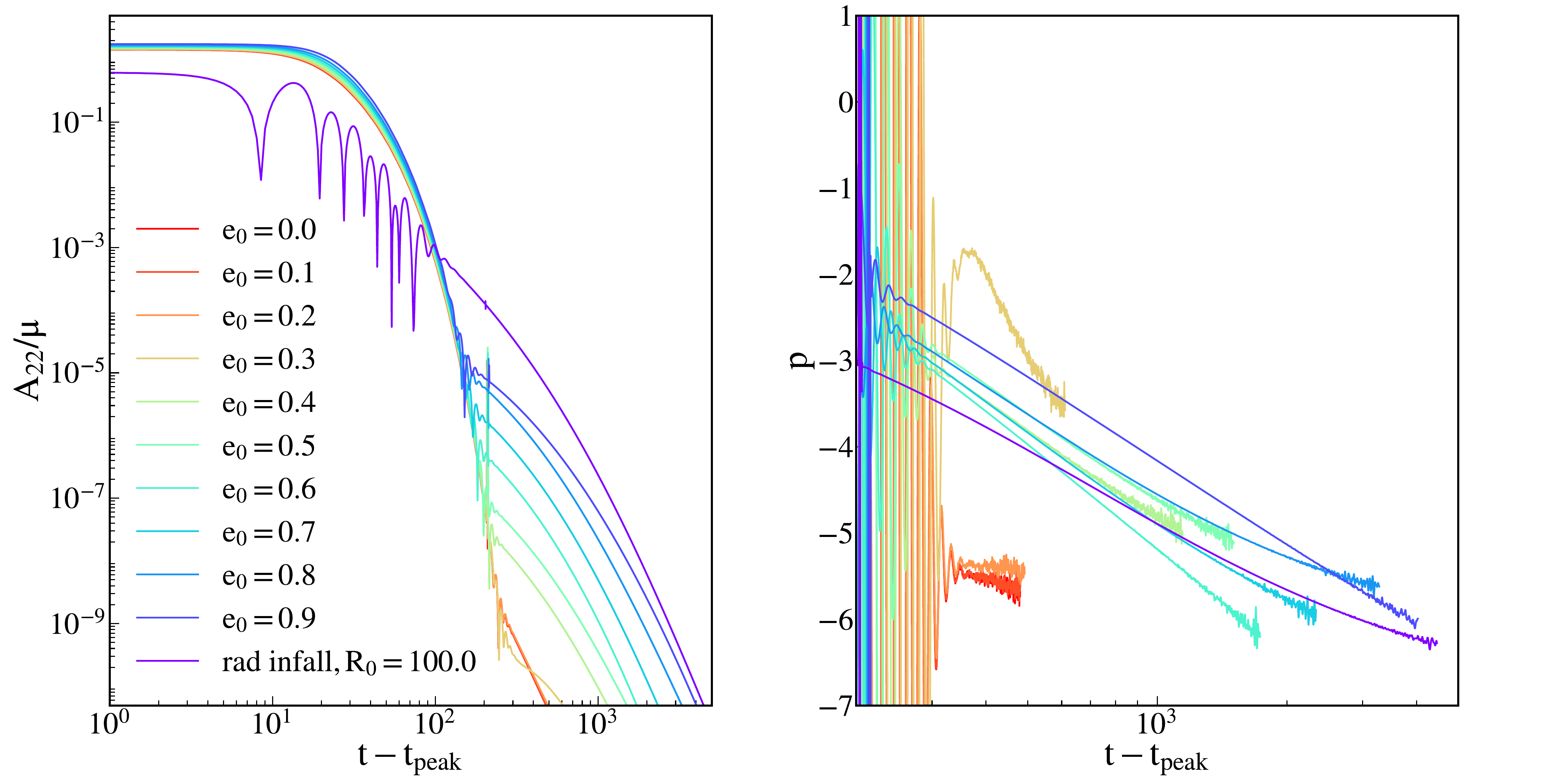}
\caption{Left: Mass-rescaled amplitude of the $(2,2)$ waveform multipole extracted at finite distance, against the observer retarded time rescaled with respect to the time of the $A_{22}$ peak. 
Right: value of the tail exponent, Eq.~\eqref{eq:p_def}.
Numerical experiments, computed integrating Eq.~\eqref{eq:RWZ_equation} with the \RWZ~code. 
These results are relative to the eccentric and quasi-circular simulations of Table~\ref{tab:sims_ecc}, each labeled by the initial eccentricity $e_0$, and a radial infall from $r_0=300$.
The observer is located at $r_{obs}=200$.
We cut the simulations for values of the amplitude $A_{44}=10^{-14}$, two order of magnitude before the numerical precision threshold dictated by double precision.
\label{fig:A22-p_vs_t-tpeak_finite_distance}}
\end{figure*}
All the results reported in the paper so far were extracted at $\mathcal{I}^+$, in terms of the retarded time $\tau$.
For what concerns real observations,
we can consider our detectors to be at $\mathcal{I}^+$ with very good approximation~\cite{Mitman:2024uss}.
However, often numerical waveforms of comparable mass mergers are extracted at finite distances.
Hence, in this appendix we analyze the tail produced in the numerical evolutions of Tabs.~\ref{tab:sims_ecc} and for the radial infall from $r_0=300$ in Table~\ref{tab:sims_enc_scatt}, as observed at finite distance.
The results are shown in Fig.~\ref{fig:A22-p_vs_t-tpeak_finite_distance}, considering
an observer placed at $r_{\rm obs}\sim 200 $ from the BH.
We observe the same scaling of the tail amplitude with the binary eccentricity as the one appearing at $\mathcal{I}^+$, Fig.~\ref{fig:A22-p_vs_t-tLR_eccs}.
For each configuration, the tail is suppressed in amplitude by approximately one order of magnitude, when observed at finite distance compared to $\mathcal{I}^+$.
The tail exponent $p$, Eq.~\eqref{eq:p_def}, is relaxing towards a smaller value than in Fig.~\ref{fig:A22-p_vs_t-tLR_eccs}.
It is not possible to determine exactly this quantity, due to the waveforms hitting the numerical double precision threshold before reaching the asymptotic relaxation regime.
However, these results are compatible with vacuum PT, i.e. a relaxation with leading behaviour $\sim t^{-2\ell-3}$ at finite distance~\cite{Price:1971fb,Leaver:1986gd,Andersson:1996cm}, compared to $u^{-\ell-2}$ ($u^{-\ell-3}$) for stationary (static) initial conditions at $\mathcal{I}^+$.

\section{Convergence tests}
\label{app:convergence_tests}

\begin{figure*}[t]
\includegraphics[width=0.95\textwidth]{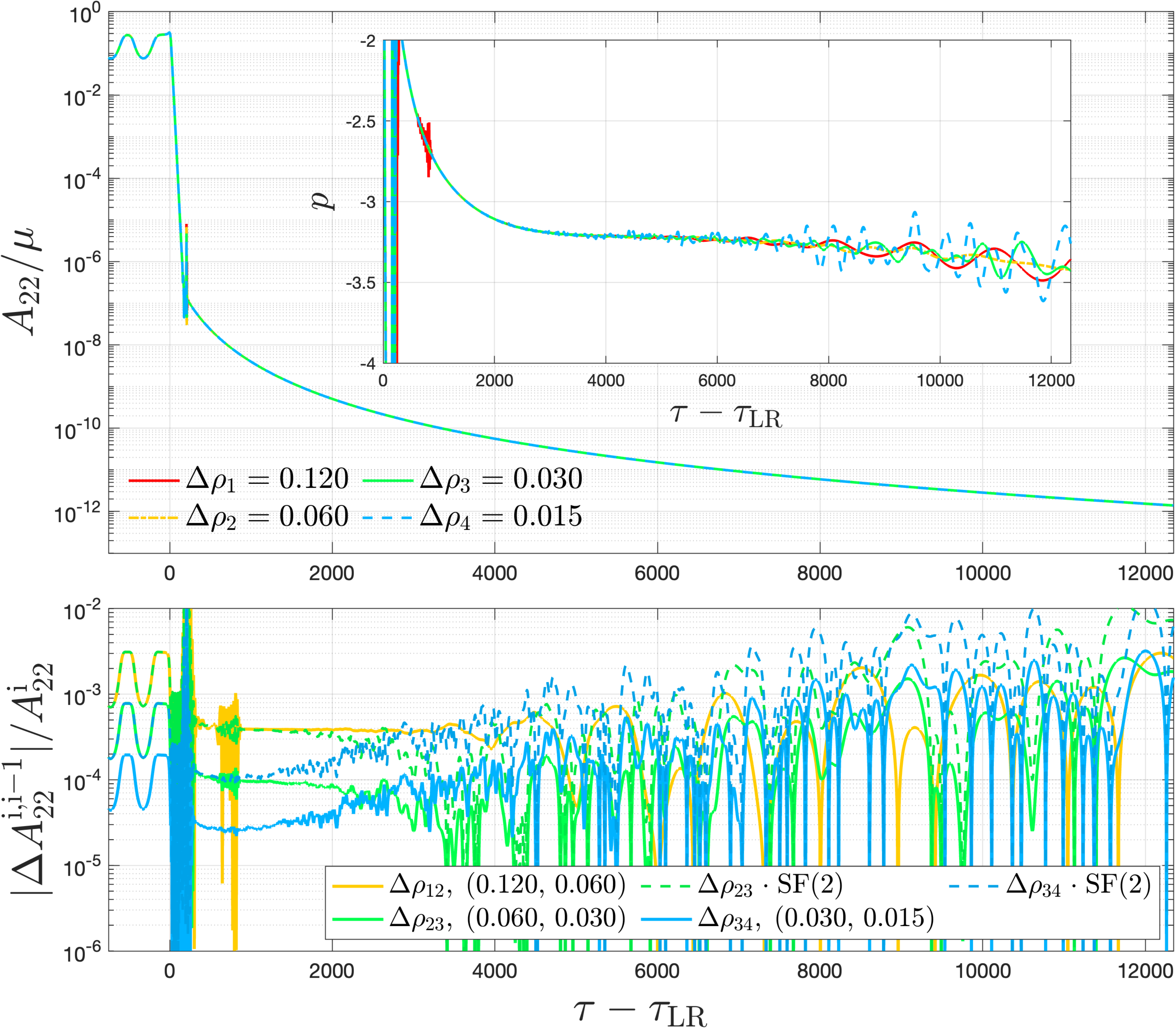}
\caption{Convergence test of the \RWZ{} code 
for the configuration
with $e_0=0.5$. Upper panel: 
amplitude of the (2,2) mode for different radial
resolutions and corresponding tail exponent (insert) 
computed according to Eq.~\eqref{eq:p_def}.
Lower panel: relative
amplitude differences among different resolutions
(solid lines). 
We also rescale the last two with a second order
rescaling factor ${\rm SF}(2)$ (dashed lines) in order to
highlight the effective second-order convergence of the code.
\label{fig:convergence_elliptic}}
\end{figure*}

We now assess the convergence of the time-domain code \RWZ{} by performing
numerical tests on an illustrative case, the initially bound configuration with $e_0=0.5$.
For all the grid configurations, we truncate our computational domain at $r_*^{\rm H} = -100$, locate future null infinity at $\rho_+ = 500$,
and perform the hyperboloidal layer matching at $\rho_{\rm match} = 400$. 
This grid setup is the one typically adopted for the runs
considered in this work, where we use a radial-step
of $\Delta \rho = 0.015$.
In this Appendix we also consider three lower resolutions, 
going up to $\Delta \rho = 0.12$.
The amplitudes of the corresponding (2,2) waveforms are shown in 
the upper panel of Fig.~\ref{fig:convergence_elliptic} for the different radial resolutions,
together with the tail decay exponents computed according to Eq.~\eqref{eq:p_def}.

To establish the convergence of the code, we consider triplets of resolutions (low/medium/high) at fixed Courant–Friedrichs–Lewy number ${\cal C}=0.5$, 
and rescale the difference between medium-high resolutions with the scaling factor $\rm SF(r)$,
defined as
\begin{equation}
{\rm SF}(r) = \frac{(\Delta\rho_{\rm L})^r - (\Delta\rho_{\rm M})^r}{(\Delta\rho_{\rm M})^r - (\Delta\rho_{\rm H})^r}.
\end{equation}
The order of convergence $r$ is determined by requiring
that the rescaled medium-high difference match the low-medium one.
We observe a 2nd order converge for the inspiral, ringdown,
and early tail.
However, the convergence starts to deteriorate from 
$2000$ after the light-ring crossing. 
Moreover, some artefacts in the data are visible in the tail decay exponent $p$ at late times, where high-frequency oscillations become particularly visible for small radial steps. 
However, all the resolutions considered provide an
accurate description of the tail, since all the relative
differences on the amplitude are well below the $1\%$ threshold.
We also performed some numerical tests considering different
grid options, finding that setups 
with larger $r_*^{\rm H}$ or
smaller $\rho_+$ provide
slightly less accurate numerical waveforms.

Finally, we highlight that the junk radiation never enters in the trajectory used in Eq.~\eqref{eq:tail_signal_analytic_expr}, by construction. In fact, as stressed in Sec.~\ref{sec:RWZ}, the fluxes used to compute the radiation-reaction effective forces, $\hat{\mathcal{F}}_{r_*}, \hat{\mathcal{F}}_{\varphi}$ in Eq.~\eqref{eq:Hamiltonian_Eqs}, are analytical.
Then, we argue that the agreement of our prediction with the late-time signal computed in the numerical evolutions, as shown in Fig.~\ref{fig:A22-p_vs_t-tLR_eccs},~\ref{fig:A22-p_vs_t-tLR_dyn_capt} and~\ref{fig:A22-p_vs_t-tLR_rad_infall}, is a test confirming the fully negligible influence of the junk on the late-time tail.

\begin{figure*}[t]
\includegraphics[width=1.0\textwidth]{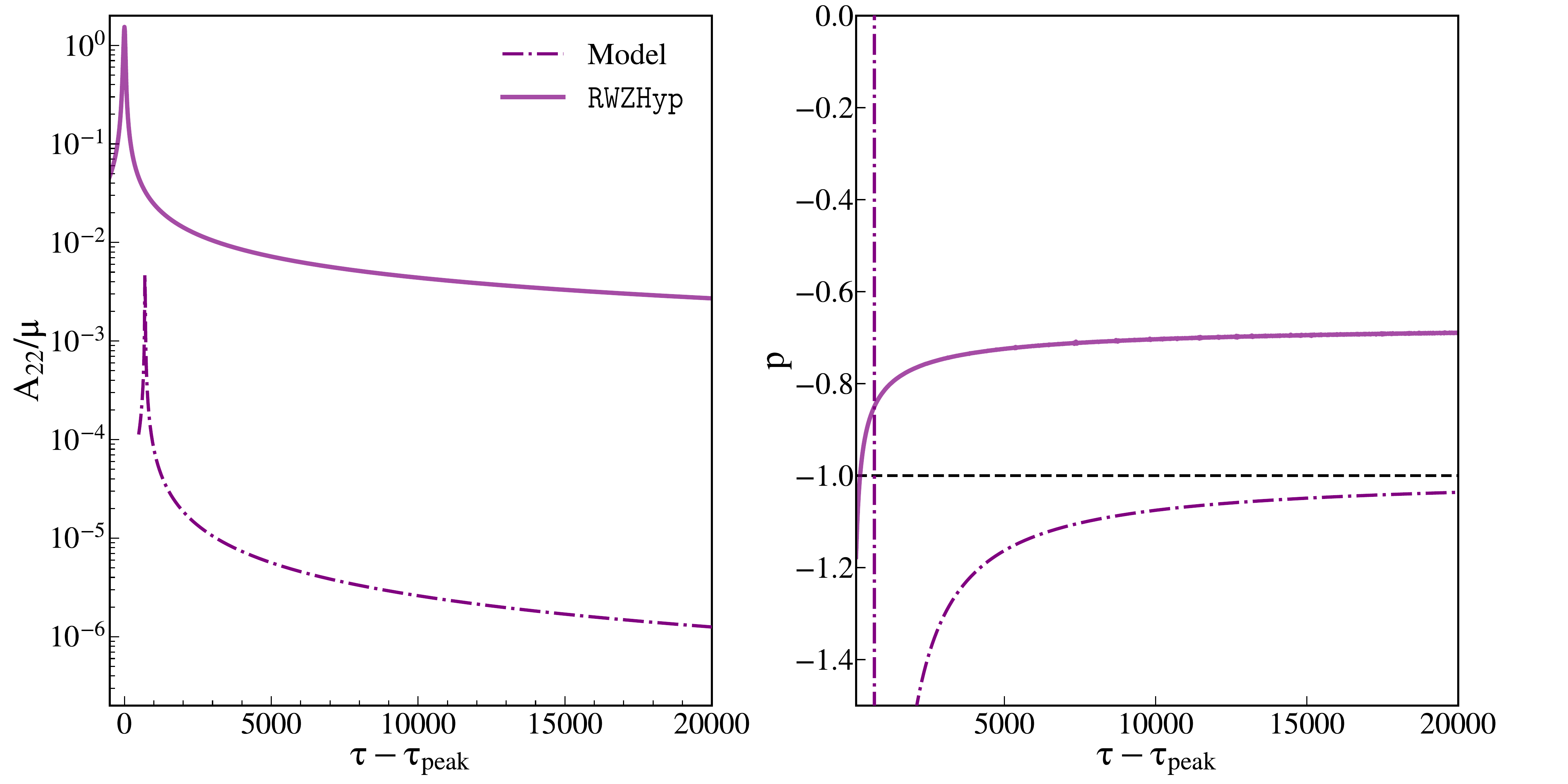}
\caption{Left: Mass-rescaled amplitude of the $(2,2)$ waveform multipole against the observer retarded time, rescaled with respect to the time of the $A_{22}$ peak. 
Right: value of the tail exponent, Eq.~\eqref{eq:p_def}.
Thick lines are results of numerical evolutions, computed integrating Eq.~\eqref{eq:RWZ_equation} with the \RWZ~code. 
Dot dashed lines are the leading order tail in Eq.~\eqref{eq:scattering_tail_with_cutoff}, normalized to remove the factor in $\Lambda$.
These results are relative a scattering simulations from $r_0=300$, with initial energy and angular momentum $E_0=1.000001$, $p_{\varphi,0}=4.070195$, evolved considering geodesics motion.
\label{fig:A22-p_vs_t-tpeak_scattering}}
\end{figure*}
\section{Scattering configurations}
\label{app:scattering}
Signals that travel well inside the flat light-cone are correctly described by Eq.~\eqref{eq:tail_signal_analytic_expr}, which instead fails to describe signals propagating on or marginally close to it.
This is manifest in the singular behaviour of the integrand, when computed at the upper bound of integration $t'+\rho_+\simeq\tau$.
If we consider systems that eventually merge, and focus on the signal emitted at asymptotically late times, we never encounter such singularity. 
For these systems, the source (hence the integrand) decays exponentially after the light-ring crossing. 
It is interesting to investigate what happens to our model, if we try to apply it to systems that do not merge. 
An example is a scattering situation; in such setting the test-particle is unbounded from the BH, and the source never vanishes.
Then, we introduce a dimensionless timescale $\Lambda$ in the upper limit of integration in Eq.~\eqref{eq:tail_signal_analytic_expr} $\tau-\rho_+\rightarrow(1-\Lambda)\left(\tau-\rho_+\right)$, that effectively select only signals travelling with velocities $\leq \Lambda$.
We consider a test-particle travelling far away from the BH, $r'\gg M$ with small constant velocity, such that the source contribution in the integral form Eq.~\eqref{eq:tail_signal_analytic_expr} is proportional to $\propto \mu|v|^2 t'^2$. The signal observed at $\mathcal{I}^+$ as predicted by Eq.~\eqref{eq:tail_signal_analytic_expr} is
\begin{equation}
\begin{gathered}
    \psi\propto \mu v^2\int_{T_{in}}^{(1-\Lambda)\left(\tau-\rho_+\right)}dt'\frac{t'^2}{\left(\tau-\rho_+-t'\right)^4}.
    \label{eq:tail_signal_scattering}
\end{gathered}
\end{equation}
The expression above can be solved analytically. We assume the observer at $\tau-\rho_+\gg T_{\rm in}$ and keep corrections up to $\mathcal{O}\left[\left(\frac{T_{\rm in}}{\tau-\rho_+}\right)^4\right]$. Then Eq.~\eqref{eq:tail_signal_scattering} becomes
\begin{equation}
\begin{gathered}
        \psi\simeq -8\pi Y_{\ell m}^*c_{\ell}\frac{1+2\ell}{\ell(\ell+1)-2}|v|^2\mu \cdot\\\left[\frac{(-1 + \Lambda)^3 }{
 3  \Lambda^3}\frac{1}{\tau-\rho_+}+\frac{T_{\rm in}^3}{3 \left(\tau-\rho_+\right)^4} \right].
\end{gathered}
\label{eq:scattering_tail_with_cutoff}
\end{equation}
There is a clear issue: as we consider signals propagating marginally close to the flat light-cone $\Lambda\rightarrow 0$, the amplitude of the $\left(\tau-\rho_+\right)^{-1}$ tail in the above expression diverges. 
In the context of the classical soft graviton theorem, logarithmic corrections to the scattering amplitude give rise to a $\tau^{-1}$ tail~\cite{Saha:2019tub,Sahoo:2021ctw}. 
Our model appears to be in agreement with this prediction. However, we regard this result as incomplete, due to the presence of the arbitrary cutoff $\Lambda$.
We leave to future work either the physical interpretation of the scale factor $\Lambda$, or a ``renormalization" procedure to get rid of this cutoff.

For completeness, we compare the leading $\left(\tau-\rho_+\right)^{-1}$ term of Eq.~\eqref{eq:scattering_tail_with_cutoff} with a numerical scattering evolution.
For this simulation we do not include radiation-reaction in the Hamiltonian equations of motion Eq.~\eqref{eq:Hamiltonian_Eqs}, thus the trajectory is geodesic.
The results are shown in Fig.~\ref{fig:A22-p_vs_t-tpeak_scattering}, where we have normalized the predicted behaviour in order to remove the cutoff $\Lambda$.
As expected, the analytical model fails completely to reproduce the correct amplitude of the signal.
The numerical evolution seems to converge towards a slower decay than $\left(\tau-\rho_+ \right)^{-1}$.
We leave a more in-depth investigation of the scattering scenario, both numerical and analytical, to future work.


\section{Tail amplitude parametrization}
\label{app:Atail_parametrization}
\begin{figure*}[t]
\includegraphics[width=1.0\textwidth]{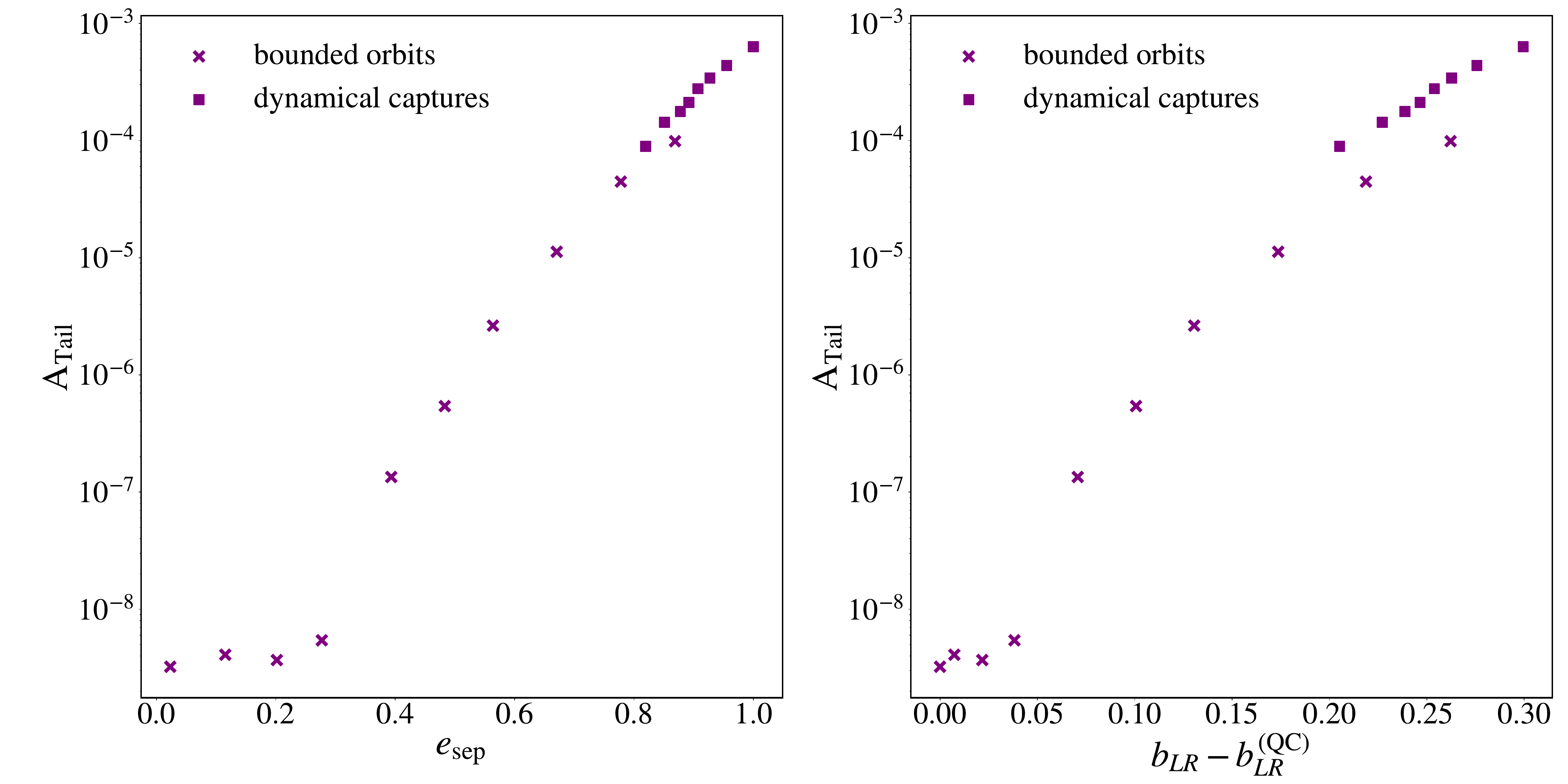}
\caption{Amplitude of the $(22)$ mode at the time at which the tail starts to dominate over the QNMs, $A_{\rm tail}$, vs the eccentricity at the separatrix $e_{\rm sep}$ (left) and the impact parameter in Eq.~\eqref{eq:impact_parameter_b_DEF} at the light-ring crossing (right). 
The results are relative to the bounded orbit simulations of Table~\ref{tab:sims_ecc} and the dynamical captures in Table~\ref{tab:sims_enc_scatt}.
\label{fig:Atail_vs_esep-bmerg}}
\end{figure*}
In Fig.~\ref{fig:Atail_vs_esep-bmerg}, we show the amplitude of the $(22)$ mode at the time in which the tail starts to dominate over the QNMs, $A_{\rm tail}$, as function of the eccentricity at the separatrix $e_{\rm sep}$ and as function of the impact parameter Eq.~\eqref{eq:impact_parameter_b_DEF} at the light-ring crossing $b_{LR}$ rescaled with respect to its value for a quasi-circular plunge.
These plots suggests $e_{\rm sep}$ to be more suited in order to parametrize $A_{\rm tail}$, while there is a double-valued behavior of $A_{\rm tail}$ in $b_{LR}$, exhibiting two branches, one for bounded orbits and one for dynamical captures.

\clearpage

\bibliography{bibliography}

\begin{thebibliography}{39}%
\makeatletter
\providecommand \@ifxundefined [1]{%
 \@ifx{#1\undefined}
}%
\providecommand \@ifnum [1]{%
 \ifnum #1\expandafter \@firstoftwo
 \else \expandafter \@secondoftwo
 \fi
}%
\providecommand \@ifx [1]{%
 \ifx #1\expandafter \@firstoftwo
 \else \expandafter \@secondoftwo
 \fi
}%
\providecommand \natexlab [1]{#1}%
\providecommand \enquote  [1]{``#1''}%
\providecommand \bibnamefont  [1]{#1}%
\providecommand \bibfnamefont [1]{#1}%
\providecommand \citenamefont [1]{#1}%
\providecommand \href@noop [0]{\@secondoftwo}%
\providecommand \href [0]{\begingroup \@sanitize@url \@href}%
\providecommand \@href[1]{\@@startlink{#1}\@@href}%
\providecommand \@@href[1]{\endgroup#1\@@endlink}%
\providecommand \@sanitize@url [0]{\catcode `\\12\catcode `\$12\catcode `\&12\catcode `\#12\catcode `\^12\catcode `\_12\catcode `\%12\relax}%
\providecommand \@@startlink[1]{}%
\providecommand \@@endlink[0]{}%
\providecommand \url  [0]{\begingroup\@sanitize@url \@url }%
\providecommand \@url [1]{\endgroup\@href {#1}{\urlprefix }}%
\providecommand \urlprefix  [0]{URL }%
\providecommand \Eprint [0]{\href }%
\providecommand \doibase [0]{https://doi.org/}%
\providecommand \selectlanguage [0]{\@gobble}%
\providecommand \bibinfo  [0]{\@secondoftwo}%
\providecommand \bibfield  [0]{\@secondoftwo}%
\providecommand \translation [1]{[#1]}%
\providecommand \BibitemOpen [0]{}%
\providecommand \bibitemStop [0]{}%
\providecommand \bibitemNoStop [0]{.\EOS\space}%
\providecommand \EOS [0]{\spacefactor3000\relax}%
\providecommand \BibitemShut  [1]{\csname bibitem#1\endcsname}%
\let\auto@bib@innerbib\@empty
\bibitem [{\citenamefont {Abbott}\ \emph {et~al.}(2016)\citenamefont {Abbott} \emph {et~al.}}]{LIGOScientific:2016aoc}%
  \BibitemOpen
  \bibfield  {author} {\bibinfo {author} {\bibfnamefont {B.~P.}\ \bibnamefont {Abbott}} \emph {et~al.} (\bibinfo {collaboration} {LIGO Scientific, Virgo}),\ }\href {https://doi.org/10.1103/PhysRevLett.116.061102} {\bibfield  {journal} {\bibinfo  {journal} {Phys. Rev. Lett.}\ }\textbf {\bibinfo {volume} {116}},\ \bibinfo {pages} {061102} (\bibinfo {year} {2016})},\ \Eprint {https://arxiv.org/abs/1602.03837} {arXiv:1602.03837 [gr-qc]} \BibitemShut {NoStop}%
\bibitem [{\citenamefont {Price}(1972)}]{Price:1971fb}%
  \BibitemOpen
  \bibfield  {author} {\bibinfo {author} {\bibfnamefont {R.~H.}\ \bibnamefont {Price}},\ }\href {https://doi.org/10.1103/PhysRevD.5.2419} {\bibfield  {journal} {\bibinfo  {journal} {Phys. Rev. D}\ }\textbf {\bibinfo {volume} {5}},\ \bibinfo {pages} {2419} (\bibinfo {year} {1972})}\BibitemShut {NoStop}%
\bibitem [{\citenamefont {Leaver}(1986)}]{Leaver:1986gd}%
  \BibitemOpen
  \bibfield  {author} {\bibinfo {author} {\bibfnamefont {E.~W.}\ \bibnamefont {Leaver}},\ }\href {https://doi.org/10.1103/PhysRevD.34.384} {\bibfield  {journal} {\bibinfo  {journal} {Phys. Rev. D}\ }\textbf {\bibinfo {volume} {34}},\ \bibinfo {pages} {384} (\bibinfo {year} {1986})}\BibitemShut {NoStop}%
\bibitem [{\citenamefont {Andersson}(1997)}]{Andersson:1996cm}%
  \BibitemOpen
  \bibfield  {author} {\bibinfo {author} {\bibfnamefont {N.}~\bibnamefont {Andersson}},\ }\href {https://doi.org/10.1103/PhysRevD.55.468} {\bibfield  {journal} {\bibinfo  {journal} {Phys. Rev. D}\ }\textbf {\bibinfo {volume} {55}},\ \bibinfo {pages} {468} (\bibinfo {year} {1997})},\ \Eprint {https://arxiv.org/abs/gr-qc/9607064} {arXiv:gr-qc/9607064} \BibitemShut {NoStop}%
\bibitem [{\citenamefont {Zenginoglu}(2010)}]{Zenginoglu:2009ey}%
  \BibitemOpen
  \bibfield  {author} {\bibinfo {author} {\bibfnamefont {A.}~\bibnamefont {Zenginoglu}},\ }\href {https://doi.org/10.1088/0264-9381/27/4/045015} {\bibfield  {journal} {\bibinfo  {journal} {Class. Quant. Grav.}\ }\textbf {\bibinfo {volume} {27}},\ \bibinfo {pages} {045015} (\bibinfo {year} {2010})},\ \Eprint {https://arxiv.org/abs/0911.2450} {arXiv:0911.2450 [gr-qc]} \BibitemShut {NoStop}%
\bibitem [{\citenamefont {Carullo}\ and\ \citenamefont {De~Amicis}(2023)}]{Carullo:2023tff}%
  \BibitemOpen
  \bibfield  {author} {\bibinfo {author} {\bibfnamefont {G.}~\bibnamefont {Carullo}}\ and\ \bibinfo {author} {\bibfnamefont {M.}~\bibnamefont {De~Amicis}},\ }\href@noop {} {\  (\bibinfo {year} {2023})},\ \Eprint {https://arxiv.org/abs/2310.12968} {arXiv:2310.12968 [gr-qc]} \BibitemShut {NoStop}%
\bibitem [{\citenamefont {Blanchet}\ and\ \citenamefont {Damour}(1992)}]{Blanchet:1992br}%
  \BibitemOpen
  \bibfield  {author} {\bibinfo {author} {\bibfnamefont {L.}~\bibnamefont {Blanchet}}\ and\ \bibinfo {author} {\bibfnamefont {T.}~\bibnamefont {Damour}},\ }\href {https://doi.org/10.1103/PhysRevD.46.4304} {\bibfield  {journal} {\bibinfo  {journal} {Phys. Rev. D}\ }\textbf {\bibinfo {volume} {46}},\ \bibinfo {pages} {4304} (\bibinfo {year} {1992})}\BibitemShut {NoStop}%
\bibitem [{\citenamefont {Blanchet}\ and\ \citenamefont {Schaefer}(1993)}]{Blanchet:1993ec}%
  \BibitemOpen
  \bibfield  {author} {\bibinfo {author} {\bibfnamefont {L.}~\bibnamefont {Blanchet}}\ and\ \bibinfo {author} {\bibfnamefont {G.}~\bibnamefont {Schaefer}},\ }\href {https://doi.org/10.1088/0264-9381/10/12/026} {\bibfield  {journal} {\bibinfo  {journal} {Class. Quant. Grav.}\ }\textbf {\bibinfo {volume} {10}},\ \bibinfo {pages} {2699} (\bibinfo {year} {1993})}\BibitemShut {NoStop}%
\bibitem [{\citenamefont {Blanchet}\ and\ \citenamefont {Sathyaprakash}(1995)}]{Blanchet:1994ez}%
  \BibitemOpen
  \bibfield  {author} {\bibinfo {author} {\bibfnamefont {L.}~\bibnamefont {Blanchet}}\ and\ \bibinfo {author} {\bibfnamefont {B.~S.}\ \bibnamefont {Sathyaprakash}},\ }\href {https://doi.org/10.1103/PhysRevLett.74.1067} {\bibfield  {journal} {\bibinfo  {journal} {Phys. Rev. Lett.}\ }\textbf {\bibinfo {volume} {74}},\ \bibinfo {pages} {1067} (\bibinfo {year} {1995})}\BibitemShut {NoStop}%
\bibitem [{\citenamefont {Poisson}(1993)}]{Poisson:1993vp}%
  \BibitemOpen
  \bibfield  {author} {\bibinfo {author} {\bibfnamefont {E.}~\bibnamefont {Poisson}},\ }\href {https://doi.org/10.1103/PhysRevD.47.1497} {\bibfield  {journal} {\bibinfo  {journal} {Phys. Rev. D}\ }\textbf {\bibinfo {volume} {47}},\ \bibinfo {pages} {1497} (\bibinfo {year} {1993})}\BibitemShut {NoStop}%
\bibitem [{\citenamefont {Poisson}\ and\ \citenamefont {Sasaki}(1995)}]{Poisson:1994yf}%
  \BibitemOpen
  \bibfield  {author} {\bibinfo {author} {\bibfnamefont {E.}~\bibnamefont {Poisson}}\ and\ \bibinfo {author} {\bibfnamefont {M.}~\bibnamefont {Sasaki}},\ }\href {https://doi.org/10.1103/PhysRevD.51.5753} {\bibfield  {journal} {\bibinfo  {journal} {Phys. Rev. D}\ }\textbf {\bibinfo {volume} {51}},\ \bibinfo {pages} {5753} (\bibinfo {year} {1995})},\ \Eprint {https://arxiv.org/abs/gr-qc/9412027} {arXiv:gr-qc/9412027} \BibitemShut {NoStop}%
\bibitem [{\citenamefont {Albanesi}\ \emph {et~al.}(2023)\citenamefont {Albanesi}, \citenamefont {Bernuzzi}, \citenamefont {Damour}, \citenamefont {Nagar},\ and\ \citenamefont {Placidi}}]{Albanesi:2023bgi}%
  \BibitemOpen
  \bibfield  {author} {\bibinfo {author} {\bibfnamefont {S.}~\bibnamefont {Albanesi}}, \bibinfo {author} {\bibfnamefont {S.}~\bibnamefont {Bernuzzi}}, \bibinfo {author} {\bibfnamefont {T.}~\bibnamefont {Damour}}, \bibinfo {author} {\bibfnamefont {A.}~\bibnamefont {Nagar}},\ and\ \bibinfo {author} {\bibfnamefont {A.}~\bibnamefont {Placidi}},\ }\href {https://doi.org/10.1103/PhysRevD.108.084037} {\bibfield  {journal} {\bibinfo  {journal} {Phys. Rev. D}\ }\textbf {\bibinfo {volume} {108}},\ \bibinfo {pages} {084037} (\bibinfo {year} {2023})},\ \Eprint {https://arxiv.org/abs/2305.19336} {arXiv:2305.19336 [gr-qc]} \BibitemShut {NoStop}%
\bibitem [{\citenamefont {Chiaramello}\ and\ \citenamefont {Nagar}(2020)}]{Chiaramello:2020ehz}%
  \BibitemOpen
  \bibfield  {author} {\bibinfo {author} {\bibfnamefont {D.}~\bibnamefont {Chiaramello}}\ and\ \bibinfo {author} {\bibfnamefont {A.}~\bibnamefont {Nagar}},\ }\href {https://doi.org/10.1103/PhysRevD.101.101501} {\bibfield  {journal} {\bibinfo  {journal} {Phys. Rev. D}\ }\textbf {\bibinfo {volume} {101}},\ \bibinfo {pages} {101501} (\bibinfo {year} {2020})},\ \Eprint {https://arxiv.org/abs/2001.11736} {arXiv:2001.11736 [gr-qc]} \BibitemShut {NoStop}%
\bibitem [{\citenamefont {Nagar}\ and\ \citenamefont {Rezzolla}(2005)}]{Nagar:2005ea}%
  \BibitemOpen
  \bibfield  {author} {\bibinfo {author} {\bibfnamefont {A.}~\bibnamefont {Nagar}}\ and\ \bibinfo {author} {\bibfnamefont {L.}~\bibnamefont {Rezzolla}},\ }\href {https://doi.org/10.1088/0264-9381/22/16/R01} {\bibfield  {journal} {\bibinfo  {journal} {Class. Quant. Grav.}\ }\textbf {\bibinfo {volume} {22}},\ \bibinfo {pages} {R167} (\bibinfo {year} {2005})},\ \bibinfo {note} {[Erratum: Class.Quant.Grav. 23, 4297 (2006)]},\ \Eprint {https://arxiv.org/abs/gr-qc/0502064} {arXiv:gr-qc/0502064} \BibitemShut {NoStop}%
\bibitem [{\citenamefont {Zerilli}(1970)}]{Zerilli:1970se}%
  \BibitemOpen
  \bibfield  {author} {\bibinfo {author} {\bibfnamefont {F.~J.}\ \bibnamefont {Zerilli}},\ }\href {https://doi.org/10.1103/PhysRevLett.24.737} {\bibfield  {journal} {\bibinfo  {journal} {Phys. Rev. Lett.}\ }\textbf {\bibinfo {volume} {24}},\ \bibinfo {pages} {737} (\bibinfo {year} {1970})}\BibitemShut {NoStop}%
\bibitem [{\citenamefont {Nagar}\ \emph {et~al.}(2007)\citenamefont {Nagar}, \citenamefont {Damour},\ and\ \citenamefont {Tartaglia}}]{Nagar:2006xv}%
  \BibitemOpen
  \bibfield  {author} {\bibinfo {author} {\bibfnamefont {A.}~\bibnamefont {Nagar}}, \bibinfo {author} {\bibfnamefont {T.}~\bibnamefont {Damour}},\ and\ \bibinfo {author} {\bibfnamefont {A.}~\bibnamefont {Tartaglia}},\ }\href {https://doi.org/10.1088/0264-9381/24/12/S08} {\bibfield  {journal} {\bibinfo  {journal} {Class. Quant. Grav.}\ }\textbf {\bibinfo {volume} {24}},\ \bibinfo {pages} {S109} (\bibinfo {year} {2007})},\ \Eprint {https://arxiv.org/abs/gr-qc/0612096} {arXiv:gr-qc/0612096} \BibitemShut {NoStop}%
\bibitem [{\citenamefont {Albanesi}\ \emph {et~al.}(2021)\citenamefont {Albanesi}, \citenamefont {Nagar},\ and\ \citenamefont {Bernuzzi}}]{Albanesi:2021rby}%
  \BibitemOpen
  \bibfield  {author} {\bibinfo {author} {\bibfnamefont {S.}~\bibnamefont {Albanesi}}, \bibinfo {author} {\bibfnamefont {A.}~\bibnamefont {Nagar}},\ and\ \bibinfo {author} {\bibfnamefont {S.}~\bibnamefont {Bernuzzi}},\ }\href {https://doi.org/10.1103/PhysRevD.104.024067} {\bibfield  {journal} {\bibinfo  {journal} {Phys. Rev. D}\ }\textbf {\bibinfo {volume} {104}},\ \bibinfo {pages} {024067} (\bibinfo {year} {2021})},\ \Eprint {https://arxiv.org/abs/2104.10559} {arXiv:2104.10559 [gr-qc]} \BibitemShut {NoStop}%
\bibitem [{\citenamefont {Bernuzzi}\ and\ \citenamefont {Nagar}(2010)}]{Bernuzzi:2010ty}%
  \BibitemOpen
  \bibfield  {author} {\bibinfo {author} {\bibfnamefont {S.}~\bibnamefont {Bernuzzi}}\ and\ \bibinfo {author} {\bibfnamefont {A.}~\bibnamefont {Nagar}},\ }\href {https://doi.org/10.1103/PhysRevD.81.084056} {\bibfield  {journal} {\bibinfo  {journal} {Phys. Rev. D}\ }\textbf {\bibinfo {volume} {81}},\ \bibinfo {pages} {084056} (\bibinfo {year} {2010})},\ \Eprint {https://arxiv.org/abs/1003.0597} {arXiv:1003.0597 [gr-qc]} \BibitemShut {NoStop}%
\bibitem [{\citenamefont {Bernuzzi}\ \emph {et~al.}(2011)\citenamefont {Bernuzzi}, \citenamefont {Nagar},\ and\ \citenamefont {Zenginoglu}}]{Bernuzzi:2011aj}%
  \BibitemOpen
  \bibfield  {author} {\bibinfo {author} {\bibfnamefont {S.}~\bibnamefont {Bernuzzi}}, \bibinfo {author} {\bibfnamefont {A.}~\bibnamefont {Nagar}},\ and\ \bibinfo {author} {\bibfnamefont {A.}~\bibnamefont {Zenginoglu}},\ }\href {https://doi.org/10.1103/PhysRevD.84.084026} {\bibfield  {journal} {\bibinfo  {journal} {Phys. Rev. D}\ }\textbf {\bibinfo {volume} {84}},\ \bibinfo {pages} {084026} (\bibinfo {year} {2011})},\ \Eprint {https://arxiv.org/abs/1107.5402} {arXiv:1107.5402 [gr-qc]} \BibitemShut {NoStop}%
\bibitem [{\citenamefont {Carullo}\ \emph {et~al.}(2024)\citenamefont {Carullo}, \citenamefont {Albanesi}, \citenamefont {Nagar}, \citenamefont {Gamba}, \citenamefont {Bernuzzi}, \citenamefont {Andrade},\ and\ \citenamefont {Trenado}}]{Carullo:2023kvj}%
  \BibitemOpen
  \bibfield  {author} {\bibinfo {author} {\bibfnamefont {G.}~\bibnamefont {Carullo}}, \bibinfo {author} {\bibfnamefont {S.}~\bibnamefont {Albanesi}}, \bibinfo {author} {\bibfnamefont {A.}~\bibnamefont {Nagar}}, \bibinfo {author} {\bibfnamefont {R.}~\bibnamefont {Gamba}}, \bibinfo {author} {\bibfnamefont {S.}~\bibnamefont {Bernuzzi}}, \bibinfo {author} {\bibfnamefont {T.}~\bibnamefont {Andrade}},\ and\ \bibinfo {author} {\bibfnamefont {J.}~\bibnamefont {Trenado}},\ }\href {https://doi.org/10.1103/PhysRevLett.132.101401} {\bibfield  {journal} {\bibinfo  {journal} {Phys. Rev. Lett.}\ }\textbf {\bibinfo {volume} {132}},\ \bibinfo {pages} {101401} (\bibinfo {year} {2024})},\ \Eprint {https://arxiv.org/abs/2309.07228} {arXiv:2309.07228 [gr-qc]} \BibitemShut {NoStop}%
\bibitem [{\citenamefont {Nagar}\ \emph {et~al.}(2021)\citenamefont {Nagar}, \citenamefont {Rettegno}, \citenamefont {Gamba},\ and\ \citenamefont {Bernuzzi}}]{Nagar:2020xsk}%
  \BibitemOpen
  \bibfield  {author} {\bibinfo {author} {\bibfnamefont {A.}~\bibnamefont {Nagar}}, \bibinfo {author} {\bibfnamefont {P.}~\bibnamefont {Rettegno}}, \bibinfo {author} {\bibfnamefont {R.}~\bibnamefont {Gamba}},\ and\ \bibinfo {author} {\bibfnamefont {S.}~\bibnamefont {Bernuzzi}},\ }\href {https://doi.org/10.1103/PhysRevD.103.064013} {\bibfield  {journal} {\bibinfo  {journal} {Phys. Rev. D}\ }\textbf {\bibinfo {volume} {103}},\ \bibinfo {pages} {064013} (\bibinfo {year} {2021})},\ \Eprint {https://arxiv.org/abs/2009.12857} {arXiv:2009.12857 [gr-qc]} \BibitemShut {NoStop}%
\bibitem [{\citenamefont {Albanesi}\ \emph {et~al.}(2024)\citenamefont {Albanesi}, \citenamefont {Rashti}, \citenamefont {Zappa}, \citenamefont {Gamba}, \citenamefont {Cook}, \citenamefont {Daszuta}, \citenamefont {Bernuzzi}, \citenamefont {Nagar},\ and\ \citenamefont {Radice}}]{Albanesi:2024xus}%
  \BibitemOpen
  \bibfield  {author} {\bibinfo {author} {\bibfnamefont {S.}~\bibnamefont {Albanesi}}, \bibinfo {author} {\bibfnamefont {A.}~\bibnamefont {Rashti}}, \bibinfo {author} {\bibfnamefont {F.}~\bibnamefont {Zappa}}, \bibinfo {author} {\bibfnamefont {R.}~\bibnamefont {Gamba}}, \bibinfo {author} {\bibfnamefont {W.}~\bibnamefont {Cook}}, \bibinfo {author} {\bibfnamefont {B.}~\bibnamefont {Daszuta}}, \bibinfo {author} {\bibfnamefont {S.}~\bibnamefont {Bernuzzi}}, \bibinfo {author} {\bibfnamefont {A.}~\bibnamefont {Nagar}},\ and\ \bibinfo {author} {\bibfnamefont {D.}~\bibnamefont {Radice}},\ }\href@noop {} {\  (\bibinfo {year} {2024})},\ \Eprint {https://arxiv.org/abs/2405.20398} {arXiv:2405.20398 [gr-qc]} \BibitemShut {NoStop}%
\bibitem [{\citenamefont {Virtanen~\textit{et al} (Contributors)}(2020)}]{scipy}%
  \BibitemOpen
  \bibfield  {author} {\bibinfo {author} {\bibfnamefont {P.}~\bibnamefont {Virtanen~\textit{et al} (Contributors)}},\ }\href@noop {} {\bibfield  {journal} {\bibinfo  {journal} {Nature Methods}\ } (\bibinfo {year} {2020})}\BibitemShut {NoStop}%
\bibitem [{\citenamefont {Bernuzzi}\ \emph {et~al.}(2012)\citenamefont {Bernuzzi}, \citenamefont {Nagar},\ and\ \citenamefont {Zenginoglu}}]{Bernuzzi:2012ku}%
  \BibitemOpen
  \bibfield  {author} {\bibinfo {author} {\bibfnamefont {S.}~\bibnamefont {Bernuzzi}}, \bibinfo {author} {\bibfnamefont {A.}~\bibnamefont {Nagar}},\ and\ \bibinfo {author} {\bibfnamefont {A.}~\bibnamefont {Zenginoglu}},\ }\href {https://doi.org/10.1103/PhysRevD.86.104038} {\bibfield  {journal} {\bibinfo  {journal} {Phys. Rev. D}\ }\textbf {\bibinfo {volume} {86}},\ \bibinfo {pages} {104038} (\bibinfo {year} {2012})},\ \Eprint {https://arxiv.org/abs/1207.0769} {arXiv:1207.0769 [gr-qc]} \BibitemShut {NoStop}%
\bibitem [{\citenamefont {Placidi}\ \emph {et~al.}(2022)\citenamefont {Placidi}, \citenamefont {Albanesi}, \citenamefont {Nagar}, \citenamefont {Orselli}, \citenamefont {Bernuzzi},\ and\ \citenamefont {Grignani}}]{Placidi:2021rkh}%
  \BibitemOpen
  \bibfield  {author} {\bibinfo {author} {\bibfnamefont {A.}~\bibnamefont {Placidi}}, \bibinfo {author} {\bibfnamefont {S.}~\bibnamefont {Albanesi}}, \bibinfo {author} {\bibfnamefont {A.}~\bibnamefont {Nagar}}, \bibinfo {author} {\bibfnamefont {M.}~\bibnamefont {Orselli}}, \bibinfo {author} {\bibfnamefont {S.}~\bibnamefont {Bernuzzi}},\ and\ \bibinfo {author} {\bibfnamefont {G.}~\bibnamefont {Grignani}},\ }\href {https://doi.org/10.1103/PhysRevD.105.104030} {\bibfield  {journal} {\bibinfo  {journal} {Phys. Rev. D}\ }\textbf {\bibinfo {volume} {105}},\ \bibinfo {pages} {104030} (\bibinfo {year} {2022})},\ \Eprint {https://arxiv.org/abs/2112.05448} {arXiv:2112.05448 [gr-qc]} \BibitemShut {NoStop}%
\bibitem [{\citenamefont {Carullo}(2024)}]{Carullo:2024smg}%
  \BibitemOpen
  \bibfield  {author} {\bibinfo {author} {\bibfnamefont {G.}~\bibnamefont {Carullo}},\ }\href {https://doi.org/10.1088/1475-7516/2024/10/061} {\bibfield  {journal} {\bibinfo  {journal} {JCAP}\ }\textbf {\bibinfo {volume} {10}},\ \bibinfo {pages} {061}},\ \Eprint {https://arxiv.org/abs/2406.19442} {arXiv:2406.19442 [gr-qc]} \BibitemShut {NoStop}%
\bibitem [{\citenamefont {Saha}\ \emph {et~al.}(2020)\citenamefont {Saha}, \citenamefont {Sahoo},\ and\ \citenamefont {Sen}}]{Saha:2019tub}%
  \BibitemOpen
  \bibfield  {author} {\bibinfo {author} {\bibfnamefont {A.~P.}\ \bibnamefont {Saha}}, \bibinfo {author} {\bibfnamefont {B.}~\bibnamefont {Sahoo}},\ and\ \bibinfo {author} {\bibfnamefont {A.}~\bibnamefont {Sen}},\ }\href {https://doi.org/10.1007/JHEP06(2020)153} {\bibfield  {journal} {\bibinfo  {journal} {JHEP}\ }\textbf {\bibinfo {volume} {06}},\ \bibinfo {pages} {153}},\ \Eprint {https://arxiv.org/abs/1912.06413} {arXiv:1912.06413 [hep-th]} \BibitemShut {NoStop}%
\bibitem [{\citenamefont {Sahoo}\ and\ \citenamefont {Sen}(2022)}]{Sahoo:2021ctw}%
  \BibitemOpen
  \bibfield  {author} {\bibinfo {author} {\bibfnamefont {B.}~\bibnamefont {Sahoo}}\ and\ \bibinfo {author} {\bibfnamefont {A.}~\bibnamefont {Sen}},\ }\href {https://doi.org/10.1007/JHEP01(2022)077} {\bibfield  {journal} {\bibinfo  {journal} {JHEP}\ }\textbf {\bibinfo {volume} {01}},\ \bibinfo {pages} {077}},\ \Eprint {https://arxiv.org/abs/2105.08739} {arXiv:2105.08739 [hep-th]} \BibitemShut {NoStop}%
\bibitem [{\citenamefont {Bishop}\ \emph {et~al.}(1996)\citenamefont {Bishop}, \citenamefont {Gomez}, \citenamefont {Lehner},\ and\ \citenamefont {Winicour}}]{Bishop:1996gt}%
  \BibitemOpen
  \bibfield  {author} {\bibinfo {author} {\bibfnamefont {N.~T.}\ \bibnamefont {Bishop}}, \bibinfo {author} {\bibfnamefont {R.}~\bibnamefont {Gomez}}, \bibinfo {author} {\bibfnamefont {L.}~\bibnamefont {Lehner}},\ and\ \bibinfo {author} {\bibfnamefont {J.}~\bibnamefont {Winicour}},\ }\href {https://doi.org/10.1103/PhysRevD.54.6153} {\bibfield  {journal} {\bibinfo  {journal} {Phys. Rev. D}\ }\textbf {\bibinfo {volume} {54}},\ \bibinfo {pages} {6153} (\bibinfo {year} {1996})},\ \Eprint {https://arxiv.org/abs/gr-qc/9705033} {arXiv:gr-qc/9705033} \BibitemShut {NoStop}%
\bibitem [{\citenamefont {Wardell}\ \emph {et~al.}(2023)\citenamefont {Wardell}, \citenamefont {Pound}, \citenamefont {Warburton}, \citenamefont {Miller}, \citenamefont {Durkan},\ and\ \citenamefont {Le~Tiec}}]{Wardell:2021fyy}%
  \BibitemOpen
  \bibfield  {author} {\bibinfo {author} {\bibfnamefont {B.}~\bibnamefont {Wardell}}, \bibinfo {author} {\bibfnamefont {A.}~\bibnamefont {Pound}}, \bibinfo {author} {\bibfnamefont {N.}~\bibnamefont {Warburton}}, \bibinfo {author} {\bibfnamefont {J.}~\bibnamefont {Miller}}, \bibinfo {author} {\bibfnamefont {L.}~\bibnamefont {Durkan}},\ and\ \bibinfo {author} {\bibfnamefont {A.}~\bibnamefont {Le~Tiec}},\ }\href {https://doi.org/10.1103/PhysRevLett.130.241402} {\bibfield  {journal} {\bibinfo  {journal} {Phys. Rev. Lett.}\ }\textbf {\bibinfo {volume} {130}},\ \bibinfo {pages} {241402} (\bibinfo {year} {2023})},\ \Eprint {https://arxiv.org/abs/2112.12265} {arXiv:2112.12265 [gr-qc]} \BibitemShut {NoStop}%
\bibitem [{\citenamefont {Islam}\ and\ \citenamefont {Khanna}(2023)}]{Islam:2023aec}%
  \BibitemOpen
  \bibfield  {author} {\bibinfo {author} {\bibfnamefont {T.}~\bibnamefont {Islam}}\ and\ \bibinfo {author} {\bibfnamefont {G.}~\bibnamefont {Khanna}},\ }\href {https://doi.org/10.1103/PhysRevD.108.044012} {\bibfield  {journal} {\bibinfo  {journal} {Phys. Rev. D}\ }\textbf {\bibinfo {volume} {108}},\ \bibinfo {pages} {044012} (\bibinfo {year} {2023})},\ \Eprint {https://arxiv.org/abs/2306.08767} {arXiv:2306.08767 [gr-qc]} \BibitemShut {NoStop}%
\bibitem [{\citenamefont {Nagar}\ \emph {et~al.}(2022)\citenamefont {Nagar}, \citenamefont {Healy}, \citenamefont {Lousto}, \citenamefont {Bernuzzi},\ and\ \citenamefont {Albertini}}]{Nagar:2022icd}%
  \BibitemOpen
  \bibfield  {author} {\bibinfo {author} {\bibfnamefont {A.}~\bibnamefont {Nagar}}, \bibinfo {author} {\bibfnamefont {J.}~\bibnamefont {Healy}}, \bibinfo {author} {\bibfnamefont {C.~O.}\ \bibnamefont {Lousto}}, \bibinfo {author} {\bibfnamefont {S.}~\bibnamefont {Bernuzzi}},\ and\ \bibinfo {author} {\bibfnamefont {A.}~\bibnamefont {Albertini}},\ }\href {https://doi.org/10.1103/PhysRevD.105.124061} {\bibfield  {journal} {\bibinfo  {journal} {Phys. Rev. D}\ }\textbf {\bibinfo {volume} {105}},\ \bibinfo {pages} {124061} (\bibinfo {year} {2022})},\ \Eprint {https://arxiv.org/abs/2202.05643} {arXiv:2202.05643 [gr-qc]} \BibitemShut {NoStop}%
\bibitem [{\citenamefont {Cardoso}\ \emph {et~al.}(2024)\citenamefont {Cardoso}, \citenamefont {Carullo}, \citenamefont {De~Amicis}, \citenamefont {Duque}, \citenamefont {Katagiri}, \citenamefont {Pereniguez}, \citenamefont {Redondo-Yuste}, \citenamefont {Spieksma},\ and\ \citenamefont {Zhong}}]{Cardoso:2024jme}%
  \BibitemOpen
  \bibfield  {author} {\bibinfo {author} {\bibfnamefont {V.}~\bibnamefont {Cardoso}}, \bibinfo {author} {\bibfnamefont {G.}~\bibnamefont {Carullo}}, \bibinfo {author} {\bibfnamefont {M.}~\bibnamefont {De~Amicis}}, \bibinfo {author} {\bibfnamefont {F.}~\bibnamefont {Duque}}, \bibinfo {author} {\bibfnamefont {T.}~\bibnamefont {Katagiri}}, \bibinfo {author} {\bibfnamefont {D.}~\bibnamefont {Pereniguez}}, \bibinfo {author} {\bibfnamefont {J.}~\bibnamefont {Redondo-Yuste}}, \bibinfo {author} {\bibfnamefont {T.~F.~M.}\ \bibnamefont {Spieksma}},\ and\ \bibinfo {author} {\bibfnamefont {Z.}~\bibnamefont {Zhong}},\ }\href {https://doi.org/10.1103/PhysRevD.109.L121502} {\bibfield  {journal} {\bibinfo  {journal} {Phys. Rev. D}\ }\textbf {\bibinfo {volume} {109}},\ \bibinfo {pages} {L121502} (\bibinfo {year} {2024})},\ \Eprint {https://arxiv.org/abs/2405.12290} {arXiv:2405.12290 [gr-qc]} \BibitemShut {NoStop}%
\bibitem [{\citenamefont {Hod}(2000)}]{Hod:1999rx}%
  \BibitemOpen
  \bibfield  {author} {\bibinfo {author} {\bibfnamefont {S.}~\bibnamefont {Hod}},\ }\href {https://doi.org/10.1103/PhysRevD.61.024033} {\bibfield  {journal} {\bibinfo  {journal} {Phys. Rev. D}\ }\textbf {\bibinfo {volume} {61}},\ \bibinfo {pages} {024033} (\bibinfo {year} {2000})},\ \Eprint {https://arxiv.org/abs/gr-qc/9902072} {arXiv:gr-qc/9902072} \BibitemShut {NoStop}%
\bibitem [{\citenamefont {Choi}\ \emph {et~al.}(2024)\citenamefont {Choi}, \citenamefont {Laddha},\ and\ \citenamefont {Puhm}}]{Choi:2024ygx}%
  \BibitemOpen
  \bibfield  {author} {\bibinfo {author} {\bibfnamefont {S.}~\bibnamefont {Choi}}, \bibinfo {author} {\bibfnamefont {A.}~\bibnamefont {Laddha}},\ and\ \bibinfo {author} {\bibfnamefont {A.}~\bibnamefont {Puhm}},\ }\href@noop {} {\  (\bibinfo {year} {2024})},\ \Eprint {https://arxiv.org/abs/2403.13053} {arXiv:2403.13053 [hep-th]} \BibitemShut {NoStop}%
\bibitem [{\citenamefont {Islam}\ \emph {et~al.}(2024)\citenamefont {Islam}, \citenamefont {Faggioli}, \citenamefont {Khanna}, \citenamefont {Field}, \citenamefont {van~de Meent},\ and\ \citenamefont {Buonanno}}]{Islam:2024vro}%
  \BibitemOpen
  \bibfield  {author} {\bibinfo {author} {\bibfnamefont {T.}~\bibnamefont {Islam}}, \bibinfo {author} {\bibfnamefont {G.}~\bibnamefont {Faggioli}}, \bibinfo {author} {\bibfnamefont {G.}~\bibnamefont {Khanna}}, \bibinfo {author} {\bibfnamefont {S.~E.}\ \bibnamefont {Field}}, \bibinfo {author} {\bibfnamefont {M.}~\bibnamefont {van~de Meent}},\ and\ \bibinfo {author} {\bibfnamefont {A.}~\bibnamefont {Buonanno}},\ }\href@noop {} {\  (\bibinfo {year} {2024})},\ \Eprint {https://arxiv.org/abs/2407.04682} {arXiv:2407.04682 [gr-qc]} \BibitemShut {NoStop}%
\bibitem [{\citenamefont {Asada}\ and\ \citenamefont {Futamase}(1997)}]{Asada:1997zu}%
  \BibitemOpen
  \bibfield  {author} {\bibinfo {author} {\bibfnamefont {H.}~\bibnamefont {Asada}}\ and\ \bibinfo {author} {\bibfnamefont {T.}~\bibnamefont {Futamase}},\ }\href {https://doi.org/10.1103/PhysRevD.56.R6062} {\bibfield  {journal} {\bibinfo  {journal} {Phys. Rev. D}\ }\textbf {\bibinfo {volume} {56}},\ \bibinfo {pages} {R6062} (\bibinfo {year} {1997})},\ \Eprint {https://arxiv.org/abs/gr-qc/9711009} {arXiv:gr-qc/9711009} \BibitemShut {NoStop}%
\bibitem [{\citenamefont {Abramowitz}\ and\ \citenamefont {Stegun}(1968)}]{abramowitz1968handbook}%
  \BibitemOpen
  \bibfield  {author} {\bibinfo {author} {\bibfnamefont {M.}~\bibnamefont {Abramowitz}}\ and\ \bibinfo {author} {\bibfnamefont {I.~A.}\ \bibnamefont {Stegun}},\ }\href@noop {} {\emph {\bibinfo {title} {Handbook of mathematical functions}}},\ Vol.~\bibinfo {volume} {55}\ (\bibinfo  {publisher} {US Government printing office},\ \bibinfo {year} {1968})\BibitemShut {NoStop}%
\bibitem [{\citenamefont {Mitman}\ \emph {et~al.}(2024)\citenamefont {Mitman} \emph {et~al.}}]{Mitman:2024uss}%
  \BibitemOpen
  \bibfield  {author} {\bibinfo {author} {\bibfnamefont {K.}~\bibnamefont {Mitman}} \emph {et~al.},\ }\href@noop {} {\  (\bibinfo {year} {2024})},\ \Eprint {https://arxiv.org/abs/2405.08868} {arXiv:2405.08868 [gr-qc]} \BibitemShut {NoStop}%
\end{thebibliography}%

\end{document}